\documentclass[12pt,a4paper]{article}
\pdfoutput=1
\usepackage[utf8]{inputenc}
\usepackage[T1]{fontenc}
\usepackage{mathrsfs}
\usepackage{style}
\usepackage[english]{babel}
\usepackage{url}
\usepackage{footnote}

\usepackage{amsmath}
\usepackage{amssymb}
\usepackage{amsthm}
\usepackage{psfrag}
\usepackage{graphicx}
\usepackage{hyperref}
\usepackage{feynmp}

\newcommand{\be}{\begin{equation}}
\newcommand{\ee}{\end{equation}}
\newcommand{\beqa}{\begin{eqnarray}}
\newcommand{\eeqa}{\end{eqnarray}}

\newcommand\m{\mu}

\renewcommand\k{\bf k}
\newcommand\x{\textbf x}

\newcommand\g{\gamma}

\newcommand\n{\nu}
\renewcommand\r{\rho}
\newcommand\s{\sigma}

\renewcommand\a{\alpha}
\renewcommand\b{\beta}

\renewcommand\l{\lambda}

\def\d{\partial}
\newcommand{\bseq}{\begin{subequations}}
\newcommand{\eseq}{\end{subequations}}

\renewcommand{\ln}{\mathop{\rm ln}\nolimits}

\title{
On the Vanishing of Love Numbers
for Kerr Black Holes
}

\author[a]{Panagiotis Charalambous\footnote{\texttt{pc2560@nyu.edu}}}
\author[a]{Sergei Dubovsky\footnote{\texttt{dubovsky@nyu.edu}}}
\author[a,b]{Mikhail M. Ivanov\footnote{\texttt{mi1271@nyu.edu}}}

\affiliation[a]{Center for Cosmology and Particle Physics, Department of Physics,
New York University,\\
New York, NY 10003, USA}
\affiliation[b]{Institute for Nuclear Research of the
Russian Academy of Sciences, \\ 
\normalsize \it  60th October Anniversary Prospect, 7a, 117312
Moscow, Russia}

\abstract{
It was shown recently that the static 
 tidal response coefficients, called Love numbers, 
 vanish identically for Kerr black holes in four dimensions.
In this work, we confirm this result and extend it to the case of 
spin-0 and spin-1 perturbations. 
We compute the static response of Kerr black holes 
to scalar, electromagnetic, and gravitational fields
at all orders in black hole spin.
We use the unambiguous and gauge-invariant definition of Love numbers 
and their spin-0 and spin-1 analogs
as Wilson coefficients of the 
point particle 
effective field theory.
This definition also allows one to clearly distinguish between conservative 
and dissipative response contributions.
We demonstrate that the behavior of Kerr black holes responses
to spin-0 and spin-1 fields
is very similar to that of 
the spin-2 perturbations. 
In particular, static 
conservative responses 
vanish identically 
for spinning black holes.
This implies that
vanishing Love numbers 
are a generic property of black holes
in four-dimensional general relativity.
We also show that the dissipative part of the 
response does not vanish
even for static perturbations 
due to frame-dragging.
}

\begin{document}

\begin{flushright}
	INR-TH-2021-001
\end{flushright}

\vspace{-1.5cm}

\maketitle
\flushbottom

\section{Introduction}

The static response of black holes to external perturbations, 
captured by the 
so-called Love numbers\footnote{The Love numbers are named after the mathematician A. E. H. Love who introduced them to 
describe the tidal deformation of the Earth in Ref.~\cite{1909RSPSA..82...73L}.}, has recently attracted significant attention both from
the observational and theoretical sides. 
On the one hand, black holes' Love numbers are measurable quantities
that can be probed with gravitational wave observations~\cite{Flanagan:2007ix,Cardoso:2017cfl}. 
On the other hand, 
they play an important role in 
the effective field theory (EFT) of binary inspirals~\cite{Goldberger:2004jt,Porto:2016pyg,Porto:2016zng,Hui:2020xxx}, where
they determine Wilson coefficients that
describe leading finite-size effects.

The tidal gravitational Love numbers of non-rotating Schwarzschild black holes
have been independently computed by Fang and Lovelace \cite{Fang:2005qq}, Damour and Nagar \cite{Damour:2009vw}, and by 
Binnington and Poisson~\cite{Binnington:2009bb}. 
Remarkably, they vanish in four dimensions in general relativity (GR), which poses a naturalness problem 
from the EFT point of view~\cite{Porto:2016zng}, and 
therefore might hint on the existence 
of a new symmetry of black holes.
Intriguingly, there exist several physical
examples where the black holes'
Love numbers do not vanish. 
In particular, the calculation of Love numbers has been extended to higher dimensions in Refs.~\cite{Kol:2011vg,Cardoso:2019vof},
which have shown that 
their identical vanishing for 
all multipoles
is a unique result taking pace only in
four dimensions. Recently this result has been 
generalized to the cases of spin-0, spin-1 and spin-2 perturbations of different parities in Ref.~\cite{Hui:2020xxx}: the static responses of
Schwarzschild black holes 
are generally non-zero for all these different types of perturbations, 
but accidentally they
vanish 
in four dimensions.
Moreover, black holes' Love numbers were found to be 
non-zero 
in certain modified gravity theories~\cite{Cardoso:2018ptl,Cardoso:2017cfl,Bernard:2019yfz}.

As of now, it has been firmly established that 
the Love numbers of all perturbing fields vanish in four dimensions for Schwarzshield
black holes~\cite{Fang:2005qq,Damour:2009vw,Binnington:2009bb,Kol:2011vg,Gurlebeck:2015xpa,Hui:2020xxx}.
However, the properties of spinning (Kerr) black holes~\cite{Kerr:1963ud} are 
still under debate.
Tidal deformations of 
slowly rotating 
black holes were studied in Refs.~\cite{Pani:2015hfa,Pani:2015nua,Landry:2015cva,Landry:2015zfa,Landry:2017piv,Poisson:2020mdi}, 
which have found that the Love numbers vanish 
for 
axisymmetric perturbations. 
Moreover, Landry and Poisson (2015)~\cite{Landry:2015zfa} have 
claimed that the Love numbers vanish 
for other types of perturbations 
at first order in black hole's spin. 
However, this result was recently questioned by Le Tiec and Casals~\cite{LeTiec:2020spy},
who argued 
that conclusions of Landry and Poisson (2015)  
might have been affected by an uncertainty introduced by the split of the gravitational potential
into the source and response parts. 
Similar concerns have been earlier raised  in the context 
of Schwarzschild black holes~\cite{Kol:2011vg,Gralla:2017djj}.
To avoid that ambiguity, Kol and Smolkin~\cite{Kol:2011vg}
have used an analytic continuation of the relevant
general relativity solutions 
into higher dimensions, which is effectively equivalent 
to promoting the orbital mode number (multipolar index)
$\ell$ to non-integer values.
Using a similar analytic continuation technique,
Le Tiec et al.~(2020)~\cite{LeTiec:2020spy,LeTiec:2020bos} have obtained non-vanishing static response coefficients and have
claimed that the Love numbers do not vanish for general 
spin-2 (tidal)
perturbations around Kerr black holes.

Recently, Chia~(2020)~\cite{Chia:2020yla} and Goldberger et al.~(2020)~\cite{Goldberger:2020fot} have pointed out 
that the ``Love numbers'' that Le Tiec et.al.~(2020)
have computed actually correspond to dissipative effects, whereas the 
conservative tidal response vanishes identically for spinning black holes. 
Analogous results have also appeared 
in Refs.~\cite{Poisson:2020mdi,Poisson:2020vap}.
All these works imply that the Love numbers defined in the classical sense
of conservative tidal deformability
are zero for Kerr black holes.

In this work, we compute analogs of the tidal Love numbers produced 
by \mbox{spin-0} and \mbox{spin-1} perturbations around Kerr black holes.
We will define Love numbers as 
Wilson coefficients of local operators in the worldline point-particle effective 
field theory. 
This will allow us to distinguish between the 
conservative response to external fields,
which is related to static Love numbers,
and the dissipative part of black hole's response.
The finite-size local EFT operators are expected to be present in the EFT 
on general grounds and 
they should have a tensorial structure 
dictated by the axial symmetry of the Kerr background. 
We introduce these couplings in the EFT
for static fields and demonstrate 
how the new tensorial Wilson coefficients 
are related to the response coefficients that we have extracted
from the solutions 
to linearized spin-0, spin-1, and spin-2
field perturbations in the Kerr background.
We show that the structure of these GR solutions is such that
the dissipative parts of scalar and electromagnetic responses do not vanish
for the Kerr black holes just like their spin-2 counterparts.
However, the EFT Wilson coefficients that capture the local (conservative) responses of spinning black holes
vanish for all bosonic perturbing fields.

On the technical side, 
we demonstrate that the analytic continuation procedure 
utilized in Refs.~\cite{LeTiec:2020bos,Chia:2020yla} allows one to avoid 
the uncertainty in the source/response split and obtain consistent 
gauge-independent results for response coefficients in the spin-0 and spin-1 cases. 
We also give an interpretation of this analytic 
continuation procedure in the 
EFT context.
We show that the subleading source corrections, which may overlap with the 
induced response contributions, 
are, in fact, produced by interactions 
between external fields and gravitational degrees
of freedom. 
This observation
allows one 
to unambiguously identify the black hole multipole
moment induced by external fields. 
Indeed, the graviton corrections to the source 
solution 
can be computed
order by order within the EFT. 
Thus, given a full GR solution, 
one can subtract 
the graviton interaction contributions from it
and hence
robustly extract the Love numbers.
This procedure is equivalent to using the analytic 
continuation $\ell \to \mathbb{R}$.

Our paper is structured as follows.
We start with a recap of the Newtonian response coefficients 
in Section~\ref{sec:prel}. Then we focus on the scalar 
response coefficients in Section~\ref{sec:spin0}, where we 
discuss in detail their calculation both in general 
relativity and in the point-particle EFT.
In Section~\ref{sec:spin1}
we compute static response of the Kerr black hole to the 
external electric field and match this result with the 
EFT calculation. We repeat the same procedure for 
the spin-2 (gravitational) perturbation
in Section~\ref{sec:spin2}. Finally, we recap 
the main general relativity calculations for all spins in Section~\ref{sec:master} and extend them to the case of non-static perturbations.
We discuss our main results and draw conclusions in Section~\ref{sec:concl}. Some additional material is presented in several appendices. Appendix~\ref{app:math} is a brief reference to key mathematical relations and conventions. In Appendix~\ref{app:NPmax} we give some details on the calculation of the 
Newman-Penrose-Maxwell scalars, 
which encapsulate the electromagnetic field
around the Kerr black hole and
which are required
to extract spin-1 response coefficients.
We explicitly compute the spin-1 magnetic response coefficients in Appendix~\ref{app:mag} --- they happen 
to identically coincide with the electric ones.
Finally, in Appendix~\ref{app:near} we comment on the 
validity of the response coefficients computed 
in the potential region approximation.

\textit{Conventions.} In what follows we will work with the metric with signature 
$(-,+,+,+)$;
Greek letters (e.g. $\mu,\nu$, etc.) will denote the 
spacetime indices; Latin letters from the middle of the  alphabet (e.g. $i,j$, etc.) will denote the spatial 3-dimensional indices;
Latin letter from the beginning of the alphabet (e.g. $a,b$, etc.) will run 
over the coordinates on the two-sphere $\mathbb{S}^2$. 
We will work in the $c=G=1$ units 
in most of the paper.

\section{Newtonian Tidal Response}
\label{sec:prel}

In this section we review the definition of tidal response
in Newtonian theory~\cite{PoissonWill2014,Binnington:2009bb,LeTiec:2020bos},
and discuss some important subtleties present for spinning bodies. 

\paragraph{Spherical bodies.}
Let us consider a non-rotating spherical body of mass $M$ and equilibrium radius $r_s$.
Now imagine that we adiabatically apply an external gravitational field $U_{\rm ext}$.
It is convenient to place the body in
the origin of the coordinate system. Then we
can  
characterize the external source in terms of the multipole moments, 
\be
\mathcal{E}_{L }(t)= -\frac{1}{(\ell-2)!}\d_{\langle i_1}...\d_{i_\ell\rangle} U_{\rm ext}\big|_{r=0}\,,
\ee
where $r$ is the distance from the origin, $\langle...\rangle$ denotes the symmetric trace-free part, 
and $L\equiv i_1...i_\ell$
is the multi-index.
The multipole moments $\mathcal{E}_{L }$ are 
 symmetric
trace-free tensors (STFs) that parametrically depend on time $t$.
In what follows we suppress this explicit parametric 
dependence of the tidal field $\mathcal{E}_L$ on time.
In response to the external field, the body
will deform and develop internal multipole moments $I_{L}$,
\be 
I_L \equiv \int_{\mathbb{R}^3}d^3x~\rho(\x)x^{\langle L\rangle}\,,
\ee 
where $\rho(\x)$ is perturbed body's density
and $x^L\equiv x^{i_1}...x^{i_\ell}$.
Summing up the two contributions, we find 
the following expression for the
total Newtonian gravitational potential
\be 
\label{eq:u1}
U=
\frac{M}{r}
-\sum_{\ell=2}
\left[\frac{(\ell-2)!}{\ell !} 
\mathcal{E}_{ L}
% x^{i_1}...x^{i_\ell} 
x^{L}
-
\frac{(2\ell-1)!!}{\ell !} 
\frac{I_{ L}
n^{L}
% n^{i_1}...n^{i_\ell}
}{r^{\ell+1}}
 \right]\,,
\ee
where $n^L\equiv n^{i_1}...n^{i_\ell}$
is the tensor product of unit direction vectors $n^i\equiv x^i/|\x|$,
and we have omitted the dipole moment $\ell=1$
since it corresponds to trivial center-of-mass translations.
At this point, it is convenient to switch to the spherical coordinates and use an expansion of the external 
source and induced multipole moments in terms of the 
spherical harmonics $Y_{\ell m}(\theta,\phi)$.
This way Eq.~\eqref{eq:u1} can be rewritten 
as follows:
\be 
\label{eq:UI}
U=
\frac{M}{r}
-\sum_{\ell=2}\sum_{m=-\ell}^\ell Y_{\ell m} 
\left[\frac{(\ell-2)!}{\ell !} \mathcal{E}_{\ell m}  r^\ell -
\frac{(2\ell-1)!!}{\ell !} 
\frac{I_{\ell m}}{r^{\ell+1}}
 \right]\,,
\ee
where the angular harmonic coefficients $\mathcal{E}_{\ell m}$
are related to the STF components via
\be
n^L\mathcal{E}_L=\sum_{m=-\ell}^\ell \mathcal{E}_{\ell m}Y_{\ell m}\,,\quad \mathcal{E}_{\ell m}=\mathcal{E}_L\oint_{\mathbb{S}^2}d\Omega~n^LY^{*}_{\ell m}\,,
\ee
and $d\Omega=\sin\theta d\theta d\phi$ denotes the measure 
on the two-sphere $\mathbb{S}^2$.

If the external gravitational field is adiabatic 
and weak, linear response theory dictates that the response multipoles should be 
proportional to the perturbing multipole moments~\cite{PoissonWill2014,Chakrabarti:2013xza,1978Icar...36..245P,1981A&A....99..126H,1973Ap&SS..23..459A},\footnote{Note that the coupling between different orbital and azimuthal 
modes is absent by virtue of linearity, i.e. weakness
of the external perturbations.}
\be
\label{eq:lin1}
\begin{split}
& I_L(t)=
\lambda_\ell \mathcal{E}_L (t)
- \nu_\ell \frac{d}{dt}\mathcal{E}_L (t) +... 
\approx \lambda_\ell \mathcal{E}_L(t-\tau') \,,
\quad 
\text{or}\quad \\
& I_{\ell m}= 
% \sum_{m}
% \lambda_{\ell m}
\lambda_{\ell }
\mathcal{E}_{\ell m}
- \nu_\ell \frac{d}{dt}\mathcal{E}_{\ell m}+... \,,
\end{split}
\ee
where we have restored the 
explicit time-dependence for clarity,
and ``$...$'' denotes non-linear corrections and  
contributions with more
time derivatives.
Here the coefficient $\lambda_{\ell }$ is referred to 
as the Newtonian tidal Love number, while $\nu_\ell$
is the dissipative response coefficient related to body's kinematic viscosity~\cite{PoissonWill2014}. 
% We have also introduced the dissipative .
This contribution captures
the fact that dissipation produces a time lag 
$\tau'$ between the 
external field and body's response~\cite{1973Ap&SS..23..459A}.\footnote{This is true for small adiabatic 
perturbations of the body's equilibrium configuration.
In the case of weak friction this description holds for an arbitrary
equation of state. Beyond the weak friction 
approximation it was formally
derived for a homogeneous incompressible body~\cite{1973Ap&SS..23..459A}.
}
Note that non-zero
viscosity triggers various dissipative effects, such as
heating of the body and the transfer of angular momentum
between the body and the source, known
as ``tidal torque,'' see e.g.~\cite{Thorne:1984mz}.
Using the frequency-space ansatz 
$\mathcal{E}_{L}\propto e^{-i\omega t}$,
Eq.~\eqref{eq:lin1}
can be written as
\be
\label{eq:lin1p5}
I_{\ell m} = 
\lambda_{\ell }
\mathcal{E}_{\ell m} + i\nu_\ell \omega \mathcal{E}_{\ell m}+... \,.
\ee
If the external tidal environment is static in body's rest frame, the viscosity contribution 
disappears.

\paragraph{Spinning bodies.}
If the test body is rotating, the definition of response 
coefficients is more intricate.
The rotating body will generally depart 
from spherical symmetry and hence it will have internal 
multipole moments even in the absence of an external perturbation.
For a moment, let us assume that the body's equilibrium configuration
can be approximated as a rigidly rotating sphere.
If the rotation is sufficiently slow, the linear response in the body's 
rotation frame takes the same form~\eqref{eq:lin1}
as for the non-rotating body~\cite{PoissonWill2014}.
However, an important 
effect appears when we switch to an inertial frame.
Let us focus on the leading frequency-dependent contribution 
$\propto \frac{d}{dt}\mathcal{E}$.
Because of rotation, the total time 
derivative in the body's rotation frame takes the following form
\be 
\label{eq:lin2p5}
\frac{d}{dt}\mathcal{E}_{i_1...i_\ell} = 
\frac{\d}{\d t}\mathcal{E}_{i_1...i_\ell}-\sum_{n=1}^\ell 
\Omega_{i_n j}\mathcal{E}_{i_1...i_{n-1}ji_{n+1}...i_\ell}
\equiv \frac{\d}{\d t}\mathcal{E}_{i_1...i_\ell}
+\kappa^{L'}_L \mathcal{E}_{L'}\,,
\ee
where $\frac{\d}{\d t}\mathcal{E}$ 
is the time derivative in a
fixed inertial frame.
Here we  introduced the angular velocity tensor $\Omega_{ij}=-\Omega_{ji}$ and defined 
\[
\kappa^{i_1...i_\ell}_{i'_1...i'_\ell}\equiv
 -\sum_{n=1}^\ell 
\Omega_{i_n i'_n}\delta^{i_1}_{i'_1}...\delta^{i_{n+1}}_{i'_{n+1}} 
\delta^{i_{n-1}}_{i'_{n-1}} ...\delta^{i_{\ell}}_{i'_{\ell}} \,.
% _{i_1...i_{n-1}ji_{n+1}...i_\ell}
\]
Note that the matrix $\kappa^{L'}_L $
is odd w.r.t. $L\leftrightarrow L'$. 
We see that in the case
of a spinning body the dissipative response contribution may not 
vanish
even if the external perturbation is purely static in a non-rotating frame, 
i.e. $\d_t \mathcal{E}_{L}=0$.
Physically, this can be 
interpreted as a result of frame-dragging, i.e. the fact that the static
sources are viewed as time-dependent by 
locally rotating observers~\cite{1972ApJ...178..347B}.\footnote{See also Ref.~\cite{Landry:2015snx}, which shows that a slowly rotating body affected by a stationary external field produces a dynamical response.
}
In this case, Eq.~\eqref{eq:lin1} takes the following form
\be 
\label{eq:lin2p6}
I_L=
\left(\lambda_\ell\delta_{LL'}
- \nu_\ell (\delta_{LL'} \d_t+
\kappa_{LL'}) \right)\mathcal{E}_{L'}\,,\quad \text{where}\quad 
\kappa_{LL'}=-\kappa_{L'L}\,.
\ee
We see that rotating bodies can generate an antisymmetric tensorial response even for static external perturbations. 
Switching to spherical harmonics and using the frequency-space ansatz $\mathcal{E}_L\propto e^{-i\omega t}$, we can recast Eq.~\eqref{eq:lin2p6} in the form similar to Eq.~\eqref{eq:lin1p5}\footnote{To get this one has to use that 
$\Omega_{ij}=\Omega \varepsilon_{ikj} \hat{z}_k$, ($\varepsilon_{ijk}$ is the Levi-Civita symbol, $\hat{z}_k=\delta^3_k$),
then
contract $\kappa_L^{L'}\mathcal{E}_{L'}$
with $n^{\langle L \rangle}$, expand $n^{\langle L \rangle}$ and $\mathcal{E}_{L'}$ over the STF tensor basis and use identities \eqref{eq:epsY} from Appendix~\ref{app:math}.}
\be 
\label{eq:ilm}
I_{\ell m} = 
\left(
\lambda_{\ell }+ i\nu_{\ell }(\omega - m\Omega) 
\right)
\mathcal{E}_{\ell m}  \,,
\ee
where $\Omega$ is body's angular velocity
and $m$ is the azimuthal (``magnetic'') harmonic number.
Note the appearance of the term $\omega-\Omega m$
is generic for rotating bodies~\cite{1971JETPL..14..180Z,1971ZhPmR..14..270Z}, it is reminiscent of 
superradiant scattering~\cite{Novikov:1989sz}.
Also note that this term 
% entering Eq.~\eqref{eq:ilm}
is clearly of non-conservative origin because it is odd
under the time reversal transformation $\omega \to -\omega$,
$m\to -m$.
If body's viscosity 
is not negligible, the dissipative contribution can survive
even if the external tidal environment if static, i.e. $\omega= 0$.
It is well known that 
$\nu_{\ell }\neq 0$ for
both static and
spinning black holes, which manifests itself
e.g. in graviton absorption and superradiance~\cite{Novikov:1989sz,Goldberger:2004jt,Goldberger:2005cd,Porto:2007qi,Porto:2016pyg}.
Hence, non-vanishing of the dissipative
response in Eq.~\eqref{eq:ilm}
for spinning black holes is to be expected.

The upshot of our discussion is that a rigidly
rotating spherically-symmetric body 
develops an antisymmetric tensorial response
to weak static external perturbations. 
These responses correspond to the imaginary part 
of harmonic-space response coefficients.
We will see that a similar picture also 
holds in a more general case when the body's equilibrium configuration
is not spherically-symmetric.
% In the following sections we will show 
% that this statement holds true in the relativistic picture,
% i.e. to all orders in spin.
In this situation it is convenient 
to use the following general ansatz for the static response
in a non-rotating frame~\cite{LeTiec:2020bos},
\be 
\label{eq:lambdalmdef}
I_L=\lambda_{LL'}\mathcal{E}_{L'}\,,\quad 
I_{\ell m} = 
\lambda_{\ell m} \mathcal{E}_{\ell m}\,.
\ee
Using the point-particle
EFT~\cite{Goldberger:2020fot},
it can be shown that
the part of $\lambda_{LL'}$
which is even under $L\leftrightarrow L'$ 
corresponds to conservative tidal deformations, whereas
the antisymmetric part of $\lambda_{LL'}$ captures
non-conservative effects such as tidal dissipation. This will 
be discussed in detail in Section~\ref{sec:spin0}.
Using the isomorphism between the STF tensors
and spherical harmonics, one can also 
separate conservative and dissipative 
responses 
at the level of relevant harmonic coefficients $I_{\ell m}$. 
In this case dissipative effects are encoded
in imaginary parts of 
harmonic-space
tidal response coefficients,
which map onto the antisymmetric w.r.t. $L\leftrightarrow L'$ response tensors, whereas the
real part of harmonic coefficients 
captures the tidal Love numbers and maps onto 
the response tensors that are even w.r.t. 
$L\leftrightarrow L'$.
This suggests that it is more
appropriate to call quantities $\lambda_{\ell m}$ 
defined in Eq.~\eqref{eq:lambdalmdef} ``tidal 
response coefficients'', and reserve the term ``Love numbers''
only for its conservative real part.

All in all, the aggregate potential produced by an external 
static perturbation \eqref{eq:UI}
takes the following form\footnote{Note that 
our response coefficients differ from those of 
Refs.~\cite{Binnington:2009bb,LeTiec:2020bos} by a factor of $2$.
}
\be 
\label{eq:NewU}
U_{\rm pert}=
-\frac{(\ell-2)!}{\ell !}\sum_{\ell=2}\sum_{m=-\ell}^\ell Y_{\ell m}\mathcal{E}_{\ell m}
r^\ell
\left[1+
k_{\ell m} \left(\frac{r}{r_s}\right)^{-2\ell-1}
 \right]\,,
\ee 
where ``pert'' means that we have subtracted the body's
internal multipole moments,\footnote{For black holes these multipole moments can be straightforwardly 
extracted from the Kerr metric, see e.g.~\cite{Thorne:1980ru}.}
$r_s$ is body's equilibrium radius 
(in our context this will be black hole's Schwarzschild radius $r_s=2M$)\footnote{This choice is only a matter of convention. We could have equally chosen this scale to be, e.g. the difference between the 
outer and inner horizons of Kerr black holes $r_+-r_-$. This would only lead to a trivial rescaling 
of response coefficients by a constant factor. }
and $k_{\ell m}$ are dimensionless tidal response coefficients in the Newtonian approximation, defined as
\be
\lambda_{\ell m}= - k_{\ell m} \frac{(\ell-2)!}{(2\ell-1)!!} r_s^{2\ell+1}\,.
\ee

\paragraph{Relativistic picture.}
So far our discussion has been entirely in the realm of the Newtonian approximation,
which is only a long-distance approximation to the full general relativity picture.
In this more general case, it is convenient to look at the temporal metric component $g_{00} = -1+2 h_{00}$ in body's local asymptotic 
rest-frame~\cite{Misner:1974qy,Thorne:1984mz},
which generalizes the Newtonian potential in general relativity.
In the case of perturbations around the Schwarzschild and Kerr black hole solutions, 
it can be written as~\cite{Yunes:2005ve,Binnington:2009bb,Landry:2015zfa,LeTiec:2020bos}\footnote{The fact that the gravitational potential 
of the Kerr metric reduces to the flat space expression \eqref{eq:NewU} follows from the fact that the Kerr spacetime 
is asymptotically flat.}
\be 
\begin{split}
\label{eq:Ufar}
h_{00}^{\rm pert}=
% \frac{M}{r}
-\frac{(\ell-2)!}{\ell !}\sum_{\ell=2}\sum_{m=-\ell}^\ell & Y_{\ell m}\mathcal{E}_{\ell m} 
r^\ell\Bigg[
% \left(\frac{r}{r_s}\right)^\ell
% \times
\left(1+a_1\frac{r_s}{r}+...\right)\\
&+k_{\ell m} 
\left(\frac{r}{r_s}\right)^{-2\ell-1}
\left(1+b_1\frac{r_s}{r}+...\right)
 \Bigg]\,,
 \end{split}
\ee 
where
$a_1$ and $b_1$ are some calculable spin-dependent
coefficients whose exact 
expressions are omitted for clarity. 
This expression asymptotes Eq.~\eqref{eq:NewU}
for
the Newtonian potential in the long-distance limit 
and hence, at first glance, provides us with a practical 
prescription to extract the tidal response coefficients from a given 
 gravitational potential produced by an external source in full general relativity.
Indeed, apparently,  
we only need to Taylor expand this potential at spatial infinity
and read off the coefficient in front of the $r^{-\ell-1}$ power
in the expansion. In what follows we will denote this procedure
by \textit{Newtonian matching}.
It is important to stress that 
the response coefficients that we have discussed so far, and which
can be extracted by means of the Newtonian matching,
are referred to as the electric type Love numbers~\cite{Binnington:2009bb}.
There also exist so-called magnetic-type response coefficients, which 
do not have counterparts in the Newtonian approximation~\cite{Binnington:2009bb,Landry:2015zfa}. 
We will discuss them in Section~\ref{sec:spin2}.

Newtonian matching for electric response coefficients
was justified 
in Refs.~\cite{Binnington:2009bb,Kol:2011vg},
which showed that it gives a gauge-independent 
result 
for the Schwarzschild background in four dimensions. 
However, from the expression \eqref{eq:Ufar}
we see that in the physical case $\ell \in \mathbb{N}/\{1\}$ there may be 
some ambiguity if the subleading corrections to the source 
appear to have the same power exponent as the response contribution, i.e. the source and response series overlap. 
We will see shortly that this ambiguity indeed takes place in
the case of the Kerr background when the perturbations
are studied in the advanced Kerr coordinates.
A similar ambiguity takes place for  
Love numbers of Schwarzschild black holes
in certain higher dimensions.
To avoid this ambiguity, Kol and Smolkin~\cite{Kol:2011vg} have suggested to 
use the analytic continuation into higher dimensions, where generically 
the overlap between the source and response series does not happen
\cite{Kol:2011vg}. Recently Le Tiec et al. (2020)~\cite{LeTiec:2020bos}
have applied a similar analytic continuation 
in four dimensions. To that end, it is enough to treat $\ell$
as a generic rational number $\ell\to \mathbb{R}$. 
In our work, we show that this approach indeed allows one to avoid
the ambiguity and moreover, is motivated from the EFT point of view.
We will also show that the EFT itself provides us with an unambiguous way to 
define the conservative part of the tidal response, i.e. the actual Love numbers.

Finally, a comment on the role of the no-hair theorems~\cite{Israel:1967wq,Carter:1971zc,Robinson:1975bv,Dubovsky:2007zi} is in order.
In our context these theorems essentially state that 
external perturbing fields cannot smoothly generate 
a non-vanishing static profile on top of the Kerr metric.
This means that when the external source is adiabatically turned on,
the generalized gravitational potential in 
Eq.~\eqref{eq:Ufar} should be uniquely defined 
by the source itself and body's response.
In other words, the no-hair theorems guarantee that 
black holes' response
is analytic in the vicinity of $\omega=0$, there is no gravitational
hysteresis (apart from a possible change of black hole's mass and spin) --- once the source
is turned off, the relevant solution 
will tend to the Kerr metric again.
Thus, it is only because of the no-hair theorems
that the decomposition~\eqref{eq:Ufar} 
is unique and the definition of the response coefficients
for black holes
is meaningful.

\section{Scalar Response Coefficients}
\label{sec:spin0}

In this section we will compute the response of a Kerr black hole 
to an external spin-0 perturbation.
In analogy with the tidal response, it will be referred to as a scalar response coefficient (SRC).
First, we will discuss the definition of SRCs in the Newtonian approximation
and in the point-particle EFT. 
Then we will compute the SRCs
by solving the scalar field equation of motion
(i.e. the spin-0 Teukolsky equation~\cite{Teukolsky:1972my,Teukolsky:1973ha,Press:1973zz,Teukolsky:1974yv})
in the Kerr background.
We will do so in two different coordinate systems and show
that the results agree 
only if we analytically continue 
the orbital number of relevant scalar perturbations
to non-integer values. We also justify this procedure
from the EFT point of view. Finally, we present an explicit
matching between the EFT and general 
relativity calculations. We will show that the 
dissipative response is captured by complex tidal
response coefficients, whereas  their real part
describes conservative deformations 
that are natural to associate with classic Love numbers.
The scalar response coefficients 
happen to be purely imaginary, and hence 
we can conclude that the scalar Love numbers
are identically zero for Kerr black holes.

\subsection{Definition from Newtonian Matching}

The response coefficients for a free scalar field can be defined in analogy 
with the Newtonian gravitation potential outside a generic
static body tidally deformed by a weak external gravitational field.
Indeed, the fully relativistic Klein-Gordon equation for
a static massless
scalar field $ \Phi$
reduces to the Poisson equation on very large scales.
Therefore,
in the asymptotic 
limit $r\to \infty$
the static scalar field 
takes the standard expression
of the Newtonian potential, 
which can be split into 
contributions from an external source and body's response
similarly to Eq.~\eqref{eq:NewU},
\be 
\label{eq:PhiU}
\Phi\Big|_{r\to \infty}=
% \frac{M}{r}
% -\frac{(\ell-2)!}{\ell !}
\sum_{\ell=1}\sum_{m=-\ell}^\ell Y_{\ell m}\mathcal{E}^{(0)}_{\ell m}  
%\left
% (\frac{r}{r_s}\right)^\ell
r^\ell
\left[1+
k^{(0)}_{\ell m} \left(\frac{r}{r_s}\right)^{-2\ell-1}
 \right]\,.
\ee 
Note that unlike the gravitation potential, 
whose multipole expansion can only start with $\ell=2$
by virtue of the equivalence principle, 
a generic scalar field is allowed to have 
a non-trivial dipole moment. For this reason we have omitted 
the normalization factors used in Eq.~\eqref{eq:NewU}.

It is known that the full solution to the static Klein-Gordon equation 
in the Schwarzschild and Kerr backgrounds
factorizes in the spherical coordinates, and hence it
can be written in the following form~\cite{Brill:1972xj,Kol:2011vg,1972ApJ...175..243P}
\be 
\label{eq:Phifar}
\Phi=
% \frac{M}{r}
% -\frac{(\ell-2)!}{\ell !}
\sum_{\ell=1}\sum_{m=-\ell}^\ell Y_{\ell m}
\mathcal{E}_{\ell m}^{(0)}
r^\ell\left[
% \left(\frac{r}{r_s}\right)^\ell
\left(1+a_1\frac{r_s}{r}+...\right)+k^{(0)}_{\ell m}
% r_s^{-\ell} 
\left(\frac{r}{r_s}\right)^{-2\ell-1}
\left(1+b_1\frac{r_s}{r}+...\right)
 \right]\,,
\ee 
where $a_1,b_1$ are some calculable constants, which 
depend on black hole's spin.
At face value, this gives us a tool to extract the SRCs
from the full GR solution: we just have to Taylor expand this solution 
in $r_s/r$ and read off the coefficient in front of the $r^{-\ell-1}$
term. However, at this point it is not clear if the Newtonian definition
of SLNs is unambiguous. 
This disadvantage can be avoided if the Love numbers are defined
within the point-particle effective field theory.

\subsection{Static Response in the EFT}

In what follows we will use the point particle effective field theory
that includes our test scalar field $\Phi$,
along with the long-wavelenght metric field $g_{\mu\nu}^L$,
and the position of the compact object $x^\mu$ (see~\cite{Porto:2016pyg,Goldberger:2004jt,Hui:2020xxx}) for
further details). Our goal here is to focus on three particular 
aspects: the EFT definition of Love numbers, a possible ambiguity 
between the source and response contributions that may appear during the 
matching of the EFT and microscopic (GR) calculations,
and the EFT extension in the presence of spin.

The core idea of the point particle EFT is that any object
acts like a point particle when viewed from large enough scales. As we get closer to that object, or as 
measurement
precision becomes better, corrections to the point particle
description become important.
These finite size effects are captured by  
higher derivative operators in the context of EFT.

We will start with the non-spinning case, which is sufficient
for our purpose to define the Love numbers in a way free of 
the arbitrariness produced by
the response/source split. 
We will re-introduce the Planck mass $M_P$
in order to keep track of mass dimensions of the  
EFT operators. 
Let us write the following action for a static scalar field coupled to gravity
\be
\begin{split}
S=S^{(2)}_{\Phi} +S^{(2)}_{h}+S_{\Phi h}+S_{\rm pp}+S_{\rm finite-size}\,,
\end{split} 
\ee
where $S^{(2)}_{\Phi}$ and $S^{(2)}_{h}$
are the bulk quadratic terms for the scalar field
and gravitation, $S_{\Phi h}$ describes
the leading interaction between them,
$S_{\rm pp}$ is the point particle worldline action,
and the part 
$S_{\rm finite-size}$ captures finite-size 
effects. Let us describe each term separately.

\paragraph{Bulk action.}
The kinetic term for the bulk scalar field in flat space is given by
\be 
S^{(2)}_{\Phi}= -\frac{1}{2}\int d^{4}x~~\d_\mu \Phi 
\d^\mu \Phi 
=-\frac{1}{2}\int d^{4}x~~(\d_i \Phi)^2\,,
\ee
where in the last equality we took the static limit.
Now let us focus on the gravitational sector. It is described 
by expanding the Einstein-Hilbert action in graviton perturbations around the flat background (see e.g.~\cite{Donoghue:2017pgk}),
\[
g_{\mu\nu}=\eta_{\mu\nu}+2h_{\mu\nu}\,.
\]
For the purposes of this section it will be enough to work in
the gauge that reproduces the Schwarzschild solution
from GR. To that end, we consider  
the following isotropic perturbations
\be 
\label{eq:gansatz}
g_{00} =- (1 + 2H_0(t,r)) 
\quad g_{rr}=(1+2H_2(t,r)) \,,
\ee
with all other components given by 
the unperturbed Minkowski metric.
The kinetic term can be extracted directly from the 
Einstein-Hilbert action
\be 
S^{(2)}_{h}=\frac{M_P^{2}}{2}\int d^{4}x\sqrt{-g}R =
 \frac{M_P^{2}}{2}\int dtd\phi d\cos\theta dr ~ 
 \left(2H^2_2  -4r(\d_r H_0)  H_2 \right)\,.
\ee
The leading interaction term between the gravitons and 
the scalar field stems from the scalar field kinetic term,
\be 
S_{\Phi h}=\frac{1}{2}\int dtd\phi d\cos\theta dr  
% ~\frac{2^{1/2} \hat h}{M_P^{(d-1)/2}} 
~r^2\left(H_2-H_0\right) (\d_r\Phi)^2
\subset 
-\frac{1}{2}\int d^{4}x \sqrt{-g} g^{\mu\nu}\d_\mu \Phi \d_\nu \Phi \,. 
\ee

\paragraph{Point particle action.}
Finally, we include the worldline action for the black hole. It starts with 
the point particle part
\be
\label{eq:Spp}
\begin{split}
& S_{\rm pp} \equiv -M\int ds=-M\int d\tau(g_{\mu\nu}\dot{x}^\mu\dot{x}^\nu)^{1/2}\\
&~~~~~= -M\int d^{4}x \int d\tau
~\left(1+H_0\right)\delta^{(4)}(x-x(\tau))\,,
\end{split} 
\ee
where $M$ is the black hole mass, $ds$ is the infinitesimal 
proper worldline 
interval,
and $\tau$ is the worldline
parameter (proper time); the overdot denotes $d/d\tau$.

\paragraph{Finite-size effects.}
As far as the finite size effects are concerned, 
it is instructive to recall some 
details of linear response theory~\cite{Porto:2007qi,Porto:2016pyg,Goldberger:2004jt,Goldberger:2005cd,Goldberger:2020fot}.    
The worldline action describing the coupling of a source 
multipole $I_L$ and the tidal field $\mathcal{E}_L(x)$ (which can be either
the gravitational tidal field or its scalar field analog 
$\mathcal{E}_{L}\propto \d_{\langle  L\rangle}\Phi$)
is given by~\cite{Porto:2016pyg}
\be
\label{eq:SIB}
S_{I\mathcal{E}}=\frac{1}{2}\int d\tau\int d^4x
\delta^{(4)}(x-x(\tau))I_L(\tau) \mathcal{E}^L(x)\,. 
\ee
The tidal field $\mathcal{E}_L$ acts like a source for $I_L$. Hence, in 
the linear approximation we can write\footnote{For clarity, we have omitted the background multipole moments. These are absent for Schwarschild black holes, 
but are present for Kerr black holes.
These moments can be easily taken into account (see e.g.~\cite{Porto:2007qi}), but they do not contribute to the tidal response
and hence are irrelevant for our discussion.
}
\be
\label{eq:Ilret0}
\langle I_L(\tau)\rangle=\int d\tau' G^{\rm ret.}{}^{L'}_L(\tau,\tau')  \mathcal{E}_{L'}(x(\tau'))\,,
\ee
where $\langle ...\rangle$ denotes ensemble-averaging w.r.t.
internal degrees of freedom and short-scale modes, and 
we have introduced the retarded Green's function as follows:
\be 
{G}^{\rm ret}{}_{L}^{L'}(\tau,\tau')=-i\langle [I^L(\tau),I^{L'}(\tau') ]\rangle \theta(\tau-\tau')\,,
\ee
where $\theta(x)$ is the Heaviside theta-function.
Now we switch to frequency space and use 
that causal Green's function are analytic around $\omega=0$.
Then, the spherical symmetry of the problem dictates the 
following general expression
for the causal Green's function
\be 
\label{eq:Gret0}
\begin{split}
{G}^{\rm ret}{}_{L}^{L'}(\omega)&=
\sum_{n=0}
\omega^{2n} 
\left(
{\hat\lambda}^{\text{loc.}}_{2n} {}_{L}^{L'}
% + i\hat \epsilon^{\text{loc.}}_{2n+1} {}^L_{L'}
% \omega
% +
% {\hat\epsilon}^{\text{non-loc.}}_{2n} {}_{L}^{L'}
% \omega^{2n} 
+ i\hat \lambda^{\text{non-loc.}}_{2n+1} {}_L^{L'}
\omega
\right)\,,
% \left(\lambda_{L}^{L'}+i\chi^L_{L'}\right)\delta(\tau-\tau')+...\,,
\end{split}
\ee
where the tensors ${\hat\lambda}^{\text{loc./non-loc.}}_{p} {}_{L}^{L'}$ must be symmetric 
under exchange $L\leftrightarrow L'$, i.e. 
\be 
{\hat\lambda}^{\text{loc./non-loc.}}_{p} {}_{L}^{L'}=
\text{const}\cdot \delta^{\langle L\rangle }_{\langle L'\rangle }\,.
\ee
The terms in the expansion \eqref{eq:Gret0}
that are symmetric under time-reversal
symmetry (i.e., are even under $\omega\to -\omega$) are dubbed local (``loc.'').
We will see shortly that they correspond to 
local terms in the effective action for 
small-wavelength fluctuations.
These terms are manifestly time reversal
invariant, and hence they correspond to conservative dynamics.
However, the terms dubbed non-local (``non-loc.'')
are not time reversal invariant and hence they
describe dissipative effects. Importantly,
in the case of the Schwarzschild metric 
they disappear in the limit of static perturbations
$\omega \to 0$.

The physical response also receives 
contributions from local operators in the worldline action.
Effectively, this leads to a renormalization
of the conservative response coefficients.
All in all, the total conservative response can be described by a set of the following local wordline operators involving only the long-wavelenght 
degrees of freedom,
\be
\label{eq:seff0}
S^{\rm eff}_{I\mathcal{E}}= 
\frac{1}{2\ell !}\int d\tau d^4x
\delta^{(4)}(x-x(\tau))~\left(\lambda_\ell~ \mathcal{E}_L\mathcal{E}^{L}
+
\lambda_{\ell(\omega^2)}~ \dot{\mathcal{E}}_L \dot{\mathcal{E}}^{L}+...
\right) \,.
\ee
This action can be rewritten in the covariant form
by using covariant derivatives along body's 4-velocity 
$v^\mu=\frac{dx^\mu}{d\tau}$
and the
projector onto directions orthogonal to $v^\mu$
\be
\label{eq:proj}
P^\mu_\nu = \delta^\mu_\nu +v^\mu v_\nu\,,\quad 
D\equiv v^\mu \nabla_\mu\,.
\ee 
As a result, the leading worldline
interaction term describing finite-size effects
in the static limit
is given by~\cite{Porto:2016pyg,Hui:2020xxx}
\be
\label{eq:LoveEFT}
\begin{split}
S_{\rm Love} = & \frac{\lambda_\ell}{2 \ell!}\int d^{4}x \int d\tau~\delta^{(4)}(x-x(\tau)) 
\left[P_{\langle \mu_1}^{\nu_1}...P_{\mu_\ell \rangle}^{\nu_\ell} \d_{{\nu_1}...{\nu_\ell}}\Phi \right]
\left[P^{\langle \mu_1}_{\sigma_1}...P^{\mu_\ell \rangle}_{\s_\ell} \d^{{\s_1}...{\s_\ell}}\Phi \right]
% \d^{({\mu_1}...{\mu_L})_T}\Phi
\,,
\end{split}
\ee
where we have introduced the multi-derivative operator
% \[
$\d_{\nu_1...\nu_\ell}\equiv \prod_{i=1}^\ell \d_{\nu_i}\,,$
% \]
whereas $\langle...\rangle$ stands for the symmetrized
traceless component. The leading non-static
effects are captured by the following action
\be
\label{eq:LoveEFT2}
\begin{split}
S_{\rm \omega^2} = & 
\frac{\lambda_{\ell (\omega^2)}}{2 \ell!}
% \int d^{d+1}x 
\int d\tau
% ~\delta^{(d+1)}(x-x(\tau)) 
~D
\left[P_{\langle \mu_1}^{\nu_1}...P_{\mu_\ell \rangle}^{\nu_\ell} \d_{{\nu_1}...{\nu_\ell}}\Phi \right]
D
\left[P^{\langle \mu_1}_{\sigma_1}...P^{\mu_\ell \rangle}_{\s_\ell} \d^{{\s_1}...{\s_\ell}}\Phi \right]\Big|_{x=x(\tau)}
% \d^{({\mu_1}...{\mu_L})_T}\Phi
\,.
\end{split}
\ee
The coupling $\lambda_\ell$ will be referred to as the EFT Love number in what follows.
Eq.~\eqref{eq:LoveEFT} can be viewed as a 
gauge-independent definition of the Love numbers, 
as the corresponding worldline operator is manifestly
covariant. We will see momentarily that it is precisely 
this operator
that generates the $r^{-\ell-1}$ term 
in the Newtonian expansion.
Importantly, we will see that the coupling $\lambda_\ell$ does not exhibit 
logarithmic running in general 
relativity in four dimensions. 
This will guarantee that result of the Newtonian matching is 
meaningful.
The situation is different for frequency-dependent 
Love numbers, which generically 
depend on distance, and hence introduce 
some ambiguity in the direct application of the Newtonian
matching. We will discuss this in Section~\ref{sec:master}.

In what follows we will work in the body's rest frame
where $v^\mu=\delta^\mu_0$, hence 
$P^\mu_0=0,P^i_j=\delta^i_j$, which removes all operators with time derivatives in Eq.~\eqref{eq:LoveEFT}. This 
gives us the following action relevant for the study
of local static response
\be
S_{\rm Love}
= \frac{\lambda_\ell}{2 \ell!}\int d^{4}x \int d\tau~\delta^{(4)}(x-x(\tau)) \d_{\langle {i_1}...{i_\ell}\rangle}\Phi 
~\d^{\langle {i_1}...{i_\ell}\rangle }\Phi\,. 
\ee

\paragraph{Static response in the EFT.}
Our goal here is to compute the static scalar field profile $\Phi$ in the presence of interactions
with gravitons and an external source $\bar\Phi$. 
As a first step, we need to compute the leading order graviton field, which describes the gravitational potential of the black hole on large scales.
In the EFT context, the black hole solution is recovered perturbatively order by 
order in the long-distance expansion controlled by $2M/(M_P^2 r)$. Let us start with the first order.
The equations of motion for the graviton modes are given by
\be
\begin{split}
& H_2=r\d_r H_0 \,,\quad \Delta H_0 = \frac{M}{2M_P^{2}}\int d\tau \delta^{(4)}(x-x(\tau))
\,.
\end{split}
\ee
Let us work in the black hole's rest frame, where the unperturbed
center-of-mass position is given by $x^i(\tau)=0$, $x^0=\tau$, such that 
\be
\Delta  H_0 = \frac{M}{2 M_P^{2}}  \delta^{(3)}(\x) \quad \Rightarrow 
\quad  H_0 = -\frac{M}{8\pi M_P^2 r}=-\frac{r_s}{2r} \,,\ee
where we have introduced the Schwarzschild radius $r_s=M/(4\pi M_P^2)$.
The second metric perturbation is given by 
\be
\label{eq:h2}
H_2 = -H_0=\frac{r_s}{2r}\,. 
\ee
Now we have to compute the scalar field profile.
The total static equation of motion for the scalar field, which includes 
the leading interaction with gravity and the tidal response 
is given by
\be
\label{eq:Phieom}
\Delta \Phi - \frac{2}{r^2}\d_r (r^2 H_2 \d_r)\Phi+(-1)^\ell\frac{\lambda_\ell}{\ell !}\d_{\langle {i_1}...{i_\ell }\rangle}\left(\d^{\langle {i_1}...{i_\ell}\rangle }\Phi \delta^{(3)}(\x) \right)=0\,.
\ee
In order to compute the black hole response, we introduce 
an external scalar field source $\bar \Phi$, satisfying the 
free Poisson equation $ \Delta \bar \Phi = 0$ in the $r\to \infty$ asymptotic.
Assuming that $\bar \Phi$ has an orbital number $\ell$, we find
\be
\label{eq:sourseasymp}
 \Delta \bar \Phi = 0\quad \Rightarrow \quad \bar \Phi = \mathcal{E}^{(0)}_{i_1...i_\ell}x^{i_1}...x^{i_\ell}\,,
\ee
where $\mathcal{E}^{(0)}_{i_1...i_\ell}$ is a symmetric trace-free tensor.
Note that the solution \eqref{eq:sourseasymp}
corresponds to $\bar \Phi=\sum_{\ell m} \mathcal{E}^{(0)}_{\ell m}r^\ell Y_{\ell m}(\theta,\phi)$ in the spherical coordinates.
We want to solve Eq.~\eqref{eq:Phieom} perturbatively expanding in 
$r_s/r$, but keep
the 
explicit $\lambda_\ell$-dependence,
\be
\Phi  =\bar \Phi  + \Phi^{(1)}_h + \Phi^{(1)}_{\rm Love}+...\,.
\ee
We will formally
retain 
corrections 
linear in $\lambda_\ell$, but 
$\lambda_\ell$ itself
does not need to be small.
Let us first compute the correction to the source coming from the interaction with the graviton. We have
\be
\begin{split}
\Delta \Phi^{(1)}_h = \frac{1}{r^2}\d_r (r^2 2H_2 \d_r) \bar \Phi \,.
\end{split} 
\ee
Expanding $\Phi^{(1)}_h$ over spherical harmonics and
using the solution from Eq.~\eqref{eq:h2} we obtain
\be
\left(\d_r^2+\frac{2}{r} \d_r-\frac{\ell(\ell+1)}{r^2}\right)\Phi^{(1)}_{h~\ell m} =\frac{r_s}{r^2}\d_r(r \d_r \bar\Phi)\,.
\ee
Note that in the above expression we can lift all restrictions 
on $\ell$ and treat it as a generic number.
Plugging our source from Eq.~\eqref{eq:sourseasymp}, this equation
can be easily solved.
The full solution including the source plus the leading graviton correction is given by
\be
\begin{split}
\Phi =  ~ & \sum_{m=-\ell}^\ell \mathcal{E}^{(0)}_{\ell m}Y_{\ell m}r^\ell \left(1-
\frac{\ell}{2}\frac{r_s}{r}\right) = \mathcal{E}^{(0)}_{i_1...i_\ell}x^{i_1}...x^{i_\ell}\left(1-
\frac{\ell}{2}\frac{r_s}{r}\right)\,.
\end{split}
\ee
We observe that the interaction with the graviton induces the sub-leading 
corrections to the source. Note that this correction is calculable, i.e. 
its strength is fixed in the EFT itself. Importantly, because the graviton 
propagator scales like $1/r$, these corrections 
are naturally organized as the following power series
\be 
\label{eq:psigrav}
\Phi \supset r^\ell\left(1+c_1\frac{r_s}{r}+c_2\left(\frac{r_s}{r}\right)^2+...\right)\,,
\ee
for any $\ell \in \mathbb{R}$. 
This gives an interpretation 
of the subleading source corrections:
these are just generated by the coupling
between the source and perturbative gravity.
Our result also justifies 
the use of the analytic continuation $\ell \to \mathbb{R}$ for the source-response split, because this indeed allows
one to isolate the series Eq.~\eqref{eq:psigrav}
and avoid a possible overlap with corrections
induced by finite-size effects.

Now let us compute the correction to the source coming from the Love interaction, which corresponds to induced multipoles.
We have
\be
\Delta \Phi^{(1)}_{\rm Love} =-(-1)^\ell
\lambda_\ell
% \frac{\lambda_\ell}{\ell!}
{\mathcal{E}^{(0)}}^{{i_1}...{i_\ell}}\d_{ {i_1}...{i_\ell}}\delta^{(3)}(\x)\,. 
\ee
This equation can be easily solved in Fourier space,
\be
\begin{split}
\Phi^{(1)}_{\rm Love} &= 
% i^\ell
(-i)^{\ell}\lambda_\ell {\mathcal{E}^{(0)}}_{{i_1}...{i_\ell}} \int \frac{d^3k}{(2\pi)^{3}}~e^{i\k\cdot \x}
\frac{k_{{i_1}}...k_{{i_\ell}}}{\textbf{k}^2}\\
% \d_{(i_{a_1}...i_{a_L})}
&=B_{\ell}\lambda_\ell {\mathcal{E}^{(0)}}_{{i_1}...{i_\ell}}x^{i_1}...
x^{i_\ell}\frac{1}{r^{2\ell+1}} 
% \left(d~x^ix^j-r^2\delta_{ij} \right)
\,,
\end{split} 
\ee
where we have introduced the following normalization constant
\be
\label{eq:Al}
B_{\ell}\equiv (-1)^{\ell}\frac{2^{\ell-2}
}{\pi^{1/2}\Gamma(1/2-\ell)}\,. 
\ee
The total solution including 
linear order correction in $\lambda_\ell$
and $r_s/r$ is given by
\be
\label{eq:Phieftfinal}
\Phi =  {\mathcal{E}^{(0)}}_{i_1...i_\ell}x^{i_1}...x^{i_\ell}\left(
\underbrace{1}_\text{source}-
\underbrace{\frac{\ell}{2}\frac{r_s}{r}}_\text{graviton interaction}
+ \underbrace{B_{\ell}\lambda_\ell \frac{1}{r^{2\ell+1}}}_\text{induced multipole}
\right)\,.
\ee
Comparing this to \eqref{eq:Phifar}, 
we can see that the EFT provides 
a tool to define the response coefficients and avoid ambiguity 
in the source/response split. The subleading corrections 
in the source expansion, which scale as $(r_s/r)^n \times r^\ell $, 
naturally correspond to diagrams produced by the interaction
between the source field and the graviton. 
These diagrams are fixed by the structure of Einstein-Hilbert and the Klein-Gordon 
actions.
Thus, all graviton corrections can be unambiguously computed
by expanding the Einstein-Hilbert equation 
in higher order operators involving the graviton
field. Hence, in principle, during the matching procedure, 
one can identify all corrections 
coming from the graviton vertices and subtract them from 
the microscopic solution obtained in GR. 
The remaining piece will 
correspond to the response coefficients. 
In practice, however, the number of diagrams to be computed 
can be very large. 
From the practical point of view, it is more convenient 
to do the analytic continuation $\ell \to \mathbb{R}$, which allows one 
to easily achieve the same goal of isolating 
non-linear corrections to the source from those generated
by the finite-size effects.

\paragraph{Inclusion of spin.} 
In order to reproduce the GR solution in the EFT,
one has to perturbatively recover the Kerr 
metric at order $a^2/r^2$, where $a\equiv J/M$ is the normalized
black hole's spin. 
To that end one has to introduce vector degrees of freedom of 
metric perturbations  
and consider their coupling to black hole's spin 
via the 
Mathisson - Papapetrou/Routhian formalism~\cite{Porto:2016pyg}.
This procedure 
has been recently 
presented in Ref.~\cite{Goldberger:2020fot}, see also Refs.~\cite{Martel:2005ir,Kobayashi:2012kh}. In order to obtain the $a^2/r^2$ corrections, we need to take into 
account the cubic interaction between the scalar and vector
graviton modes. 
The details of this calculation
are not essential for our discussion. 
Once we obtain the following 
graviton perturbation\footnote{We assume a gauge consistent with the Boyer-Lindquist coordinates~\cite{1967JMP.....8..265B}.}
\be
h_{\phi\phi}=\frac{a^2}{r^2} \cdot r^2\sin^2\theta  \,,
\ee
it can be coupled to the scalar field through the kinetic term,
\be
\label{eq:Phimsq}
\begin{split}
 &\int d^{4}x \sqrt{-g} g^{\mu\nu}\d_\mu \Phi \d_\nu \Phi
 \supset \int d^4x h^{\phi\phi}(\d_{\phi}\Phi)^2
% \\&
 \propto 
\int dt d\phi d\theta dr ~r^2\sin\theta~
 \left[\frac{a^2m^2}{r^4}\Phi^2 \right]\,,
 \end{split}
\ee
where in the last equation we used the expansion
over spherical harmonics $\Phi\propto e^{im\phi}$.
Then, we can easily account for perturbations that are 
produced by the interactions between the source and the gravitons 
at the leading order in spin. 
Varying the action~\eqref{eq:Phimsq} over $\Phi$ and using the 
perturbative expansion
\be
\Phi=\bar \Phi  + \Phi_h^{(1)}+ \Phi_a^{(1)} \,,
\ee
we obtain that the correction due to black hole spin $\Phi_a^{(1)}$ satisfies the following 
differental equation
\be
\left(\d_r^2+\frac{2}{r} \d_r-\frac{\ell(\ell+1)}{r^2}\right)\Phi^{(1)}_a =
\frac{a^2}{r^4}\d_\phi^2\bar \Phi\,.
\ee
Using our ansatz for the source \eqref{eq:sourseasymp},
we obtain
the following net expression for the source interacting with the Kerr black hole at the leading orders in black hole's mass
and spin
\be
\label{eq:Phisourcea}
\begin{split}
\Phi =  ~ & \sum_{m}{\mathcal{E}^{(0)}}_{\ell m}Y_{\ell m}r^\ell \left(1-
\frac{\ell}{2}\frac{2M}{r}
+\frac{m^2 a^2}{ 4 \ell-2 }\frac{1}{r^2}
\right) 
\,.
\end{split}
% c_{a_1...a_L}x^{a_1}...x^{a_L}\left(1-
% r^{d-2}_s\frac{L}{2(d-2)}\frac{1}{r^{d-2}}\right)\,.
\ee
Iterating this procedure at higher orders in $r_s/r$, we can reconstruct 
all power series responsible for the interaction between the source and the graviton.

\paragraph{Finite size operators in the presence of spin.}
The structure of finite-size operators becomes more complicated
due to the spin, because now we can produce 
new tensor structures using the spin vector $s^i$ and the Levi-Civita
antisymmetric symbol $\epsilon_{ijk}$.
Therefore, now 
we can write the following general expression~\cite{Goldberger:2020fot} for the causal 
response function introduced in Eq.~\eqref{eq:Ilret0}
\be 
\label{eq:Gret}
\begin{split}
{G}^{\rm ret}{}_{L}^{L'}(\omega)&=
\sum_{n=0}
\omega^{2n} 
\left(
{\hat\lambda}^{\text{loc.}}_{2n} {}_{L}^{L'}
+ i\hat \epsilon^{\text{loc.}}_{2n+1} {}^L_{L'}
\omega
+
{\hat\epsilon}^{\text{non-loc.}}_{2n} {}_{L}^{L'}
+ i\hat \lambda^{\text{non-loc.}}_{2n+1} {}^L_{L'}
\omega
\right)\,.
\end{split}
\ee
Let us now specify symmetry properties of different response tensors in this expression.
From the axial symmetry 
of the problem, the tensors 
that are even and odd w.r.t. $L\leftrightarrow L'$
must be constructed from even and odd numbers
of spin vectors,
respectively.\footnote{The only available 
tensor structures are $\delta_{ij},\epsilon_{ijk}$
and $s_j=s z_j$, where $z_j=\delta^3_j$.
Hence, any tensor that is odd w.r.t. $L\leftrightarrow L'$ (but is still STF w.r.t. multi-indices $L$
and $L'$)
has to look like 
$\epsilon_{ii'k}s_k$ times 
a tensor that is even w.r.t. remaining multi-index
exchange $(L-1)\leftrightarrow (L-1)'$.}
The local terms in Eq.~\eqref{eq:Gret}
must be symmetric w.r.t. time reversal invariance 
(which includes the spin flip), which now corresponds
to simultaneous exchange $L\leftrightarrow L', \omega \to -\omega$. 
These terms correspond to 
local operators in the point-particle EFT
in the body's rotation frame. 
In contrast, the non-local terms
correspond to internal
(gapless)
worldline degrees of freedom that
capture dissipation~\cite{Goldberger:2020fot} and hence they must be odd w.r.t. 
exchange $L\leftrightarrow L', \omega \to -\omega$.
This implies that tensors $\hat\lambda^{\rm loc./non-loc.}_{p} {}_{L}^{L'}$
and $\hat\epsilon^{\rm loc./non-loc.}_{p} {}_{L}^{L'}$
in Eq.~\eqref{eq:Gret}
must be even and odd w.r.t. $L\leftrightarrow L'$,
respectively. 
Going back to proper time we get,
\be 
\begin{split}
\langle I_L\rangle&=
\sum_{n}
(-1)^n
\left(
{\hat\lambda}^{\text{loc.}}_{2n} {}_{L}^{L'}
+  \hat \epsilon^{\text{loc.}}_{2n+1} {}^L_{L'}
\frac{d}{d\tau}
+
{\hat\epsilon}^{\text{non-loc.}}_{2n} {}_{L}^{L'}
+  \hat \lambda^{\text{non-loc.}}_{2n+1} {}^L_{L'}\frac{d}{d\tau}
\right)\frac{d^{2n}}{d\tau^{2n}}\mathcal{E}_{L'}\,.
\end{split}
\ee
Unlike the Schwazschild black hole, 
the terms with time derivatives here do not
vanish in the static limit (w.r.t. a fixed inertial frame) because of 
the rotation of the body. Hence, if we want to capture
the effects of spin to all orders, we need to keep track of all 
powers of frequency here. This problem can be solved 
if we rewrite Eq.~\eqref{eq:Gret}
in a fixed inertial frame. To that end we can 
expand time derivatives in body's rotation frame as
\be 
\frac{d}{d\tau} \mathcal{E}_{L'}=
\d_t \mathcal{E}_{L'}
-\ell \Omega_{(i'_1 q}
\mathcal{E}_{q~i'_2...i'_\ell)}\,,
\ee
where $\d_t$ is the time derivative in the fixed
inertial (source) frame,
$\Omega_{ij}$ is the antisymmetric
angular velocity tensor, and we have symmetrized 
the rightmost term w.r.t. its free indices. 
To simplify the argument, let us neglect terms with 
partial time derivatives $\d_t$, which is reasonable since we are interested in the static
limit.

In this limit, applying an even number of time 
derivative in the body's local rotating frame 
will produce a tensor built out of the same even number
of the angular velocity tensors. Multiplying
an original response matrix by this tensor will not 
change its parity properties w.r.t. multi-index exchange,
\be 
{\hat\lambda}^{\text{loc.}}_{2n} {}_{L}^{L'}
\frac{d^{2n}}{d\tau^{2n}}\to 
{\lambda}^{\text{loc.}}_{2n} {}_{L}^{L'}~(\text{even}~L\leftrightarrow L')\,,\quad 
{\hat\epsilon}^{\text{non-loc.}}_{2n} {}_{L}^{L'}
\frac{d^{2n}}{d\tau^{2n}}\to 
{\epsilon}^{\text{non-loc.}}_{2n} {}_{L}^{L'}~(\text{odd}~L\leftrightarrow L')\,,
\ee
where ${\lambda}_{2n}$ and ${\epsilon}_{2n}$ are response
tensors written in the inertial frame.
However, applying an odd number of 
time derivatives produces a tensor
that contains the same odd number of the angular
velocity tensors. This tensor is odd 
w.r.t. to the exchange of its multi-indices,
and hence it will change the parity
of the corresponding response matrices 
 w.r.t. 
exchange $L\leftrightarrow L'$,
\be 
{\hat\lambda}^{\text{non-loc.}}_{2n+1} {}_{L}^{L'}
\frac{d^{2n+1}}{d\tau^{2n+1}}\to 
{\lambda}^{\text{non-loc.}}_{2n} {}_{L}^{L'}~(\text{odd}~L\leftrightarrow L')\,,
\quad 
{\hat\epsilon}^{\text{loc.}}_{2n+1} {}_{L}^{L'}
\frac{d^{2n+1}}{d\tau^{2n+1}}\to 
{\epsilon}^{\text{loc.}}_{2n+1} {}_{L}^{L'}~(\text{even}~L\leftrightarrow L')\,,
\ee
where ${\lambda}_{2n+1}$ and ${\epsilon}_{2n+1}$ are 
again response
tensors in the inertial frame.
We see that after we have changed the frame, 
all terms of non-local origin can be collected
into a new antisymmetric response matrix
$\kappa {}^{L'}_{L}$, 
whereas all local terms effectively sum up into  
a new symmetric matrix $\lambda {}^{L'}_{L}$.
Importantly, the transition to the 
inertial frame did not mix the properties of the 
response matrices written in Eq.~\eqref{eq:Gret} --- the local response is still captured by a matrix 
that is even w.r.t. $L \leftrightarrow L',\omega\to -\omega$.
All in all, we can write
\be 
\label{eq:IL}
\begin{split}
\langle I_L\rangle
&=\left(\lambda {}^{L'}_{L} 
+ \kappa {}^{L'}_{L}\right)
\mathcal{E}_{L'} + 
\left(\Lambda{}^{L'}_{L}
+\Lambda'{}^{L'}_{L}\right) \d_t\mathcal{E}_{L'}+...\,,\quad \text{where}\\
&
\lambda{}^{L'}_{L} =\lambda{}^{L}_{L'}\,,\quad  
\kappa{}^{L'}_{L} =
-\kappa{}^{L}_{L'} \,, \quad  \Lambda{}^{L'}_{L} = -\Lambda{}^{L'}_{L}\,,\quad 
\Lambda'{}^{L'}_{L}
=\Lambda'{}^{L}_{L'}\,,
\end{split}
\ee
and ``$...$'' stands for operators that involve
more than one partial time derivatives.
% where $\Omega_H=a/(2Mr_+)$ is the angular velocity of the Kerr black hole horizon.
We stress that
the instantaneous contribution proportional to $\kappa^L_{L'}$, in fact, stems from
non-local operators, which was nicely explained
in Ref.~\cite{Goldberger:2020fot}.
The corresponding 
induced multipole $I_L\propto \kappa_L^{L'}\mathcal{E}_{L'}$
describes dissipative effects such as tidal torques
or mass loss/accretion~\cite{Goldberger:2020fot,LeTiec:2020bos}. 
In contrast, the part of the 
response proportional to $\lambda_{L'}^{L}$
in Eq.~\eqref{eq:IL}
captures the local static deformation 
produced by external fields, it is 
indistinguishable from the effect of local operators in the point particle action.

All in all, the leading-order local finite-size effects are described 
by the following generalized response operator~\eqref{eq:LoveEFT},
\be
\label{eq:lovea}
S^{a}_{\rm Love}
= \frac{1}{2 \ell!}\int d^{4}x \int d\tau~\delta^{(4)}(x-x(\tau)) {\l^{(0)}}^{{j_1}...{j_\ell}}_{{i_1}...{i_\ell}}\d_{\langle {j_1}...{j_\ell}\rangle }\Phi 
~\d^{\langle {i_1}...{i_\ell} \rangle }\Phi\,, 
\ee
which accounts for the violation of the spherical 
symmetry by the spinning black hole background.
The coupling Eq.~\eqref{eq:lovea} can always be
recast in the manifestly covariant form by means 
of the projection operator~\eqref{eq:proj}.
Note that $\lambda$ is a symmetric trace-free tensor w.r.t.
upper and lower sets of indices,
\be
\label{eq:lamdasym}
{\l^{(0)}}^{{j_1}...{j_\ell}}_{{i_1}...{i_\ell}}
= {\l^{(0)}}^{\langle {j_1}...{j_\ell}\rangle }_{\langle {i_1}...{i_\ell}\rangle }\,,
\ee
but its trace w.r.t. the contraction of lower 
and upper indices does not vanish. 
We will refer to $\lambda^{(0)}$
and its analogs for higher spins
as ``Love tensor'' in what follows.

The calculation of the response of the scalar 
field induced by this operator is identical to one 
presented above, and it yields
\be
\Phi^{(1)}_{\rm Love}
=B_\ell r^{-\ell -1}
{\mathcal{E}^{(0)}}_{L'} n^L {\l^{(0)}}^{L'}_L\,,
% \frac{}{r^{1+\ell}}\,.
\ee
where the constant $B_\ell$ is given in Eq.~\eqref{eq:Al}.

\subsection{Microscopic Computation}

To find the SRCs  $k^{(0)}_{\ell m}$, we need to 
solve the vacuum Klein-Gordon equation in the Kerr background, assuming 
that the scalar field varies very slowly in time.
It is instructive to carry out our microscopic (i.e. general relativity) calculation
in two different coordinate systems. Let us start with 
the advanced Kerr coordinates, which are manifestly regular 
at the horizon.

\paragraph{Advanced Kerr coordinates.} 

The interval of the Kerr spacetime in the advanced Kerr coordinates is given by~\cite{Chandrasekhar:1985kt}
\be
\begin{split}
ds^2=
& -\left(1-\frac{2Mr}{\Sigma} \right)dv^2+2dvdr - \frac{4Mra}{\Sigma} \sin^2 \theta dv
d\tilde{\phi}
 - 2 a \sin^2\theta dr d\tilde{\phi} \\
&+\Sigma d\theta^2 
+ \left(r^2 +a^2 + \frac{2Mr}{\Sigma} a^2 \sin^2\theta \right)  \sin^2 \theta d\tilde{\phi}^2 \,,
\end{split}
\ee
where $a\equiv J/M$ is the reduced spin parameter and $\Sigma \equiv r^2+a^2\cos^2\theta$.
It is well known that in the static case ($\omega=0$) the Klein-Gordon equation for the massless scalar field
factorizes into 
usual scalar spherical harmonics
in the Kerr background~\cite{Brill:1972xj,1973JETP...37...28S}. 
In the advanced Kerr coordinates 
this decomposition
takes the following form
\be
\Phi =\sum_{\ell m} \mathcal{E}^{(0)}_{\ell m} \tilde R_{\ell m}(r)Y_{\ell m}(\theta,\tilde\phi)\,. 
\ee
To match the source boundary condition at infinity, we impose $\tilde R_{\ell m}\to r^\ell$ at $r\to \infty$ and demand this function to be smooth at the external black hole 
horizon.
The equation defining the radial mode function 
$\tilde{R}_{\ell m}$
takes the following form
    \be
    \label{eq:Teuks0k}
    \begin{split}
        &x\left(1+x\right)\tilde{R}_{\ell m}^{\prime\prime}\left(x\right) + \left[\left(1+2x\right)+2im\gamma\right]\tilde{R}_{\ell m}^{\prime}\left(x\right) 
        -\ell \left(\ell+1\right)\tilde{R}_{\ell m}\left(x\right) = 0\,,
    \end{split}
    \ee 
where ${}'\equiv \d/\d x$ and we have defined
\be
    x\equiv \frac{r-r_{+}}{r_{+}-r_{-}}\;\;,\;\;\gamma\equiv\frac{a}{r_{+}-r_{-}}\,,\quad 
    r_{\pm }= M\pm \sqrt{M^2-a^2}\,.
\ee
Note that $r_+$ and $r_-$
are the outer and inner horizons of the Kerr black hole, respectively.
In what follows we will be mostly focusing 
on $r_+$ and we will refer to it simply as ``black hole's horizon.''
We will also use the following notation
\be
r_{sa} \equiv r_+-r_{-} \,.
\ee
The solution of Eq.~\eqref{eq:Teuks0k} regular at the horizon ($x\to 0$) is given by
\be 
\label{Rtilde}
\tilde{R}_{\ell m} =\text{const}\cdot {}_2F_1\left(\ell+1,-\ell;1+2im\gamma,-x\right)\,,
\ee
where $ {}_2F_1$ stands for the Gauss hypergeometric function
(see Appendix~\ref{app:math} for more detail).
In the physical case $\ell \in \mathbb{N}$, the function $\tilde{R}_{\ell m} $ is a polynomial
in $x\propto r$ and hence it does not contain any decaying power 
of $r$. Thus, by looking at Eq.~\eqref{eq:Phifar} one may be tempted 
to conclude that the SRCs are zero for the Kerr background.
However, we have to ensure that this is not a result of a cancellation between 
 body's response and the subleading source contributions due to graviton
interactions. 
To that end, let us consider an analytic continuation $\ell\to \mathbb{R}$, in which case the solution 
Eq.~\eqref{Rtilde} can be Taylor-expanded
at spatial infinity as (see Appendix~\ref{app:math})
\be 
\label{Rexpan}
\begin{split}
&\tilde{R}_{\ell m}=
  \text{const}\cdot \Bigg(\frac{\Gamma(1+2im\g)\Gamma(2\ell+1)}{\Gamma(\ell+1)\Gamma(1+\ell+2im\g)}
x^{\ell}\cdot
{}_2F_1\left(-\ell,-\ell-2im\g,-2\ell,-x^{-1}\right) \\
& +
\frac{\Gamma(1+2im\g)\Gamma(-2\ell-1)}{\Gamma(-\ell)\Gamma(-\ell+2im\g)}
x^{-\ell-1}\cdot
{}_2F_1\left(\ell+1,\ell+1-2im\g,2\ell+2,-x^{-1}\right)\Bigg)\\
&
\xrightarrow[x\to \infty]{}
% \overrightarrow{x\to \infty }
% \quad \frac{\Gamma(1+2im\g)\Gamma(2\ell+1)}{\Gamma(\ell+1)\Gamma(1+\ell+2im\g)}
r_{sa}^\ell x^{\ell}
\left(
1 + 
\frac{\Gamma(-2\ell-1)\Gamma(\ell+1)\Gamma(1+\ell+2im\g)}{\Gamma(2\ell+1)\Gamma(-\ell)\Gamma(-\ell+2im\g)}
% \frac{\Gamma(1+2im\g)\Gamma(-2\ell-1)}{\Gamma(-\ell)\Gamma(-\ell+2im\g)}
x^{-2\ell-1}\right) 
% \quad \text{at} \quad x\to \infty 
\,.
\end{split}
\ee
Since the first distinctive contribution in Eq.~\eqref{Rexpan} scales as $r^\ell$ at infinity, it is natural 
to associate it with the external source and 
its corrections produced by non-linear gravitational interactions. 
The second distinctive contribution in
Eq.~\eqref{Rexpan} scales as $r^{-\ell-1}$ at infinity,
and hence it is natural to interpret it as black hole's response. Comparing Eq.~\eqref{Rexpan} with the (post-)Newtonian expansion formula \eqref{eq:Phifar}, 
we find that for a general multipolar index $\ell$ the response
coefficients are given by
\be
\label{eq:ks0}
k^{(0)}_{\ell m} =  \frac{\Gamma(-2\ell-1)\Gamma(\ell+1)\Gamma(1+\ell+2im\g)}{\Gamma(2\ell+1)\Gamma(-\ell)\Gamma(-\ell+2im\g)}
\left(\frac{r_{sa}}{r_s}\right)^{2\ell +1 }\,.
\ee 
The scalar tidal response coefficients extracted by means of the analytic continuation 
do not vanish even in physical limit $\ell \to \mathbb{N}$.
In this case Eq.~\eqref{eq:ks0} can be simplified
\be
\label{eq:ks0v2}
\begin{split}
k^{(0)}_{\ell m} =  
& -im\gamma
\frac{(\ell!)^2}{(2\ell)!(2\ell+1)!}
 \prod_{n=1}^\ell(n^2+4m^2\gamma^2)
% \frac{\Gamma(-2\ell-1)\Gamma(\ell+1)\Gamma(1+\ell+2im\g)}{\Gamma(2\ell+1)\Gamma(-\ell)\Gamma(-\ell+2im\g)}
\left(\frac{r_{sa}}{r_s}\right)^{2\ell +1 }\\
= & -\frac{im\chi}{2}
\frac{(\ell!)^2}{(2\ell)!(2\ell+1)!}
 \prod_{n=1}^\ell(n^2(1-\chi^2)+m^2\chi^2)\,,
\end{split}
\ee 
where in the last line we have introduced
the dimensionless spin $\chi\equiv a/M$
and used Eq.~\eqref{eq:G2sinh}.
Note that the expression \eqref{eq:ks0v2} vanishes
in the limit $\chi\to 0$, reproducing the 
well-established result that the scalar response coefficients
of non-spinning black holes are zero~\cite{Kol:2011vg,Hui:2020xxx}.
Importantly, the response coefficients \eqref{eq:ks0v2}
are purely imaginary. 
As discussed before,
they correspond
to dissipative effects and not to the classic 
conservative
static response coefficients which we will refer to as
scalar
Love numbers.

All in all, we have obtained that
the radial solution~\eqref{Rtilde}
in the advanced Kerr coordinates
is a polynomial without any decaying power of $r$,
and, at the same time, the response coefficients are non-zero.
The only possibility to reconcile these two facts is that 
the GR corrections to the source and the 
induced response happened to exactly cancel one another 
in the advanced Kerr coordinates
in the
physical limit $\ell \to \mathbb{N}$.
This is exactly what happens.
To see this, 
we expand the relevant source solution at infinity as follows
\be
\begin{split}
& x^\ell{}_2F_1\left(-\ell,-\ell-2im\g,-2\ell,-x^{-1}\right)  \\
&=
x^{\ell}
\sum_{n=0}^{\infty}\frac{\Gamma(-\ell+n)\Gamma(-\ell-2im\gamma+n)\Gamma(-2\ell)}{\Gamma(-\ell)\Gamma(-\ell-2im\gamma)\Gamma(-2\ell+n)}(-1)^n\frac{x^{-n}}{n!}\,.
    \end{split}
\ee
Let's focus on the $n=2\ell+1$'th term in the hypergeometric series above. 
This term scales like $r^{-\ell - 1}$ just like the response
contribution.
We have
\be 
\begin{split}
& 
% \frac{\Gamma(1+2i\gamma m)\Gamma(2\ell+1)}{\Gamma(\ell+1)\Gamma(1+\ell+2im\gamma)} 
x^{2\ell+1}{}_2F_1\left(-\ell,-\ell-2im\g,-2\ell,-x^{-1}\right)\\
&\supset
% x^{-\ell-1}
\frac{\Gamma(-2\ell-1)\Gamma(\ell+1)\Gamma(1+\ell-2im\g)}{\Gamma(2\ell+1)\Gamma(-\ell)\Gamma(-\ell-2im\g)}
=im\gamma
\frac{(\ell!)^2}{(2\ell)!(2\ell+1)!}
 \prod_{n=1}^\ell(n^2+4m^2\gamma^2)\,,
\end{split}
\ee
where we took the physicacl limit
$\ell\to\mathbb{N}$ in the second line.
This exactly equals minus the coefficient
in front of the response part in Eq.~\eqref{Rexpan},
hence the subleading source contribution exactly cancels 
the response in the advanced Kerr coordinates.

The upshot of our discussion is that 
the naive identification of the response coefficients from the 
solution to the Klein-Gordon 
equation 
may be ambiguous due to uncertainty 
in the source/response split.
This ambiguity can be removed by means of the analytic 
continuation $\ell \to \mathbb{R}$~\cite{LeTiec:2020bos}.
This is explicitly confirmed by a calculation
in the Boyer-Lindquist coordinates, to which we proceed now.

\paragraph{Boyer-Lindquist coordinates.}
The line element of the Kerr spacetime in the 
Boyer-Lindquist coordinates~\cite{1967JMP.....8..265B} is given by
\be
\begin{split}
ds^2= &-\left(1-\frac{2Mr}{\Sigma}\right)dt^2
- \left(\frac{4 M a r \sin^2\theta}{\Sigma} \right)dtd\phi
+\frac{\Sigma}{\Delta}dr^2 +\Sigma d\theta^2\\
& +\sin^2\theta\left(r^2+a^2+\frac{2 M a^2r\sin^2\theta}{\Sigma}\right)d\phi^2\,,
\end{split}
\ee
where $\Delta\equiv r^2-2Mr + a^2$, $\Sigma \equiv r^2+a^2\cos^2\theta$.
The Boyer-Lindquist and advanced Kerr coordinates are related via
\be
\begin{split}
    & d\upsilon = dt + dr\left(1+\frac{2Mr}{\Delta}\right) \,,
    \quad d\tilde{\phi} = d\phi + dr\frac{a}{\Delta}\,.
\end{split}
\ee
The static scalar field equation factorizes in these coordinates 
as follows~\cite{Brill:1972xj}
\be
\Phi =\sum_{\ell m} 
{\mathcal{E}}^{(0)}_{\ell m} R_{\ell m}(r)Y_{\ell m}(\theta,\phi)\,,
\ee
where the radial mode function $R_{\ell m}$ satisfies the following equation:
\be
\label{eq:Kerrfull}
\begin{split}
&
x (x+1) R''_{\ell m}(x)
+
(2 x+1) R'_{\ell m}(x)
+
\left(-\left(\ell^2+\ell\right)
+\frac{\gamma^2 m^2}{
% \left(4 a^2-1\right) 
x (x+1)} \right) R_{\ell m}=0\,.
\end{split}
\ee
We are looking for a solution which is smooth at the black hole's 
horizon 
and has a singularity at spatial infinity. 
Identifying this solution in the 
Boyer–Lindquist coordinates
is not evident, as these coordinates 
are singular at the Kerr horizon.
Nevertheless, it can be shown that the regularity at the horizon in the 
Boyer-Lindquist coordinates
corresponds to the following condition, obtained by Press~(1972)~\cite{1972ApJ...175..243P} and Teukolsky~(1973)~\cite{Teukolsky:1973ha}
\be 
R_{\ell m}=\text{const}\cdot (r-r_{+})^{+im\gamma}\quad \text{as}\quad r\to r_{+}\,.
\ee 
The constant should be chosen such that $R_{\ell m}/r^\ell \to 1$
at $r\to \infty$.
The solution satisfying these boundary conditions is given by 
\be 
\label{eq:Rfinal}
\begin{split}
& R_{\ell m} = \text{const}\cdot 
    \left(\frac{x}{1+x}\right)^{im\gamma}{}_2F_1\left(\ell+1,-\ell,1+2im\gamma,-x\right)\,,\\
    &
\xrightarrow[x\to \infty]{}
% \overrightarrow{x\to \infty }
% \quad \frac{\Gamma(1+2im\g)\Gamma(2\ell+1)}{\Gamma(\ell+1)\Gamma(1+\ell+2im\g)}
r_{sa}^\ell\cdot
\left(
x^{\ell} + 
\frac{\Gamma(-2\ell-1)\Gamma(\ell+1)\Gamma(1+\ell+2im\g)}{\Gamma(2\ell+1)\Gamma(-\ell)\Gamma(-\ell+2im\g)}
% \frac{\Gamma(1+2im\g)\Gamma(-2\ell-1)}{\Gamma(-\ell)\Gamma(-\ell+2im\g)}
x^{-\ell-1}
% r_{sa}^{-\ell}
\right) 
% \quad \text{at} \quad x\to \infty 
\,,
\end{split}
\ee 
where  in the last step we used an analytic continuation
of the hypergeometric function at spatial infinity
and retained only the leading asymptotics.
Assuming that $\ell \in \mathbb{R}$, we can extract the SRCs just 
like in Eq.~\eqref{Rexpan} and find the same expression~\eqref{eq:ks0v2}. 
However, in contrast to the advanced Kerr coordinates, 
the part of the solution containing the power $r^{-\ell-1}$ does not get canceled by 
the graviton corrections to the source   
even in the physical case $\ell \to \mathbb{N}$.
This happens 
due to the presence of
the prefactor 
$\left(\frac{x}{1+x}\right)^{im\gamma}$.
 
We see that the coefficient in front of the power $r^{-\ell -1}$
depends on a choice of coordinates in the physical case 
$\ell \to \mathbb{N}$. 
If we were to use the naive Newtonian matching, 
we would find coordinate-dependent SRCs.
The agreement between the different coordinate systems 
is restored if we use the analytic continuation $\ell \to \mathbb{R}$.

Finally, we note that non-vanishing of SRCs for Kerr black holes was, in fact,
first discovered by Press in 1972~\cite{1972ApJ...175..243P},
who also argued that they had to be purely imaginary in order for the solution to satisfy the complex regularity
condition at the black hole horizon.\footnote{Note that the results presented in that work seem to be  
affected by an insignificant typo
% , which does not alter main conclusions of that paper. 
Ref.~\cite{1972ApJ...175..243P} presents the 
following solution to the radial models of Klein-Gordon equation in the Kerr background (their Eq.~(10))
\be
\label{eq:P72}
\text{Press'72:}\quad R_{\ell m} =(r-r_{-})^{-im\gamma}
(r-r_+)^{im\gamma} {}_2F_1(\ell+1,-\ell,1+2im\gamma;(r-r_-)/(r_{+} -r_{-}))\,,
\ee
which, in fact, does not satisfy the radial Teukolsky 
equation~\cite{Teukolsky:1973ha}. The actual solution may be written 
in one of the following equivalent forms
\be 
\begin{split}
R_{\ell m} & =(r-r_{-})^{-im\gamma}
(r-r_+)^{im\gamma} {}_2F_1(\ell+1,-\ell,1-2im\gamma;(r-r_-)/(r_{+} -r_{-}))\,,\\
& =(r-r_{-})^{-im\gamma}
(r-r_+)^{im\gamma} {}_2F_1(\ell+1,-\ell,1+2im\gamma;(r_+-r)/(r_{+} -r_{-}))\,,
\end{split}
\ee
which differs from Eq.~\eqref{eq:P72}
either by the argument or by the sign in front of the complex part 
of the third order parameter of the hypergeometric function.
This typo 
has resulted in a sign difference for response
coefficients 
compared to our result. 
Correcting this typo, 
Eqs.~(10),~(15c) of Press~(1972) imply
\be
\text{Press'72:}\quad \text{Im}R_{\ell m}\Big|_{\ell=1,m=1}= -\frac{a }{3}
\left(\frac{M}{r}\right)^2\quad \text{as}\quad r\to \infty\,,
\ee
which coincides with our expression \eqref{eq:Rfinal}
at linear order in $a$.
}
Moreover, Press~(1972) has also shown that the SRCs
capture 
the spin down produced by 
the perturbing scalar field.
A similar connection was recently discussed 
in Le Tiec et al.~(2020)~\cite{LeTiec:2020bos} in the context of gravitational perturbations.

\subsection{Matching to the EFT}

To match the microscopic and the EFT calculation we
need to compare the coefficients in front of the $1/r^{\ell+1}$ power from the two calculations.
We have
\be
\label{eq:match0}
B_\ell n^{\langle L\rangle } \lambda^{L'}_L\mathcal{E}_{ L' } 
=r_s^{2\ell+1}\sum_{m=-\ell}^\ell \mathcal{E}^{(0)}_{\ell m} 
Y_{\ell m}k^{(0)}_{\ell m}
\ee
Rewriting the r.h.s. of this equation in the basis 
of the constant STF tensors $\mathscr{Y}^{L}_{\ell m}$
on $\mathbb{S}^2$ as~\cite{Thorne:1980ru}
\be 
\sum_{m=-\ell}^\ell \mathcal{E}^{(0)}_{\ell m} 
Y_{\ell m}k^{(0)}_{\ell m}=
\sum_{m=-\ell}^\ell 
k^{(0)}_{\ell m}
\mathscr{Y}^{L*}_{\ell m} n^{\langle L\rangle}
\frac{4\pi \ell!}{(2\ell+1)!!}
\mathscr{Y}^{L'}_{\ell m}
\mathcal{E}^{(0)}_{L'}\,,
\ee
we arrive at the following equation
\be
\left(\lambda^{(0)} {}^{i'_1...i'_{\ell}}_{i_1...i_\ell} -
\frac{r_s^{2\ell+1}}{B_\ell }
\frac{4\pi \ell!}{(2\ell+1)!!}
\sum_{m=-\ell}^\ell 
k^{(0)}_{\ell m}
{\mathscr{Y}^*}^{i_1...i_\ell}_{\ell m} 
% n^L
\mathscr{Y}^{i'_1...i'_{\ell}}_{\ell m}
% \mathcal{E}_{L'}
\right)n_{\langle i_1}...n_{i_\ell \rangle} \mathcal{E}^{(0)}_{i'_1...i'_{\ell}}= 0\,,
\ee
where we have restored the explicit tensorial indices.
In what follows we will refer to the constant 
STF tensors $\mathscr{Y}^{L}_{\ell m}
$ as ``Thorne tensors''
\cite{Thorne:1980ru}.
They allow us to write the following expression 
for the scalar response tensor
\be
\label{eq:chifinals0}
{\lambda^{(0)}}^{i'_1...i'_{\ell}}_{i_1...i_\ell} =
\frac{r_s^{2\ell+1}}{B_\ell }
\frac{4\pi \ell!}{(2\ell+1)!!}
\sum_{m=-\ell}^\ell 
k^{(0)}_{\ell m}
{\mathscr{Y}^*}^{i_1...i_\ell}_{\ell m} 
% n^L
\mathscr{Y}^{i'_1...i'_{\ell}}_{\ell m}\,,
% \mathcal{E}_{L'}
% n_{i_1...i_\ell} \mathcal{E}_{i'_1...i'_{\ell}}
\ee
which is valid up 
to terms antisymmetric in $i'_1...i'_{\ell}$ and 
$i_1...i_{\ell}$, and up to a Kronecker delta in 
any combination of $i_{p}$ and $i'_q$.
Note that the expression~\eqref{eq:chifinals0}
does not impose any restrictions on the symmetry 
properties of ${\lambda^{(0)}}^L_{L'}$ w.r.t. multi-index
exchange $L \leftrightarrow L'$. Hence, our matching 
procedure based on the expression~\eqref{eq:match0}
computes both dissipative and conservative responses.

If the Newtonian SRCs $k^{(0)}_{\ell m}$ do not depend 
on the magnetic number $m$, i.e. $k^{(0)}_{\ell m}=k^{(0)}_{\ell}$,  which is the case 
for Schwarzschild black holes, then the 
sum over the STF tensors can be explicitly taken,
$\frac{4\pi \ell!}{(2\ell+1)!!}
\sum_{m=-\ell}^\ell 
{\mathscr{Y}^*}^{L}_{\ell m} 
% n^L
\mathscr{Y}^{L'}_{\ell m}=\delta_{LL'}$. 
In this case we reproduce the expression 
for SLNs obtained in Hui et al.~\cite{Hui:2020xxx}
for the Schwarzschild black holes (upon identification
$\hat{L}\to \ell$ and $D\to 4$)
\be
{\lambda^{(0)}}^L_{L'}=
\lambda_\ell \delta^{L}_{L'}\,,\quad 
\lambda_\ell=
(-1)^\ell \frac{\pi^{1/2}\Gamma(1/2-\ell)}{2^{\ell-2}}
\frac{\Gamma(-2\ell-1)\Gamma(\ell+1)^2}{\Gamma(2\ell+1)\Gamma(-\ell)^2}r_s^{2\ell+1}\,.
\ee
These response coefficients vanish identically 
in the physical case $\ell\in \mathbb{N}$.

Let us now explicitly compute 
the response matrix \eqref{eq:chifinals0}
for the $\ell=1$ and $\ell=2$ sectors.
Using formulas
from Appendix~\ref{app:math}, we obtain the following expression for $\ell=1$
\be
{\lambda^{(0)}}^i_j=\frac{a r_s^2}{12 B_1}
\left(
\begin{array}{ccc}
 0 & 1 & 0 \\
 -1& 0 & 0 \\
 0 & 0 & 0 \\
\end{array}
\right)\,,
\ee
where for simplicity we have retained only the terms 
linear in black hole's spin $a$.
This matrix is antisymmetric, which means that the 
corresponding dipole
worldline coupling vanishes 
because it contracts two gradients of the scalar field.
Hence, the EFT Love tensor is zero in this case
even though the Newtonian response coefficients are not.
Thus, the Kerr black hole's response is purely dissipative.

Now let us consider the quadrupolar sector $\ell=2$. 
The corresponding Love tensor
computed from Eq.~\eqref{eq:chifinals0} takes the following form
\be
\label{eq:lambda0q}
{\lambda^{(0)}}^{ij}_{kl}
=-(4\pi)\frac{\chi M^5}{135}
\left[
4(1-\chi^2)^2I^{(1)}_{ij,kl}
+5\chi^2(1-\chi^2)^2I^{(3)}_{ij,kl}
+\chi^4I^{(5)}_{ij,kl}
\right] \,,
\ee
where we have introduced 
the dimensionless black hole spin $\chi=a/M$
and 
used the following real-valued tensors (defined for any $n$)
\be 
\label{eq:Is}
I^{(2n-1)}_{L,L'}\equiv 
\frac{8\pi \ell!}{(2\ell+1)!!}\sum_{m=-\ell}^\ell
m^{2n-1}\text{Im}(\mathscr{Y}^* {}^{\ell m}_L
\mathscr{Y}^{\ell m}_{L'})
=
\frac{1}{2}\left(
\begin{array}{ccc}
 -2^{2n-1}\textbf{M}_{12}& 2^{2n-1}\textbf{M}_{11} & -\textbf{M}_{23} \\
 2^{2n-1}\textbf{M}_{11}& 2^{2n-1}\textbf{M}_{12} & \textbf{M}_{13} \\
 -\textbf{M}_{23} & \textbf{M}_{13} &\textbf{0} \\
\end{array}
\right)\,,
\ee
which are composed of the STF basis matrices given by~\cite{LeTiec:2020spy,LeTiec:2020bos}
\be 
\label{eq:Ms}
\begin{split}
\textbf{M}_{11}=
\left(
\begin{array}{ccc}
1& 0 & 0 \\
 0& -1 & 0 \\
 0 & 0 &0 \\
\end{array}
\right)\,,\quad 
\textbf{M}_{12}=
\left(
\begin{array}{ccc}
0& 1 & 0 \\
 1& 0 & 0 \\
 0 & 0 &0 \\
\end{array}
\right)\,,\quad 
\textbf{M}_{13}=
\left(
\begin{array}{ccc}
0& 0 & 1 \\
 0& 0 & 0 \\
 1 & 0 &0 \\
\end{array}
\right)\,,
\quad 
\textbf{M}_{23}=
\left(
\begin{array}{ccc}
0& 0 & 0 \\
 0& 0 & 1 \\
 0 & 1 &0 \\
\end{array}
\right)\,,
\end{split}
\ee
and $\textbf{0}$
is a trivial $3\times 3$ matrix.
For small spin the response matrix takes the following simplified form:
\be 
{\lambda^{(0)}}^{ij}_{kl}=-\frac{4\pi}{3}\frac{\gamma r_s^5}{180}I^{(1)}_{ij,kl}+\mathcal{O}(\gamma^2)
=-(4\pi)\frac{4\chi M^5}{135}I^{(1)}_{ij,kl}
+\mathcal{O}(\chi^2)\,.
\ee
We observe that
the tensor ${\lambda^{(0)}}^{ij}_{kl}$
is antisymmetric w.r.t. the upper and lower groups 
of indices, i.e. 
${\lambda^{(0)}}^{ij}_{kl}=-{\lambda^{(0)}}^{kl}_{ij}$.
This means that the local quadrupole 
worldline operator 
vanishes as well. This result can be extended 
to higher order multipoles.

The antisymmetry
of the corresponding matrices stems from the fact
that the scalar response numbers are purely imaginary. 
To see this, let us consider the following general ansatz
for response coefficients
consistent with the reality requirement for the scalar field
multipole
expansion~\eqref{eq:PhiU}~\cite{LeTiec:2020bos}:
\be
\label{eq:ansatz}
k^{(0)}_{\ell m}=k_{\ell 0} + \chi\sum_{n=1}^\infty k_{\ell n}(\chi)
(im)^n\,,
\ee
where $k_{\ell n}$ are real spin-dependent functions. 
Plugging this into Eq.~\eqref{eq:chifinals0} we obtain:
\be 
\label{eq:lgeneric}
\begin{split}
\lambda^{(0)} {}_L^{L'}=\frac{r_s^{2\ell+1}}{A_1^\ell}\Bigg(
k_{\ell 0}\delta^L_{L'}
+\chi\sum_{n=1}^\infty (-1)^n\left[
k_{\ell (2n-1)}I_{L,L'}^{(2n-1)}
+k_{\ell (2n)}R_{L,L'}^{(2n)}
\right]
\Bigg)\,,
\end{split}
\ee
where $I_{L,L'}$ are the antisymmetric tensors (w.r.t. exchange $L\leftrightarrow L'$) introduced 
in Eq.~\eqref{eq:Is}, whereas $R_{L,L'}$ are new fully symmetric 
tensors defined as follows~\cite{LeTiec:2020bos}:
\be 
\label{eq:Rs}
R^{(2n)}_{L,L'}\equiv 
\frac{8\pi \ell!}{(2\ell+1)!!}\sum_{m=-\ell}^\ell
m^{2n}\text{Re}(\mathscr{Y}^* {}^{\ell m}_L
\mathscr{Y}^{\ell m}_{L'})=
\frac{1}{2}\left(
\begin{array}{ccc}
 2^{2n}\textbf{M}_{11}& 2^{2n}\textbf{M}_{12} & \textbf{M}_{13} \\
 2^{2n}\textbf{M}_{12}& -2^{2n}\textbf{M}_{11} & \textbf{M}_{23} \\
 \textbf{M}_{13} & \textbf{M}_{23} &\textbf{0} \\
\end{array}
\right)\,.
\ee
Comparing Eq.~\eqref{eq:ansatz} with Eq.~\eqref{eq:lgeneric} we see that the real part of the response coefficient 
generates a symmetric part of $\lambda^{(0)}$ (i.e. even w.r.t. $L \leftrightarrow L'$), and eventually contributes to
the symmetric Love tensor implying non-trivial
local EFT operators.  
However, the imaginary part of the response coefficients 
generates an antisymmetric part of the response, 
\be
I_{L'}\propto \kappa_{L'L} \mathcal{E}_{L}\,,\quad \kappa_{L'L}=-\kappa_{LL'}\,,
\ee 
which can be identified with a quasi-local contribution
given in Eq.~\eqref{eq:IL}. 
Since the response that we have found is purely imaginary,
we conclude that (a) the Love numbers vanish identically,
(b) the tidal response of Kerr black holes 
to static scalar perturbations is entirely dissipative.
Note that our antisymmetric quadrupolar response tensor~\eqref{eq:lambda0q}
coincides (up to a numerical factor) 
with the tensors that describe the 
black hole's torque 
obtained in Refs.~\cite{LeTiec:2020bos,Goldberger:2020fot}.

Finally, we compare the first corrections
to the source due to 
the graviton interaction. 
We will focus on the first $r_s/r$ correction 
and the first non-trivial spin contribution.
Taylor expanding Eq.~\eqref{eq:Rfinal} we obtain
\be
\label{eq:Phisource}
\Phi
\subset  
r^\ell\left(1-\frac{\ell}{2}\frac{r_s}{r}
+\frac{a^2 m^2 }{(4 \ell-2) r^2}+...
\right)\,,
\ee
which agrees with the EFT 
calculation \eqref{eq:Phisourcea}.

% \newpage 

\section{Spin-1 Response Coefficients}
\label{sec:spin1}

In this section we extend our static response calculation to the spin-1 field
and compute the response of a Kerr
black hole to a long-wavelength electromagnetic
perturbation. 
Similar calculations were done 
for Schwarzschild black holes
in four dimensions in Ref.~\cite{Damour:2009va} 
and in a general number of dimensions in Ref.~\cite{Hui:2020xxx}. We will start with the definition of spin-1 response coefficients and Love numbers.
Then we will compute the electromagnetic field around
the 
Kerr black hole by means of the 
Newton-Penrose formalism~\cite{1962JMP.....3..566N,1963JMP.....4..998N}, and extract the vector response coefficients
from this solution. 
Finally, we will match our general relativity  calculations to the EFT, which
will help us fix the relevant tensor Wilson coefficients.
As in the scalar field case, our matching procedure
will imply the vanishing of the EFT Love tensor, and hence
the spin-1 response will be identified to be purely dissipative.

\subsection{Definition}

\paragraph{EFT Love numbers.}
The local worldline 
EFT for the electromagnetic field 
to zeroth order in metric perturbations
is given by the following action~\cite{Hui:2020xxx}
\be 
\label{eq:eftEB}
\begin{split}
&S^{\rm em}_{\rm EFT}=S_{\rm pp}-\frac{1}{4}\int d^{4}x~F_{\mu \nu}F^{\mu \nu}\\
&
+\sum_{\ell =1}\frac{1}{2\ell!}
\int d^{4}x
\int d\tau \delta^{(4)}(x-x(\tau))
% \left[
{\lambda^{(1)}}^L_{L'}
( \d_{\langle i_1}...\d_{i_{\ell-1}}E_{i_\ell\rangle})
(\d^{\langle i'_1}...\d^{i'_{\ell-1}}E^{i'_\ell\rangle})\\
&+\sum_{\ell =1}\frac{1}{4\ell!}
\int d^{4}x
\int d\tau \delta^{(4)}(x-x(\tau))
% \left[
\tilde{\lambda}^{(1)}{{ }}^L_{L'}
\langle \d_{(i_1}...\d_{i_{\ell-1}}B_{i_{\ell}\rangle j})
(\d^{\langle i'_1}...\d^{i'_{\ell-1}}B^{i'_{\ell}\rangle j})
% \right]
~\,,
% \right]
% \,,
\end{split}
\ee
where $F_{\mu\nu}\equiv 2\d_{[\mu} A_{\nu]} $
is the Maxwell tensor, $A_\mu$ is the $U(1)$ gauge 
potential,
$S_{\rm pp}$ is the usual point-particle action \eqref{eq:Spp}, and we have introduced the electric field vector and the magnetic tensor as
\be 
E_i = F_{0 i}\,,\quad B_{ij}=F_{ij}\,.
\ee
They can be defined in a manifestly covariant way
by means of the body's 4-velocity and the projector 
$P_{\m}^{\nu}=\delta^\n_\m + v_\m v^\n$,
\be 
E_\nu = F_{\mu \nu}v^\mu \,,\quad
B_{\m \nu} = P_{\m}^\s P^\rho_\n F_{\sigma \rho}\,.
\ee 
The Wilson coefficients $\lambda^{(1)}{}^L_{L'}$ and $\tilde{\lambda}{}^{(1)}{}^L_{L'}$
are the electric and magnetic Love tensors,
respectively.
In the static limit
in the particle's rest frame
\be 
E_i = F_{\mu i}v^\mu=-\d_i A_0 \,,
% \quad \text{in the static limit.}
\ee
which implies that
the EFT for the electric field  
only depends on a scalar field 
$A_0$, and effectively it reduces
to the EFT for a massless scalar field
that we have studied in the previous section.
In order to study the magnetic field 
it is convenient to employ the transverse gauge $\d_i A_i=0$,
in which case the  
kinetic term for the electromagnetic 
field takes the following form
in the zero-frequency limit
\be 
-\frac{1}{4}\int d^4x F^{\mu\nu}
F_{\mu\nu}\to 
\frac{1}{2}\int d^4x~\left[(\d_i A_0)^2
-\d_i A_j \d_i A_j \right]\,.
\ee

\paragraph{Definition \`{a} la Newtonian matching.}
It is also useful to introduce a definition of 
electromagnetic
response coefficients
in the way similar to the gravitational potential in the Newtonian approximation.
This will prove convenient 
to extract the response coefficients
from the general relativity solution that we will obtain
in the
harmonic space.
To that end we can use the fact that in the static limit the electric field is 
fixed by $A_0$. 
The equation of motion 
for $A_0$
reduces to the Poisson equation
in the long-distance limit,
\be 
\nabla^2_i A_0= 0\,.
\ee 
In the static limit $A_0$ becomes gauge-independent,\footnote{Indeed, the relevant
$U(1)$ gauge transformation $A_\mu\to A_{\mu}+\d_\mu \alpha$ does not alter $A_0$ in the stationary limit $\d_t\to 0$. } and hence we can use it to define response coefficients just like in the case of the Newtonian
gravitational potential.
Since $A_0$ transforms as a scalar under rotations, it can be written as a series over the scalar 
spherical harmonics\footnote{Recall that $A_0$ and $A_r$ transform as scalars under the 
$SO(3)$ group transformations, whereas the vector $A_a~(a=(\theta,\phi))$ has two distinctive contributions, which transform as a scalar
and as a vector under $SO(3)$~\cite{Hui:2020xxx}. 
}
\be
\label{eq:Loves1}
\begin{split}
& A_0=
% \frac{(\ell-2)!}{\ell !}
\sum_{\ell=1}\sum_{m=-\ell}^\ell Y_{\ell m}\a_{\ell m} r^\ell \left[1
% \left(\frac{r}{r_s}\right)^\ell
% \times\left(1+\frac{r_s}{r}+...\right)
+k^{(1)}_{\ell m}
% r_s^\ell 
\left(\frac{r}{r_s}\right)^{-2\ell-1}
% \left(1+\frac{r_s}{r}+...\right)
 \right]\,,
%  &\d_{[j} A_{i]}=-\frac{(\ell-2)!}{\ell !}\sum_{\ell=2}\sum_{m=-\ell}^\ell \d_{[j}Y^{i]}_{\ell m}b_{\ell m} \left[\left(\frac{r}{r_s}\right)^\ell\times\left(1+\frac{r_s}{r}+...\right)+k_{\ell m}
% \left(\frac{r}{r_s}\right)^{-\ell-1}
% \left(1+\frac{r_s}{r}+...\right)
%  \right]\,,
 \end{split}
\ee 
where $\alpha_{\ell m}$ are source harmonic
coefficients, which satisfy $\alpha_{\ell m}^*=(-1)^m\alpha_{\ell (-m)}$ such that $A_0$ is real.
Hence, we can use an analog of the Newtonian matching 
supplemented with the analytic continuation $\ell \to \mathbb{R}$
to extract electric response coefficients.
Note that we have neglected the background electric monopole contribution
because we consider neutral black holes
in this paper.

One can define the magnetic response using 
an expansion for the vector part of the gauge potential similar to \eqref{eq:Loves1}. It is easiest to do that 
at the level of the angular component of the electromagnetic 
tensor~\cite{Hui:2020xxx}
\be 
\label{eq:Lovemag}
F_{ab}=2\nabla_{[a} A_{b]}=2
% -2
% \frac{(\ell-2)!}{\ell !}
\sum_{\ell=1}\sum_{m=-\ell}^\ell \nabla_{[a}Y^{\rm RW}_{b]~\ell m}
\frac{\beta_{\ell m}}{\sqrt{\ell(\ell+1)}} r^{\ell + 1} \left[1
-\frac{\ell+1}{\ell}
\tilde{k}^{(1)}_{\ell m}
% r_s^\ell
 \left(\frac{r}{r_s}\right)^{-2\ell-1}
% \left(1+\frac{r_s}{r}+...\right)
 \right]\,,
\ee
where $[a,b]$ denotes antisymmetrization, $\beta_{\ell m}$ are magnetic source coefficients, satisfying $\b_{\ell m}^*=(-1)^m\b_{\ell (-m)}$, and $Y^{\rm RW}_{b~\ell m}$ are the 
Regge-Wheeler transverse vector spherical harmonics~\cite{Regge:1957td}, see Appendix~\ref{app:math}
for detail.
The normalization factor $-(\ell+1)/\ell$ is inserted 
for convenience. 
With this factor the multipole expansion for the 
magnetic field $B^i =\frac{1}{2}\epsilon^{ijk}F_{jk}$
and the electric 
field $E^i$ take very similar forms, e.g. for the radial component we have
\be
\left(
\begin{array}{c}
 E^r \\
 B^r \\
\end{array}
\right) =\sum_{\ell=1}\sum_{m=-\ell}^\ell\left(
\begin{array}{c}
\ell \alpha_{\ell m} \\
 \b_{\ell m} \\
\end{array}
\right)~Y_{\ell m}r^{\ell -1}\left(1-\frac{\ell+1}{\ell}
\left(
\begin{array}{c}
 k^{(1)}_{\ell m} \\
 \tilde{k}^{(1)}_{\ell m} \\
\end{array}
\right)\left(\frac{r_s}{r}\right)^{2\ell+1}
\right)\,.
\ee
From this expression our convention becomes 
natural as we expect  
the magnetic and electric response coefficients
to coincide in 4 dimensions due to the electric-magnetic duality~\cite{Hui:2020xxx,Porto:2007qi}, $k_{\ell m}^{(1)}=\tilde k_{\ell m}^{(1)}$.
Hence,
we will focus on the electric field in the main body of the paper and present an explicit calculation of the magnetic response 
in Appendix~\ref{app:mag}.

\subsection{Newman-Penrose Formalism}

To compute $A_\mu$ in the Kerr black hole background, we will work 
within the Newman-Penrose (NP) formalism~\cite{1963JMP.....4..998N,1962JMP.....3..566N}. In this formalism, the electromagnetic 
tensor is represented by 3 complex scalars $\Phi_0,\Phi_1,\Phi_2$ as
\be
F_{\mu \nu}= 2\left[\Phi_1 (n_{[\mu} l_{\nu]}
+m_{[\mu }m^*_{\nu]})
+\Phi_2 l_{[\mu} m_{\nu]}
+\Phi_0 m^*_{[\mu} n_{\nu]}
\right] \quad +\quad \text{c.c.}\,,
\ee
where $l_\m,n_\m,m_\m$ 
($m^*_\m$ is the complex conjugate of $m_\m$) are the 
so-called Newman-Penrose null tetrades. Their explicit 
expressions in the Boyer-Lindquist coordinates are given in
Kinnersley~(1969)~\cite{Kinnersley:1969zza},
    \be
    \label{eq:NPtetr}
    \begin{split}
        \ell^{\mu} &= \left(\frac{\left(r^2+a^2\right)}{\Delta},1,0,\frac{a}{\Delta}\right)\,, \quad 
        n^{\mu} = \left(\frac{r^2+a^2}{2\Sigma},-\frac{\Delta}{2\Sigma},0,\frac{a}{2\Sigma}\right) \\
        m^{\mu} &= \frac{1}{\sqrt{2}\left(r+ia\cos\theta\right)}\left(ia\sin\theta,0,1,\frac{i}{\sin\theta}\right) \,.
        % \quad 
        % \bar{m}^{\mu} = \frac{1}{\sqrt{2}\left(r-ia\cos\theta\right)}\left(-ia\sin\theta,0,1,-\frac{i}{\sin\theta}\right)\,.
    \end{split}
    \ee
The quantities $\Phi_0,\Phi_1,\Phi_2$ will be referred to as Maxwell-Newmann-Penrose (MNP) scalars in what follows.
The components that we will need for the Newtonian matching 
are $F_{0r}$ and $F_{\theta \phi}$. They are given by
\be
\begin{split} 
& F_{0r}=-2\text{Re$\Phi_1$} 
+\frac{a \sin \theta (  a\cos \theta
\text{Re}\Phi_0
+r \text{Im}\Phi_0 )
}{\sqrt{2} 
\Sigma
}
+\frac{a \sin \theta \sqrt{2}( 
a\cos \theta\text{Re}\Phi_2-
r \text{Im}\Phi_2
)
}{
\Delta
}\,,\\
& F_{\theta \phi}= 
2 \text{Im$\Phi $}_1 \left(a^2+r^2\right) \sin \theta 
 +\frac{a r \text{Re$\Phi $}_0 \sin ^2\theta  
\Delta
 }{\sqrt{2} \Sigma}
 -\sqrt{2} a r \text{Re$\Phi $}_2 \sin ^2\theta 
\\
&-\sqrt{2} a^2 \text{Im$\Phi$}_2 \sin ^2\theta  
\cos \theta 
-\frac{a^2 \text{Im$\Phi $}_0 \sin ^2\theta  
\cos \theta  \Delta
}{\sqrt{2} \Sigma }\,,
\end{split}
\ee 
where $\Delta= r^2-2Mr+a^2$, $\Sigma=r^2+a^2\cos^2\theta$.
The quantities $\Phi_0$ and $\tilde\Phi_2$, defined as
\[
\tilde \Phi_2\equiv 
\frac{(r-ia\cos\theta)^2}{(r_+-r_-)^2}\Phi_2\,,
\]
are separable solutions of the Teukolsky equations
for spin weights $s=+1$ and $s=-1$, 
respectively~\cite{Teukolsky:1972my,Teukolsky:1973ha}
(see also Refs.~\cite{1976GReGr...7..959B,1977CzJPh..27..127B,1980PhRvD..22.2933B}).
Assuming that the external source is located
at spatial infinity, the leading asymptotic behaviors 
of these MNP scalars are given by~\cite{Teukolsky:1973ha,1976GReGr...7..959B}
\be
\begin{split}
\Phi_0\sim \Phi_1 \sim \Phi_2\sim ~ r^{\ell-1} 
+ \text{const} \cdot r^{-\ell -2} \,,
\end{split}
\ee
where ``$\text{const}$'' is some calculable 
constant.
Hence, in the asymptotic limit $r\to \infty$ 
both the magnetic and electric response coefficients
can be extracted from a single MNP scalar~$\Phi_1$
\be
 F_{\theta \phi}\Big|_{r\to \infty}=
2 \text{Im$\Phi $}_1 r^2 \sin \theta\,, \quad 
F_{0r} \Big|_{r\to \infty}= -2 \text{Re$\Phi $}_1 \,.
\ee

\subsection{From Maxwell-Newman-Penrose Scalars to Response Coefficients}

In order to compute $\Phi_1$
for the Kerr metric, we will follow
the algorithm proposed in
Bi$\check{\text{c}}$\'ak and Dvo$\check{\text{r}}$\'ak~(1976)~\cite{1976GReGr...7..959B}.
We will use the Maxwell equations 
rewritten in terms of the Newman-Penrose
quantities, which are presented in Appendix~\ref{app:NPmax}.
As a first step, we compute $\tilde \Phi_2$.
In the Kerr background it factorizes as
\be
 \tilde \Phi_2
 =\sum_{\ell=1}^\infty \sum_{m=-\ell}^\ell a_{\ell m}R^{(2)}_{\ell m}(r)~{}_{-1}Y_{\ell m}(\theta,\phi)\,,
\ee 
where ${}_{-1}Y_{\ell m}$
are the spin-weighted spherical harmonics 
with weight $-1$,
and $a_{\ell m}$ are the source-dependent constant harmonic coefficients. Note that 
we do not impose any 
restrictions on 
$a_{\ell m}$ --
they are generic complex numbers because $\Phi_2$ 
is complex.
The radial function 
${}^{(2)}R_{\ell m}(r)$ satisfies the
$s=-1$
Teukolsky equation~\cite{Teukolsky:1972my,Teukolsky:1973ha},
\be 
\label{eq:Teuksm1}
\begin{split}
&\left(-\ell^2-\ell+ \frac{\gamma  m (\gamma  m-i (2 x+1))}{x (x+1)}\right)~R^{(2)}_{\ell m}(x)
% -\frac{a m R(x) \left(\gamma m+i \sqrt{1-4 \gamma^2} s (2 x+1)\right)}{\left(4 a^2-1\right) x (x+1)}
+ x (x+1)R''^{(2)}_{\ell m}(x)=0\,,
\end{split}
\ee
supplemented with the following smoothness boundary condition 
at the horizon ($r\to r_{+}$)~\cite{Teukolsky:1973ha}
\be 
R''^{(2)}_{\ell m}(x)=\text{const}\cdot x^{1+im\gamma}\,,\quad \text{as}\quad x\to 0\,.
\ee
The relevant solution is given by
\be
\begin{split} 
&R^{(2)}_{\ell m}(x)
% =\left(\frac{x}{x+1}\right)^{2i\gamma m}x(x+1){}_2F_1(\ell+2,1-\ell,2-2i\gamma m;1+x)\\
=\left(\frac{x}{x+1}\right)^{+i\gamma m}x(x+1)~{}_2F_1(\ell+2,1-\ell,2+2i\gamma m;-x)\,,
% &\frac{d}{dx}(y_{\ell m})=-(1-2i\gamma m)
% \left(\frac{x}{x+1}\right)^{2i\gamma m}
% {}_2F_1(\ell+1,-\ell,1-2i\gamma m,1+x) 
\end{split} 
\ee
which can be conveniently written as $R_{\ell m}(x)\equiv \left(\frac{x}{x+1}\right)^{-i\gamma m}y_{\ell m}(x)$ with
\be
\label{eq:ydef}
\begin{split} 
% &\,,\\
% &\text{where}\quad 
% =\left(\frac{x}{x+1}\right)^{2i\gamma m}x(x+1){}_2F_1(\ell+2,1-\ell,2-2i\gamma m;1+x)\\
y_{\ell m}\equiv \left(\frac{x}{x+1}\right)^{2i\gamma m}x(x+1)~{}_2F_1(\ell+2,1-\ell,2+2i\gamma m;-x)\,.
% &\frac{d}{dx}(y_{\ell m})=-(1-2i\gamma m)
% \left(\frac{x}{x+1}\right)^{2i\gamma m}
% {}_2F_1(\ell+1,-\ell,1-2i\gamma m,1+x) 
\end{split} 
\ee
Now we can compute the second MNP scalar $\Phi_0$,
which also factorizes in the spherical coordinates
\be
\Phi_0 
=
\sum_{\ell=1}^\infty \sum_{m=-\ell}^\ell
a_{\ell m}
R^{(0)}_{\ell m}(r)~{}_{+1}
Y_{\ell m}(\theta,\phi)\,,
% R_{\ell m}(r)~{}_{+1}Y_\ell^m(\theta,\phi)\,,
\ee
where ${}_{+1}
Y_{\ell m}$ are the spin-weighted spherical harmonics
with weight $s=+1$.
The radial part $
R^{(0)}_{\ell m}$ can be extracted from $
R^{(2)}_{\ell m}$
using the Maxwell equations in the NP formalism (see Appendix~\ref{app:NPmax} for more detail)
\be
\label{eq:R0throughR2}
R^{(0)}_{\ell m}=\frac{2r_{sa}^2}{\ell(\ell+1)}
\left(\frac{d}{dr}+\frac{iam}{\Delta}\right)^2 
R^{(2)}_{\ell m}=\left(\frac{x}{1+x}\right)^{-i\gamma m}
\frac{2}{\ell(\ell+1)}\frac{d^2}{dx^2} y_{\ell m}\,.
\ee
Once $\Phi_0$ and $\Phi_2$
are determined, we can use the remaining two Maxwell equations
in the NP formalism to extract $\Phi_1$. 
This quantity does not fully factorize in the Kerr
background
\be
\label{eq:Phi1final}
\begin{split}
\Phi_1=&\frac{\sqrt{2}(r_+-r_-)^2}{(r-ia\cos\theta)^2}\sum_{\ell=1 }\sum_{m=-\ell}^\ell \frac{a_{\ell m}}{\ell(\ell+1)}\left(\frac{x}{x+1}\right)^{-i\gamma m}\\
&\cdot \Bigg\{
[(\ell+1)\ell]^{1/2}
\Bigg[\frac{(r-ia\cos\theta)}{r_{sa}}\frac{d}{dx}(y_{\ell m})
-y_{\ell m}
\Bigg]{}Y_{\ell m}-ia \sin\theta \frac{d}{dx}(y_{\ell m})\cdot {}_{+1}Y_{\ell m}
\Bigg\}\,,
\end{split} 
\ee
where we have neglected the monopole contribution
that corresponds to a shift of black hole's charge. 
The leading asymptotic 
in the limit $x\to \infty$, which is relevant for 
the Newtonian matching, reads
\be
\label{eq:Phi1exp}
\Phi_1\Big|_{x\to \infty} =\frac{\sqrt{2}}{x^2}
\sum_{\ell=1 }\sum_{m=-\ell}^\ell
\frac{a_{\ell m}}{(\ell(\ell+1))^{1/2}}
% \left(\frac{x}{x+1}\right)^{-i\gamma m}\\&
\cdot 
% \Bigg\{
% [(\ell+1)\ell]^{1/2}
\left[x\frac{d}{dx}(y_{\ell m})
-y_{\ell m}
\right]Y_{\ell m}\,.
\ee
As anticipated, $\Phi_1$ factorizes
in the asymptotic limit $r\to \infty$.
Recall that we are eventually interested in $A_0$, which is related to $\Phi_1$ via 
\[
\d_r A_0=-F_{0r}=2\text{Re}{\Phi_1}\,.\]
At this point we can rewrite $a_{\ell m}$
as,
\be
\label{eq:alphatoa} 
 \frac{2\sqrt{2}}{(\ell(\ell+1))^{1/2}}a_{\ell m}r_{sa}=\left(\alpha_{\ell m} + i  \beta_{\ell m}\right)\,,
\ee
where $\alpha_{\ell m}$ and $\beta_{\ell m}$
satisfy the reality conditions $\alpha^*_{\ell m}=(-1)^m\alpha^*_{\ell (-m)}$ and $\beta^*_{\ell m}=(-1)^m\beta^*_{\ell (-m)}$.
The harmonic coefficients
$\alpha_{\ell m}$ and $\beta_{\ell m}$ capture
electric and magnetic parts of the Maxwell tensor,
respectively.
To obtain $A_0$, we integrate Eq.~\eqref{eq:Phi1exp} as
follows
\be
\label{eq:alm}
\begin{split}
% \Bigg\
 A_0 & =2\text{Re}\int^r dr' \Phi_1(r')= \text{Re}
 \sum_{\ell =1}\sum_{m=-\ell}^\ell \frac{2
 \sqrt{2}a_{\ell m}r_{sa}}{(\ell(\ell+1))^{1/2}}
Y_{\ell m} 
 \int^x \frac{dx'}{x'^2}
 \left[x'\frac{d}{dx'}(y_{\ell m}(x'))
-y_{\ell m}(x')
\right]\\
&=\sum_{\ell =1}\sum_{m=-\ell}^\ell Y_{\ell m}
\alpha_{\ell m}
\frac{y_{\ell m}}{x}  = \sum_{\ell =1}\sum_{m=-\ell}^\ell\alpha_{\ell m}A^{(0)}_{\ell m}(r) ~
Y_{\ell m}\,,
% &\frac{d}{dx}(y_{\ell m})=-(1-2i\gamma m)
% \left(\frac{x}{x+1}\right)^{2i\gamma m}
% {}_2F_1(\ell+1,-\ell,1-2i\gamma m,1+x)
\end{split} 
\ee
where in the last line we used that the sum
$\sum_{\ell m}\alpha_{\ell m}Y_{\ell m} y_{\ell m}$ is 
real, which is a consequence of $\alpha_{\ell m}^*=\alpha_{\ell (-m)}$, 
$Y^*_{\ell m}=(-1)^mY_{\ell (-m)}$,
and $y_{\ell m}^* = y_{\ell (-m)}$ (see Eq.~\eqref{eq:ydef}).
Expanding the radial mode functions at spatial infinity
we obtain
\be
\begin{split}
A^{(0)}_{\ell m}(r) & = \text{const}\cdot 
% \frac{2\sqrt{2}}{(\ell(\ell+1))^{1/2}}
(x+1){}_2F_1(\ell+2,1-\ell,2 + 2i\gamma m,-x)\\
& \xrightarrow[x\to \infty]{} r_{sa}^{\ell}x^{\ell}
\left(1+\frac{\Gamma (-2 \ell-1) \Gamma (\ell+2) 
\Gamma (\ell+2 i m\gamma +1)}{\Gamma (1-\ell) \Gamma (2 \ell+1) \Gamma (-\ell+2i m\gamma )}x^{-2\ell-1}\right) \,. 
\end{split}
\ee
Now we can compare this result with the large-distance
approximation~\eqref{eq:Loves1} and read off the following 
electromagnetic response coefficients
\be
\label{eq:ks1}
\begin{split}
 k^{(1)}_{\ell m} & =\frac{\Gamma (-2 \ell-1) 
 \Gamma (\ell) 
\Gamma (\ell+2 i m\gamma +1)}{\Gamma (-1-\ell) \Gamma (2 \ell+1) \Gamma (-\ell+2i m\gamma )}\left(\frac{r_{sa}}{r_s}\right)^{2\ell+1}\\
% k^{(1)}_{\ell m} =  
& = im\gamma
\frac{(\ell+1)!(\ell-1)!}{(2\ell)!(2\ell+1)!}
 \prod_{n=1}^\ell(n^2+4m^2\gamma^2)
% \frac{\Gamma(-2\ell-1)\Gamma(\ell+1)\Gamma(1+\ell+2im\g)}{\Gamma(2\ell+1)\Gamma(-\ell)\Gamma(-\ell+2im\g)}
\left(\frac{r_{sa}}{r_s}\right)^{2\ell +1 }\\
& =  \frac{im\chi}{2}
\frac{(\ell+1)!(\ell-1)!}{(2\ell)!(2\ell+1)!}
 \prod_{n=1}^\ell(n^2(1-\chi^2)+m^2\chi^2)\,,
\end{split}
\ee 
where we replaced \mbox{$\Gamma(\ell+2)/\Gamma(1-\ell)\to \Gamma(\ell)/\Gamma(-\ell-1)$} in the first line,
then assumed the physical values $\ell \in \mathbb{N}$,  
and finally used 
Eq.~\eqref{eq:G2sinh}.
As in the scalar case, the electromagnetic 
response coefficients are purely imaginary,
which means that the static Love numbers must 
vanish. We will confirm that shortly.

It is instructive to take the limit $\gamma\to 0$
and compare our resulting expression with the 
Schwarzschild black hole
electromagnetic
Love numbers $k_S$ computed in Ref.~\cite{Hui:2020xxx}.
This work defined electromagnetic Love numbers w.r.t. 
the scalar mode $\Psi_S$, defined as
\be 
\Psi_S = \frac{r^2}{\sqrt{\ell(\ell+1)}} \d_r A_0
\ee
% in the limit $r\to \infty$ 
in four dimensions.
This means that
we need to differentiate $A_0$ w.r.t. 
the radial coordinate $r$ to obtain $\Psi_S$,
which produces an additional factor $-(\ell+1)/\ell$ in front 
of the Love number. With this factor taken into account, we have 
\be 
\begin{split}
k_S\equiv-\frac{(\ell+1)}{\ell} k^{(1)}_{\ell m}\Big|_{\gamma = 0}&=
-\frac{(\ell+1)}{\ell}
\frac{\Gamma (-2 \ell-1) \Gamma (\ell+2) 
\Gamma (\ell +1)}{\Gamma (1-\ell) \Gamma (2 \ell+1) \Gamma (-\ell )}\\
&=
\frac{\Gamma (-2 \ell-1) \Gamma (\ell+2) 
\Gamma (\ell)}{\Gamma (1-\ell) \Gamma (2 \ell+1) \Gamma (-\ell -1)}\,,
\end{split}
\ee
where we have used $\Gamma(x)x=\Gamma(x+1)$.
This expression exactly coincides with the electromagnetic 
Love numbers given in Ref.~\cite{Hui:2020xxx}
after the identification $\hat{L}\to \ell$
and $D\to 4$. It vanishes once we take 
the physical limit $\ell\to \mathbb{N}$.

\subsection{Matching to the EFT}

In this section, we perform an explicit matching 
of the worldline point-particle effective field theory
that includes electromagnetism
to the results of the full GR calculation.
This will allow us to extract the electric polarizability operator in the EFT 
from the electric response coefficients that we have previously found 
in this section. 
The calculation of the magnetic susceptibilities 
can be easily performed in the same fashion.
We present this calculation in Appendix~\ref{app:mag}
for completeness.

The calculation of the electric Love numbers
is identical to the scalar field Love number matching.
Introducing an external background source as 
\be 
\bar A_0=\bar{\alpha}_{i_1...i_\ell}x^{i_1}...x^{i_\ell}\,,
\ee
where $\bar{\alpha}_{i_1...i_\ell}$ is an STF tensor,
and solving the equation of motion for $A_0$ just like 
in the scalar field case we obtain
\be 
\label{eq:alm2}
A_0=\sum_{\ell m}\bar{\alpha}_{\ell m}r^\ell Y_{\ell m}
% \left[1
+{\lambda^{(1)}}^L_{L'} \bar{\alpha}_L n^{L'} (-1)^{\ell+1} \frac{2^{\ell-2}}{\pi^{1/2}\Gamma(1/2-\ell)}r^{-\ell-1}
% \right]
\,.
\ee
In principle, we could match directly
Eq.~\eqref{eq:alm} and Eq.~\eqref{eq:alm2} as $A_0$ 
is gauge-independent in the static limit.
However, generally it is more appropriate to match the components of 
the electric tensor, such as $E_{r}$, in order to ensure 
that the result in gauge-independent. 
Acting on Eqs.~\eqref{eq:alm} and Eq.~\eqref{eq:alm2} with one derivative w.r.t. the radial coordinate 
$r$, matching the two results, 
and rewriting 
the sum over the spherical harmonics in terms 
of the Thorne STF tensors we obtain
the following expression for the electromagnetic response matrix
\be
\label{eq:chifinals1}
{\lambda^{(1)}}^{i'_1...i'_{\ell}}_{i_1...i_\ell} =
-\frac{r_s^{2\ell +1}}{B_\ell }
\frac{4\pi \ell!}{(2\ell+1)!!}
\sum_{m=-\ell}^\ell 
k^{(1)}_{\ell m}
{\mathscr{Y}^*}^{i_1...i_\ell}_{\ell m} 
% n^L
\mathscr{Y}^{i'_1...i'_{\ell}}_{\ell m}\,,
% \mathcal{E}_{L'}
% n_{i_1...i_\ell} \mathcal{E}_{i'_1...i'_{\ell}}
\ee
where $B_\ell$ is a constant given in Eq.~\eqref{eq:Al}.
As in the scalar case, we see that the requirement 
that $\lambda^{(1)}{}^{L'}_L$ is even w.r.t. exchange 
$L\leftrightarrow L'$ has disappeared in the 
expression~\eqref{eq:chifinals1}, and hence we can interpret
it as a general expression for response coefficients
that includes conservative and dissipative effects
on the same footing.

Plugging $\gamma=0$, we find that this 
expression reduces the Love 
numbers for the Schwarzschild black holes given in
Eq.~(5.30) of Ref.~\cite{Hui:2020xxx} (upon
identification $\hat L\to \ell,~D\to 4$, and up to a sign),
\be
\begin{split}
& {\lambda^{(1)}}_{L'}^L=\lambda_\ell^{(E)}\delta^L_{L'}\,,\quad \text{where}\\
&\lambda_\ell^{(E)}= (-1)^{\ell+1} 
\frac{\pi^{1/2}\Gamma(1/2-\ell)}{2^{\ell-2}}\frac{\Gamma (-2 \ell-1) \Gamma (\ell+2) 
\Gamma (\ell +1)}{\Gamma (1-\ell) \Gamma (2 \ell+1) \Gamma (-\ell )}
r_{s}^{2\ell+1}\,.
\end{split}
\ee
This Wilson coefficient vanishes for physical values 
of the orbital number $\ell \in \mathbb{N}$.

Now let us get back to the expression for the electromagnetic 
response tensor \eqref{eq:chifinals1}.
As in the scalar case, we see that the electric response tensors are antisymmetric for all $\ell$'s as a result of vanishing 
of the real part of $k^{(1)}_{\ell m}$. 
For instance, in the 
quadrupolar sector we have 
\be 
\begin{split}
{\lambda^{(1)}}^{ij}_{kl}
& = -(4\pi)\frac{\chi M^5}{45}\frac{1}{2}
\left[
4(1-\chi^2)^2I^{(1)}_{ij,kl}
+5\chi^2(1-\chi^2)^2I^{(3)}_{ij,kl}
+\chi^4I^{(5)}_{ij,kl}
\right]\\
% \ee
% \be 
& 
% {\chi^{(1)}}^{ij}_{kl}=
=-\frac{4\pi}{3}\frac{\gamma r_s^5}{120}I^{(1)}_{ij,kl}
+\mathcal{O}(\gamma^2)
=-(4\pi)\frac{2\chi M^5}{45}I^{(1)}_{ij,kl}
+\mathcal{O}(\chi^2)
\,,
\end{split}
\ee
where we used the dimensionless
spin $\chi=a/M$ and took the $\gamma \to 0$
limit in the last line.
The 
STF basis tensors $I^{(1)},I^{(3)},I^{(5)}$ are defined in Eqs.~(\ref{eq:Is},\ref{eq:Ms}). We see that just like 
in the scalar case, the local electromagnetic worldline EFT couplings
vanish even though the imaginary electric 
response coefficients do not. We conclude that the spin-1 response is purely 
dissipative.

\section{Spin-2 Response Coefficients}
\label{sec:spin2}

For completeness, in this section we present 
the computation of 
the static response of Kerr black hole
to the external gravitational perturbation.
This calculation has been discussed in detail in 
Refs.~\cite{LeTiec:2020bos,Chia:2020yla}, and some
important technical results were previously obtained in Refs.~\cite{Poisson:2004cw,Yunes:2005ve}.
Our main novel result here will be an explicit 
matching of the spin-2 Kerr black hole response 
coefficients to the worldline EFT Wilson coefficients along the lines of the previous sections. 

\subsection{Definition}

\paragraph{EFT Love numbers.}
The local worldline EFT of gravitational perturbations 
is built out of various operators constructed 
from the Weyl tensor~\cite{Porto:2016pyg,Hui:2020xxx}.
In four dimensions this tensor has two distinctive
components,
\be 
E_{\mu\sigma}=C_{\mu \nu \s \r}v^\nu v^\r\,,\quad 
B_{\mu\nu\sigma}=P^{\m'}_\mu P^{\n'}_\nu P^{\s'}_\sigma C_{\r \mu' \nu' \s' }v^\r \,.
\ee
In the body's rest frame these components reduce to
\be 
E^{(2)}_{ij}\equiv C_{0i0j}\,,\quad 
B^{(2)}_{ijk}\equiv C_{0ijk}
\ee
Note that magnetic tensor can also be dualized as
$B_{\mu\sigma}=\frac{1}{2}\epsilon_{\mu \alpha \b \nu}
C^{\a \b}_{~~~\s \r}v^\nu v^\r$, but here we will not
do that in order to match the convention
of Ref.~\cite{Hui:2020xxx}.
The most generic quadratic action for $E^{(2)}_{ij}$
and $B^{(2)}_{ijk}$ is given by
\be 
\begin{split}
& S^{\rm grav}_{\rm EFT}=S_{\rm pp}+\int d^{4}x~h\mathcal{D}^2 h\\
& +\sum_{\ell =2}\frac{1}{2\ell !}
\int d^{4}x
\int d\tau \delta^{(4)}(x-x(\tau))
% \left[
{\lambda^{(2)}}_L^{L'} \d_{\langle i_1}...\d_{i_{\ell-2}}
E^{(2)}_{i_{\ell-1} i_{\ell}\rangle}
\d^{\langle i'_1}...\d^{i'_{\ell-2}}
{E^{(2)}}^{i'_{\ell-1} i'_{\ell}\rangle}\\
& +\sum_{\ell =2}\frac{1}{4\ell !}
\int d^{4}x
\int d\tau \delta^{(4)}(x-x(\tau))
% \left[
{\tilde{\lambda}{}^{(2)}}_L^{L'} \d_{\langle i_1}...\d_{i_{\ell-2}}
B^{(2)}_{i_{\ell-1} i_{\ell}\rangle j }
\d^{\langle i'_1}...\d^{i'_{\ell-2}}
{B^{(2)}}^{i'_{\ell-1} i'_{\ell}\rangle j}\,,
\end{split}
\ee
where $\int d^{4}x~h\mathcal{D}^2 h$
denotes the graviton kinetic term, whose explicit expression can be found e.g. in Refs.~\cite{Donoghue:2017pgk,Hui:2020xxx}.
The tensorial Wilson coefficients 
${{\lambda}{}^{(2)}}_L^{L'}$
and 
${\tilde{\lambda}{}^{(2)}}_L^{L'}$
will be referred to as 
the electric and magnetic spin-2
Love tensors, respectively.

\paragraph{Response coefficients in the Newtonian limit.}
The general spin-2 tidal response coefficients
can be related to the harmonic expansion 
of the Newtonian potential in the 
large distance limit.
Their calculation relies on the curvature 
Weyl scalar, defined as
\be
\psi_0 \equiv C_{\a\b\g\delta} l^\a m^\b l^\g m^\delta,
\ee
where $C_{\a\b\g\delta} $ is the Weyl tensor projected
onto the Newman-Penrose null tetrades~\cite{1963JMP.....4..998N,1962JMP.....3..566N}.
In the Newtonian limit the Weyl scalar takes the following form,
\be
\label{eq:phi0_nablas}
 \psi_0 = -2m^i m^j \nabla_i \nabla_jU\,,
\ee
where $U$ is the Newtonian potential, and $\nabla_i$ is covariant derivative
of the 3-dimensional euclidean spacetime.
Plugging the expression for the Newtonian potential 
\eqref{eq:NewU} into Eq.~\eqref{eq:phi0_nablas}, we find
\be
\begin{split}
\label{eq:phi0New0}
\psi_0 \Big|_{r\to \infty}= &
\sum_{\ell =2}^\infty\sum_{m=-\ell}^\ell
\sqrt{\frac{(\ell+2)(\ell+1)}{\ell(\ell-1)}}r^{\ell-2}
\mathcal{E}_{\ell m} \left[
1+k_{\ell m}\left(\frac{r_s}{r}\right)^{2\ell+1}
\right]{}_{+2}Y_{\ell m}(\theta,\phi)\,,
\end{split}
\ee 
where ${}_{+2}Y_{\ell m}$ denotes the $s=+2$ spin-weighted 
spherical harmonics. In the relativistic regime this expression can be generalized
as follows~\cite{Poisson:2004cw,Yunes:2005ve,LeTiec:2020bos}:
\be
\begin{split}
\label{eq:phi0New}
\psi_0 \Big|_{r\to \infty}= &
\sum_{\ell =2}^\infty\sum_{m=-\ell}^\ell
\sqrt{\frac{(\ell+2)(\ell+1)}{\ell(\ell-1)}}r^{\ell-2}
\left(\mathcal{E}_{\ell m} + i\frac{\ell+1}{3}
\mathcal{B}_{\ell m}
\right)\left[
1+k_{\ell m}\left(\frac{r_s}{r}\right)^{2\ell+1}
\right]{}_{+2}Y_{\ell m}(\theta,\phi)\,,
\end{split}
\ee 
where $\mathcal{E}_{\ell m},\mathcal{B}_{\ell m}$ 
are the spherical harmonic
coefficients of the electric-type and magnetic-type tidal
tensors, defined by means of the Weyl tensor as follows:
\be
\begin{split}
&\mathcal{E}_L\equiv \frac{1}{(\ell-2)!}\nabla_{\langle i_3...i_\ell}C_{0|i_1|0|i_2\rangle }\,,\quad \mathcal{B}_L\equiv \frac{3}{2(\ell-2)!(\ell+1)!}
\nabla_{\langle i_3...i_\ell}\epsilon_{jk|i_1}C_{i_2|0jk\rangle }\,.
\end{split} 
\ee
The electric-type tidal tensor is a relativistic generalization
of the Newtonian tidal tensor discussed in Section~\ref{sec:prel}.
Note that the response coefficients are the same for magnetic-type 
and electric-type perturbations as a consequence
of the gravitational electric-magnetic duality,
which takes place for fluctuations around 
Kerr black holes in four dimensions~\cite{Porto:2007qi}.

\subsection{From Weyl Scalar to Response Coefficients}

Eq.~\eqref{eq:phi0New} can be used to extract the 
Newtonian response coefficients from the full general 
relativity calculation. Indeed, the Weyl scalar
$\psi_0$ factorizes in the Kerr background
as~\cite{LeTiec:2020bos}
\be
 \psi_0=\sum_{\ell =2}^\infty\sum_{m=-\ell}^\ell
\sqrt{\frac{(\ell+2)(\ell+1)}{\ell(\ell-1)}}
\left(\mathcal{E}_{\ell m} + i\frac{\ell+1}{3}
\mathcal{B}_{\ell m}
\right)
 R^{s=+2}_{\ell m}(r){}_{+2}Y_{\ell m}(\theta,\phi)\,,
\ee
where the radial function $R^{s=+2}_{\ell m}$
satisfies the following differential equation
\be 
\begin{split}
&\left[\left(-\ell^2-\ell+6+ \frac{\gamma  m (\gamma  m+i 2 (2 x+1))}{x (x+1)}\right)
+3 (2 x+1) \frac{d}{dx}+x (x+1) \frac{d^2}{dx^2}\right]
R^{s=+2}_{\ell m}(x)
=0\,.
\end{split}
\ee
The solution smooth at the black hole horizon
must satisfy the following boundary condition
in the Boyer-Lindquist coordinates~\cite{Teukolsky:1973ha},
\be
 R^{s=+2}_{\ell m}=\text{const}\cdot x^{-2+im\gamma}\quad \text{as}\quad r\to r_+~(x\to 0)\,.
\ee
The desired radial function is given by
\be 
\begin{split}
&R^{s=+2}_{\ell m}=\text{const}\cdot (1+x)^{-im\gamma-2}
x^{im\gamma-2}\, _2F_1(-\ell-2,\ell-1;-1+2 i \gamma m;-x)\,.
\end{split}
\ee 
Taylor-expanding this function at spatial infinity we find 
\be
R^{s=+2}_{\ell m}= x^{\ell -2}r_{sa}^{\ell-2}\left(1+
x^{-2\ell-1}
\frac{\Gamma (-2 \ell-1) \Gamma (\ell-1) \Gamma (\ell +2 \gamma i m+1)}{\Gamma (-\ell-2) \Gamma (2 i \gamma m-\ell)\Gamma (2 \ell +1)}
\right)\,,
\ee
which provides us with the following gravitational 
response coefficients
\be 
\label{eq:s2}
\begin{split}
&k_{\ell m}\equiv k_{\ell m}^{(2)}=
\frac{\Gamma (-2 \ell-1) \Gamma (\ell-1) \Gamma (\ell +2 \gamma i m+1)}{\Gamma (-\ell-2) \Gamma (2 i \gamma m-\ell)\Gamma (2 \ell +1)}\left(\frac{r_{sa}}{r_s}\right)^{2\ell+1}
\,.
\end{split}
\ee 
Note that this expression coincides with Eqs.~(4.29,~4.40) of Ref.~\cite{LeTiec:2020bos}.
For the physical case $\ell \in \mathbb{N}$ we have
\be
\label{eq:ks2v2}
\begin{split}
k^{(2)}_{\ell m} =  
& -im\gamma
\frac{(\ell-2)!(\ell+2)!}{(2\ell)!(2\ell+1)!}
 \prod_{n=1}^\ell(n^2+4m^2\gamma^2)
\left(\frac{r_{sa}}{r_s}\right)^{2\ell +1 }\\
= & -\frac{im\chi}{2}
\frac{(\ell-2)!(\ell+2)!}{(2\ell)!(2\ell+1)!}
 \prod_{n=1}^\ell(n^2(1-\chi^2)+m^2\chi^2)\,,
\end{split}
\ee 
where in the last line we expressed the result in terms
of the dimensionless black hole spin $\chi=a/M$.
Importantly, the spin-2 response coefficients 
are purely imaginary just like their spin-0 and spin-1
counterparts. This means that the tidal spin-2
Love tensors must vanish identically.

\subsection{Matching to the EFT}

We will focus on the electric part
of the spin-2 perturbations captured by \mbox{$E^{(2)}_{ij}\equiv C_{0i0j}$} in what follows.
The calculation of the magnetic part 
can be carried out in a similar fashion.
To match the electric-type Love numbers, it is sufficient
to consider only the following scalar graviton modes
\be
\label{eq:h00gauge}
g_{00}=-1+2h_{00}\,,\quad g_{ij}=\delta_{ij}(1+2\tilde{h}_{00})\,,
\ee  
which corresponds to Newtonian gauge.
In this gauge the gravity kinetic term takes the following form
\be 
\int d^{4}x~h\mathcal{D}^2 h = \frac{1}{16\pi }\int d^4x \left[4 h_{00}\Delta \tilde{h}_{00}-2\tilde{h}_{00}\Delta \tilde{h}_{00}\right]\,.
\ee
The field $\tilde{h}_{00}$ does not appear in $S_{\rm pp}$ at zeroth order
in particle's displacement from the center of mass position.
Thus, in this approximation it 
can be integrated out from the action
by means of its equation of motion $\tilde{h}_{00}=h_{00}$, which gives us
\be 
\int d^{4}x~h\mathcal{D}^2 h = 
\frac{1}{8\pi }\int d^4x  h_{00}\Delta h_{00}\,.
\ee
Then the electric part of the Weyl tensor takes the following form
\be
E^{(2)}_{ij}= - \d_i \d_j  h_{00}\,.
\ee
All in all, in the static limit the EFT takes the same form
as the EFT for a scalar field, modulo a factor 
$4\pi$ in the graviton kinetic term
\be 
\begin{split}
&
S^{\rm grav}_{\rm EFT}=S_{\rm pp}+
\frac{1}{8\pi }
\int d^{4}x~h_{00}\Delta  h_{00}\\
&
+\sum_{\ell =2}\frac{1}{2\ell !}
\int d^{4}x
\int d\tau \delta^{(4)}(x-x(\tau))
% \left[
{\lambda^{(2)}}^{i_1...i_\ell}_{i'_1...i'_\ell} 
(\d_{\langle i_1}...\d_{i_{\ell}\rangle}
h_{00})
(\d^{\langle i'_1}...\d^{i'_{\ell}\rangle }
h_{00})
\,,
\end{split}
\ee
where $S_{\rm pp}$ is the standard point-particle action~\eqref{eq:Spp}.
Repeating the scalar field calculation
for a fixed multipolar index $\ell$, we can easily obtain the following static response
\be 
\label{eq:h00}
% \bar h_{00}
h_{00}
=\sum_{m=-\ell}^\ell
{\bar{\mathcal{E}}
% ^{(E)}
}_{\ell m}r^{\ell}
Y_{\ell m}
% \left(
+
{\lambda^{(2)}}^L_{L'}n^{L'}\bar{\mathcal{E}}_L\cdot 
% \lambda^{(C_E)}_\ell 
(-1)^\ell \frac{8\pi }{2}\frac{2^{\ell-2}}{\pi^{1/2}\Gamma(1/2-\ell)}r^{-\ell-1}
% \right)
\,,
\ee
where 
${\bar{\mathcal{E}}}_{\ell m}$ are
spherical modes of the background source
and $\bar{\mathcal{E}}_L$ is the corresponding 
constant
STF tensor.
In order to be rigorous and ensure that the result for
response coefficients
is gauge-independent, we need to match gauge invariant quantities from both sides. The simplest such quantity is 
the $rr$ component of the electric part of the Weyl tensor~\cite{Hui:2020xxx},
\be 
C_{0r0r}=E^{(2)}_{rr}=-\d^2_r h_{00}\,.
\ee
Taking two derivatives w.r.t. $r$ in Eq.~\eqref{eq:h00}
and in the formula for the Newtonian potential~\eqref{eq:NewU}, and matching the two expressions we obtain
\be
\label{eq:chifinals2}
{\lambda^{(2)}}^{i'_1...i'_{\ell}}_{i_1...i_\ell} =
\frac{2 r_s^{2\ell+1}}{8\pi B_\ell }
\frac{4\pi \ell!}{(2\ell+1)!!}
\sum_{m=-\ell}^\ell 
k^{(2)}_{\ell m}
{\mathscr{Y}^*}^{i_1...i_\ell}_{\ell m} 
% n^L
\mathscr{Y}^{i'_1...i'_{\ell}}_{\ell m}\,,
% \mathcal{E}_{L'}
% n_{i_1...i_\ell} \mathcal{E}_{i'_1...i'_{\ell}}
\ee
where $k^{(2)}_{\ell m}$ are given in Eq.~\eqref{eq:ks2v2},
$B_\ell$ a constant is given in Eq.~\eqref{eq:Al}.\footnote{Note that the same result can be obtained by a direct
matching of $h_{00}$ from~\eqref{eq:h00} 
and the Newtonian potential~\eqref{eq:NewU} 
because our choice of Newtonian gauge~\eqref{eq:h00gauge} 
is precisely the one that reproduces
the Newtonian limit at large distances.}

Plugging $\gamma=0$, we find that 
expression \eqref{eq:chifinals2}
coincides with the Love 
number for the Schwarzschild black holes given in
Eq.~(5.50) of Ref.~\cite{Hui:2020xxx} 
for generic $\ell \in \mathbb{R}$ (upon
identification $\hat{L}\to \ell,~D\to 4$, and modulo the conventional factor $8\pi$). However, this expression vanishes in the physical case $\ell \in \mathbb{N}/\{1\}$,
\be
\begin{split}
& {\lambda^{(2)}}^{ i'_1...i'_{\ell}}_{i_1...i_\ell}=\lambda_\ell^{(C_E)}
\delta^{ \langle i'_1...i'_{\ell}\rangle}_{\langle i_1...i_\ell \rangle}\,,\\
&\lambda_\ell^{(C_E)}= \frac{2}{8\pi }(-1)^\ell 
\frac{\pi^{1/2}\Gamma(1/2-\ell)}{2^{\ell-2}}
\frac{\Gamma (-2 \ell-1) \Gamma (\ell-1) 
\Gamma (\ell +1)}{\Gamma (-\ell-2) \Gamma (-\ell)\Gamma (2 \ell +1)}
r_{s}^{2\ell+1}=0\quad \text{if}\quad \ell \in \mathbb{N}/\{1\}\,.
\end{split}
\ee

Using explicit formulas for the Thorne tensors from Appendix~\ref{app:math}, we obtain the following expression for the EFT quadrupolar worldline tensor coupling in terms of the dimensionless spin parameter $\chi=a/M$: 
\be 
\label{eq:l2}
{\lambda^{(2)}}^{ij}_{kl}
=-
% (8\pi)
\frac{2\chi M^5}{45}
\left[
4(1-\chi^2)^2I^{(1)}_{ij,kl}
+5\chi^2(1-\chi^2)^2I^{(3)}_{ij,kl}
+\chi^4I^{(5)}_{ij,kl}
\right]\,,
\ee
where the 
STF basis tensors $I^{(1)},I^{(3)},I^{(5)}$ are defined in Eqs.~(\ref{eq:Is},\ref{eq:Ms}).
For small spin this expression simplifies as,
\be 
\label{eq:l2small}
{\lambda^{(2)}}^{ij}_{kl}=
-2\frac{4\pi}{3(8\pi)}\frac{\gamma r_s^{5}}{30}I^{(1)}_{ij,kl}
+\mathcal{O}(\chi^2)
=-
% (8\pi)
\frac{8\chi M^5}{45}I^{(1)}_{ij,kl}
+\mathcal{O}(\chi^2)\,.
\ee
Just like in the case of spin-0 and spin-1 perturbations, the worldline finite-size operators vanish
even though the Newtonian response coefficients do not.
Note that our
expressions for the response matrices~(\ref{eq:l2},\ref{eq:l2small})
coincide with those presented in Le Tiec et al.~(2020)~\cite{LeTiec:2020bos} and those obtained in Goldberger et al.~(2020)~\cite{Goldberger:2020fot} (up to a conventional numerical factor). The antisymmetric response captured 
by these matrices is responsible for the dissipative 
effect of tidal torques.

\section{Master Formula for Black Hole Response
Coefficients}
\label{sec:master}

In this section we demonstrate that black hole's response coefficients
for any perturbing boson field can be extracted directly 
from Teukolsky equations for relevant 
Newman-Penrose scalars. Then we will present the response
coefficients for time-dependent perturbations.

\subsection{Static Responses}

An important observation is that
the Kerr black hole response coefficients for all fields
can be extracted directly from the solution 
to the radial Teukolsky equation for a generic spin weight $s$~\cite{Teukolsky:1973ha},
\be 
\label{eq:teuks1}
\begin{split}
\Bigg[s^2+s-\ell^2-\ell+ &
\frac{(\gamma  m)^2 +i\gamma  m s (2 x+1)}{x (x+1)}\\
% -\frac{a m R(x) \left(\gamma m+i \sqrt{1-4 \gamma^2} s (2 x+1)\right)}{\left(4 a^2-1\right) x (x+1)}
&+(s+1) (2 x+1) \frac{d}{dx}+x (x+1) \frac{d^2}{dx^2}\Bigg]R(x)=0\,.
\end{split}
\ee
This solution needs to satisfy the following
boundary condition at the future horizon
\be 
R=\text{const}\times (r-r_{+})^{i\gamma m-s}\,,\quad \text{as}\quad r\to r_+\,,
\ee 
which ensures that the in-falling observes 
sees only the so-called ``non-special'' fields (= fields that are not singular and not identically equal to zero).
Moreover, these boundary condition guarantees that the energy momentum flux
flows strictly into the black hole~\cite{Teukolsky:1973ha,Teukolsky:1974yv,Teukolsky:1972my}.
The desired solution can be easily 
constructed~\cite{1974JETP...38....1S,Page:1976df},
\be 
\label{eq:teuk2}
\begin{split}
&R=\text{const}\cdot (1+x)^{-im\gamma-s}
x^{im\gamma-s}\, _2F_1(-\ell-s,\ell+1-s;1+2 i \gamma m-s;-x)\,.
\end{split}
\ee 
Taylor-expanding this solution at spatial infinity $x\to \infty$ (see Appendix~\ref{app:math} for the relevant analytic continuation formula) we find 
\be
\label{eq:Rexpan}
R=\text{const}\cdot x^{\ell-s}r_{sa}^{\ell}\left(1+
x^{-2\ell-1}
\frac{\Gamma (-2 \ell-1) \Gamma (\ell-s+1) \Gamma (\ell +2 \gamma i m+1)}{\Gamma (-\ell-s) \Gamma (2 i \gamma m-\ell)\Gamma (2 \ell +1)}
\right)\,,
\ee
which provides us with the following dimensionless response coefficients
\be 
\label{eq:Lgen}
\begin{split}
k_{\ell m}^{(s)}&=
\frac{\Gamma (-2 \ell-1) \Gamma (\ell-s+1) \Gamma (\ell +2 \gamma i m+1)}{\Gamma (-\ell-s) \Gamma (2 i \gamma m-\ell)\Gamma (2 \ell +1)}
% r_{sa}^{2\ell+1}
\left(\frac{r_{sa}}{r_s}\right)^{2\ell+1}\\
&=(-1)^{s+1}\frac{im\chi}{2}
\frac{(\ell+s)!(\ell-s)!}{(2\ell)!(2\ell+1)!}
 \prod_{n=1}^\ell(n^2(1-\chi^2)+m^2\chi^2)\,,
\end{split}
\ee 
where 
% $r_{sa}=r_+-r_-=2\sqrt{M^2-a^2}$, $\gamma=a/r_{sa}$, 
$\chi=J/M^2$ is black hole's 
dimensionless spin, and $r_s=2M$ is the Schwarzschild radius.
This expression recovers the scalar response coefficients
for $s=0$ (see Eq.~\eqref{eq:ks0}), the spin-1 (electromagnetic) response coefficients
for $s=1$ (see Eq.~\eqref{eq:ks1}), and the spin-2 (gravitational) response coefficients for $s=2$ (see Eq.~\eqref{eq:s2}).

\subsection{Time-Dependent Responses}

Using the Teukolsky equation, we can
also obtain expressions for non-static responses 
to all orders in frequency, which we
present here for completeness.
Let us consider a generic 
field $\psi$ of spin weight $s$ that factorizes 
in the Kerr background as~\cite{Teukolsky:1972my,Teukolsky:1973ha}
\be
\psi = e^{-i\omega t+im\phi} R(r)S(\theta)\,.
\ee
The functions $R$ and $S$ satisfy the following
frequency-dependent equations
\be 
\label{eq:RTeuk}
\begin{split}
&\Bigg[
% \left(
\frac{\left(\omega  \left(a^2+r^2\right)-a m\right) \left(\omega  \left(a^2+r^2\right)-a m+2 i s (M-r)\right)}{a^2+r (r-2 M)}-a^2 \omega ^2+2 a m \omega -A+4 i r s \omega 
% \right)
\\
&~~~~~~~~~~~~~~~~~~~~~ +\left(a^2+r (r-2 M)\right) \frac{d^2}{dr^2}
+2 (s+1) (r-M)\frac{d}{dr}
\Bigg]R(r)=0\,,\\
&\frac{1}{\sin\theta}
\frac{d}{d\theta}\left(\sin\theta\frac{d S(\theta)}{d\theta}
\right)+\Bigg(
a^2\omega^2\cos^2\theta -\frac{m^2}{\sin^2\theta}
-2a\omega s\cos\theta\\
&~~~~~~~~~~~~~~~~~~~ -\frac{2ms\cos\theta}{\sin^2\theta}
-s^2\cot^2\theta
+s+A-a^2\omega^2-2am\omega \Bigg)S(\theta)=0\,,
\end{split}
\ee
where 
$A$ denotes angular eigenvalues. For small $a\omega$ they are given by 
\be 
\label{eq:Aeig}
A=(\ell-s)(\ell+s+1)-a\omega\frac{2ms^2}{\ell(\ell+1)}+\mathcal{O}(a^2\omega^2)\,.
\ee
This expression provides a rational behind the 
the analytic continuation
$\ell \to \mathbb{R}$: 
the angular eigenvalues are actually always 
non-integer for non-zero frequencies~\cite{1973JETP...37...28S,1974JETP...38....1S,Page:1976df}.

The purely incoming boundary condition for $R$ at the 
black hole horizon has the following form in the Boyer-Lindquist coordinates~\cite{Teukolsky:1973ha,1974JETP...38....1S}
\be 
R=\text{const}\times (r-r_{+})^{iQ-s}\,,\quad \text{as}\quad r\to r_+\,,
\ee
where we have introduced 
\be
Q\equiv \gamma m-\frac{2Mr_+}{r_+-r_-}\omega
=\frac{am-2Mr_+\omega}{r_+-r_-}
\,. 
\ee
The solution of the Teukolsky equation 
at finite frequency that satisfies the purely
incoming boundary condition at the horizon can be obtained in the form of a series
over hypergeometric functions, see Refs.~\cite{Mano:1996vt,Mano:1996mf,Mano:1996gn,Sasaki:2003xr},
\be
\label{eq:ex0}
\begin{split}
R= & e^{ -i\frac{\epsilon x}{\sqrt{1-\chi^2}}} 
 x^{-s-i\epsilon/2 -i\tilde{Q}}(1+x)^{i\epsilon/2 + i\tilde{Q}} 
\Bigg(
\sum_{n=-\infty}^\infty a^\nu_n
\frac{\Gamma(1-s+2iQ)\Gamma(2n+2\nu+1)}{\Gamma(n+\nu+1+2i\tilde{Q})\Gamma(n+\nu+1-s-i \epsilon)}
\\
& \times x^{\nu + n}  {}_2F_1(-n-\nu+2i\tilde{Q},-n-\nu+s+i\epsilon,-2n-2\nu,-x^{-1})
\\
& + \sum_{n=-\infty}^\infty a^{-\nu-1}_n
\frac{\Gamma(1-s+2iQ)\Gamma(2n-2\nu-1)}{\Gamma(n-\nu+2i\tilde{Q})\Gamma(n-\nu-s-i \epsilon)}\\
&\times
x^{-\nu-1+n}
{}_2F_1(-n+\nu+1+2i\tilde{Q},-n+\nu+1+s+i\epsilon,2\nu+2-2n,-x^{-1})
\Bigg)\,,
\end{split}
\ee
where we have used $Q=\tilde{Q}-\epsilon/2$ along with
\be
\label{eq:nudef}
\begin{split}
&\epsilon\equiv r_s \omega\,,\quad \tilde{Q} =  \frac{m \chi - r_s\omega}{2\sqrt{1-\chi^2}}\,,\quad 
\nu =  \ell  
+ \Delta \ell\,,\\
% &~~~~~~+\frac{((\ell+1)^2-s^2)^2}{(2\ell +1)(2\ell+2)(2\ell+3)}
% -\frac{(\ell^2-s^2)^2}{(2\ell-1)2\ell(2\ell+1)}
% \Bigg] +\mathcal{O}(\epsilon^3)\,,\\
&\Delta \ell = 
\frac{\epsilon^2}{2\ell+1}\Bigg[
-2-\frac{s^2}{\ell(\ell+1)}
+\frac{((\ell+1)^2-s^2)^2}{(2\ell +1)(2\ell+2)(2\ell+3)}
-\frac{(\ell^2-s^2)^2}{(2\ell-1)2\ell(2\ell+1)}
\Bigg] +\mathcal{O}(\epsilon^3)\,,
\end{split}
\ee
The quantity $\nu$ is called ``renormalized angular momentum.'' The coefficients $a^\nu_n$ 
and $a^{-\nu-1}_n$ satisfy certain recursion 
relations that are given in Refs.~\cite{Mano:1996vt,Mano:1996mf,Mano:1996gn,Sasaki:2003xr}. In general they depend on $\epsilon$ parametrically and they are suppressed in the low-frequency limit, e.g. $a_n^\nu=\mathcal{O}(\epsilon^{|n|})$ for $n\geq -\ell$.
In what follows we will extract the part of 
the solution that has the desired source 
and response asymptotics at large distances.
We will work at linear order in $\epsilon=r_s \omega$,
in which case we will need only the following coefficients,
\be 
\begin{split}
& a^{-\nu-1}_0=a^{\nu}_0=1\,,\quad a_{-1}^\nu=
a_{1}^{-\nu-1}=i\epsilon\frac{(\ell+s)^2(\ell  - 2i\tilde{Q})}{2\ell^2 (2\ell+1)}\sqrt{1-\chi^2}+\mathcal{O}(\epsilon^2)\,,\\
& a_{1}^\nu=
a_{-1}^{-\nu-1}=i\epsilon\frac{(\ell-s+1)^2(\ell+1 + 2i\tilde{Q})}{2(\ell+1)^2 (2\ell+1)}\sqrt{1-\chi^2}+\mathcal{O}(\epsilon^2)\,.
\end{split}
\ee
The relevant solution that scales 
as $r^{\nu-s}(1+\mathcal{O}(r^{-2\nu-1}))$
at large distances in the small-frequency limit
is given by:
\be
\label{eq:ex}
\begin{split}
& R\Big|_{r\to \infty }
\supset \text{const}\times e^{ -i\frac{\omega r}{1-\chi^2}} 
r^{\nu-s} \Bigg( 1 + 
\varkappa^{(s)}_{\nu m} 
\left(\frac{r_{sa}}{r}\right)^{2\ell+1} \Bigg) 
\Bigg[\frac{\Gamma(2\nu+1)}{\Gamma(\nu+1+2i\tilde{Q})\Gamma(\nu+1-s-i \epsilon)}\\
&-a_1^\nu \frac{\Gamma(2\nu+3)(-1-\nu+2i\tilde{Q})
(-1-\nu+s+i\epsilon)
}{\Gamma(\nu+2+2i\tilde{Q})\Gamma(\nu+2-s-i \epsilon)(-2\nu-2)}\\
&-a_1^\nu(\nu+1)\frac{r_+}{r_{sa}}
\frac{\Gamma(2\nu+3)
}{\Gamma(\nu+2+2i\tilde{Q})\Gamma(\nu+2-s-i \epsilon)}
\Bigg]+\mathcal{O}(\epsilon^2)\,,
% \\
% & + 
% {}_2F_1(\nu+1+2i\tilde{Q},\ell+1+s+i\epsilon,2\nu+1,-x^{-1})\,,
\end{split}
\ee
where we have used the frequency-dependent 
response coefficient $k^{(s)}_{\nu m} \equiv \varkappa^{(s)}_{\nu m}
\left({r_{sa}}/{r_s}\right)^{2\ell+1}$,
\be
\label{eq:kappas}
\begin{split}
 \varkappa^{(s)}_{\nu m}\equiv &
\frac{\Gamma (\nu-s+1 - i\epsilon) 
\Gamma (\nu +2i\tilde{Q}  +1)\Gamma (-2 \nu-1)}{\Gamma (2 \nu +1)\Gamma (-\nu-s- i\epsilon) \Gamma (2i\tilde{Q}-\nu)}
% \times \Bigg(
\\
&\times \Bigg(1-
a^{-\nu-1}_1
\frac{(\nu+2i\tilde{Q})(2\nu+1)}{(\nu-2i\tilde{Q})}
-a_1^\nu \frac{(2\nu+1)(1+\nu-2i\tilde{Q})
}{(\nu+1+2i\tilde{Q})}
% \frac{\Gamma (\nu-s+1 - i\epsilon) 
% \Gamma (\nu +2i\tilde{Q}  +1)(\nu+2i\tilde{Q})(\nu+s+i\epsilon)\Gamma (-2 \nu + 1)}{\Gamma (2 \nu +1)(2\nu)
% \Gamma (1-\nu-s- i\epsilon) \Gamma (1-\nu+2i\tilde{Q})}
% \\
% &-a_1^\nu 
% \frac{\Gamma (\nu-s+1 - i\epsilon)
% \Gamma (\nu +2i\tilde{Q}  +1)\Gamma (-2 \nu-1)}{\Gamma (2 \nu +1)\Gamma (-\nu-s- i\epsilon) \Gamma (2i\tilde{Q}-\nu)}
\\
&+a_1^\nu
\frac{r_+}{r_{sa}}
\frac{2(\nu+1)^2(2\nu+1)
}{(\nu+1+2i\tilde{Q})(\nu+1-s)}
+a^{-\nu-1}_1\frac{r_+}{r_{sa}}
\frac{2\nu^2(2\nu+1)
}{(\nu-2i\tilde{Q})(\nu+s)}\Bigg)
+\mathcal{O}(\epsilon^2)\,.
% \frac{\Gamma (\nu-s+1) 
% \Gamma (\nu +2i\tilde{Q}  +1)\Gamma (-2 \nu+1)}{\Gamma (2 \nu +1)\Gamma (1-\nu-s) \Gamma (2i\tilde{Q}+1-\nu)}
% \frac{\Gamma (\nu-s+1 - i\epsilon)
% \Gamma (\nu +2i\tilde{Q}  +1)\Gamma (-2 \nu-1)}{\Gamma (2 \nu +1)\Gamma (-\nu-s- i\epsilon) \Gamma (2i\tilde{Q}-\nu)}
% \\
% &
% \frac{}{}
% \frac{\Gamma(2\nu+3)}{\Gamma(\nu+2+2i\tilde{Q})\Gamma(\nu+2-s-i \epsilon)}
% \,.
% \frac{}{}
% \frac{}{}
% \Bigg)\,.
% \frac{\Gamma (-2 \nu-1) \Gamma (\nu-s+1 - i\epsilon) 
% \Gamma (\nu +2i\tilde{Q}  +1)}{\Gamma (-\nu-s- i\epsilon) \Gamma (2i\tilde{Q}-\nu)\Gamma (2 \nu +1)} \,,
\end{split}
\ee
For $\omega=0$ the response coefficient
that appears in Eq.~\eqref{eq:kappas} reduces to Eq.~\eqref{eq:Lgen}.
We will expand now this coefficient to linear order 
in $\epsilon$, while retaining black hole's spin 
to all orders.
The $\epsilon\to 0$ limit of Eq.~\eqref{eq:kappas} is 
complicated by the presence of a pole 
in the gamma functions. The ambiguity 
associated with this pole
can be eliminated if we formally consider
$\Delta \ell$ and $\epsilon$
as independent parameters, and use
the expression Eq.~\eqref{eq:nudef}
only after regularizing the singularity.
The presence of the pole also generates a finite logarithmic contribution, see Appendix~\ref{app:near}.
We obtain
\be
\label{eq:ksf}
\begin{split}
 &\varkappa^{(s),~\text{finite}}_{\nu m} = \Bigg[\frac{i \gamma m }{\sinh(2\pi \gamma m)}
 \sinh\left\{2\pi \frac{r_sr_+}{r_{sa}}  (\omega-m\Omega)\right\}
 -2(r_s\omega)\gamma m \ln x\\
 &-\frac{ma\omega(2\ell+1)}{2(\ell+1)\ell} \Bigg(
 \ell^2+\ell+s^2+
2im\gamma \left(2s +\frac{s^2}{\ell(\ell+1)}\right)
 \Bigg)
 +
 (2\ell+1)\gamma m r_{+}\omega 
\Bigg]\\
&\times (-1)^{s}
% \frac{im\chi}{2}
\frac{(\ell+s)!(\ell-s)!}{(2\ell)!(2\ell+1)!}
% \left(\frac{r_{sa}}{r_s}\right)^{2\ell+1}
 \prod_{n=1}^\ell(n^2+4(\gamma m)^2)
 % & -2(-1)^s\epsilon \tilde{Q}\frac{(\ell+s)!(\ell-s)!}{(2\ell)!(2\ell+1)!}
 % % \left(\frac{r_{sa}}{r_s}\right)^{2\ell+1}
 % \left(\prod_{n=1}^\ell(n^2+4\tilde{Q}^2)\right)\times \ln x
 +\mathcal{O}((r_s\omega)^2,(r_s\omega)^2 \ln x)\,,
\end{split} 
\ee
where $\Omega$ is the black hole's angular velocity 
$\Omega\equiv a/(r_+^2+a^2)=a/(2Mr_+)$,
and it is useful to recall that $x=(r-r_+)/r_{sa}$.
The first important observation is that 
if we 
expand the response coefficient at leading
order in black hole's spin
and frequency of the external perturbation, we will find
that 
Eq.~\eqref{eq:ksf}
matches the Newtonian expression~\eqref{eq:ilm} with
the
vanishing static Love number $\lambda_\ell = 0$,
but a non-zero dissipative part,
\be 
k^{(s)}_{\nu m} =i r_s\left(\omega - m\Omega\right)
(-1)^{s}
\frac{(\ell+s)!(\ell-s)!(\ell!)^2}{(2\ell)!(2\ell+1)!}
+\mathcal{O}(\omega \Omega, \omega^2,\Omega^2)\,.
\ee
The dissipative imaginary
response
part vanishes for the locking frequency $\omega= m\Omega$.
Therefore, 
at leading order in black hole's spin
and frequency of the external perturbation 
the Kerr black holes behave like rigidly rotating 
dissipative
spheres.

The second important 
observation is that 
generically the conservative response coefficients 
$\mathcal{O}(m\Omega)$ is not zero. Indeed, at face value, Eq.~\eqref{eq:ksf} implies that the following time-dependent 
worldline operator does not vanish
\be
\int d\tau \mathcal{E}^L \dot{\mathcal{E}}^{L'}\Lambda^{(\omega \Omega)}_{L L'}\,,
\ee
where
the spin-dependent 
coefficient
$\Lambda^{(\omega \Omega)}_{L L'}$ 
% and $\lambda^{(\omega^2)}_{L L'}$
is odd  w.r.t. time reversal.
The third important observation is that for non-zero $\omega$ the tidal response coefficients
exhibit classical renormalization-group running. 
This means that only the logarithmic part of 
the conservative frequency-dependent Love number appearing in 
Eq.~\eqref{eq:ksf} 
is universal 
and independent of the renormalization scheme.
This situation can be contrasted with the $\omega=0$ 
Love numbers,
which do not receive any logarithmic contributions and hence do 
not run with distance~\cite{Porto:2016zng,Chakrabarti:2013lua}.\footnote{This is true in four dimensions. In certain spacetime dimensions the static Love numbers also exhibit renormalization-group running~\cite{Kol:2011vg,Hui:2020xxx}.}
Fourth,  Eq.~\eqref{eq:ex0} can be used to extract 
the Love numbers that depend on the 
the frequency squared.
They will be interesting to compare 
with recent results on frequency-dependent Love numbers given in Ref.~\cite{Poisson:2020vap}.
 We leave this question for future work.

Finally, let us comment on the
near-field approximation, which has been recently
used to compute Love numbers 
from the Teukolsky equation~\cite{Chia:2020yla}.
We present this calculation for a generic spin $s$
perturbation
in Appendix~\ref{app:near}. 
There we show that the near-field
expansion does not exactly map 
onto the small frequency expansion.
As a result, the leading order 
near-field approximation
does not fully capture the $\mathcal{O}(\omega \Omega)$
corrections to the tidal response coefficients.

\section{Discussion and Conclusions}
\label{sec:concl}

In this work we have computed the static
response of Kerr black holes to external 
electromagnetic and scalar perturbations in 
four dimensions. 
This complements the analysis of Refs.~\cite{LeTiec:2020bos,Chia:2020yla},
which have calculated the response of Kerr black holes to spin-2 
(gravitational) perturbations.
Our main results are summarized in the master
formula~\eqref{eq:Lgen}, which displays
the Kerr black hole static response 
coefficients
for a perturbing field with generic integer spin $s$.
Importantly, all responses are purely dissipative, i.e.
the Love numbers for spinning black holes
identically vanish
for spin-0, spin-1, and spin-2 fields to all orders in 
black hole spin.
We have also extended our results to 
leading frequency-dependent 
effects, which also include 
the running of response coefficients, 
see Eq.~\eqref{eq:ksf}.

We have used the gauge-invariant definition of Love numbers
as Wilson coefficients in the 
point-particle effective field theory (EFT).
To that end we have introduced 
local finite-size operators in the EFT 
and have 
extracted the relevant Wilson coefficients
by matching the EFT and full GR calculations.
We have also shown that the EFT allows one to clearly
separate dissipative and conservative responses. 
The key ingredient of our matching procedure is the 
analytic continuation of relevant static response 
solutions to non-integer values of the 
orbital multipole number $\ell$.
We have explicatively shown that this procedure allows one
to extract the response coefficients
from full general relativity solutions
in a coordinate-independent fashion.
Moreover, we have interpreted this procedure 
in the EFT context and have shown that it
helps to separate non-linear gravity corrections 
to perturbing sources (i.e. source-graviton EFT
diagrams)
from corrections generated by
the induced multipole moments. 
Curiously, we have found that the subleading
source corrections exactly cancel
the response part in the advanced Kerr coordinates.
It will be interesting to understand the origin
of this cancellation in the future.

We have demonstrated that 
spinning black holes 
are very similar to the static ones 
from the EFT point of view:
both of them can be described with a single point-particle 
term in the worldline action in the static limit.
Our analysis suggests that the case of spinning black holes
may be useful to understand the vanishing 
of tidal Love numbers in four dimensions and a possible EFT
naturalness
problem related to that.
In particular, we have shown 
that this problem may be addressed at the level of 
the massless scalar field. This is 
a very simplistic model, yet it captures
many qualitative details of Love number 
calculations relevant for both static 
and spinning black holes.
This suggests that the scalar field toy model 
may play an important role in elucidating  
the nature of vanishing of local 
finite-size EFT operators for
black holes.

Our analysis can be extended in multiple ways.
First, one can compute Love numbers and
the relevant Wilson coefficients 
of the point-particle EFT for spinning black holes in spacetime 
dimensions greater than four.
The properties of higher dimensional
spinning black holes are known to differ significantly from their four dimensional counterparts
(see e.g.~\cite{Frolov:2008jr,Caldarelli:2010xz}), 
and hence we can expect interesting consequences for 
response coefficients there.
Second, one can study the relationship
between dissipative spin-0 and spin-1 
response coefficients that we have computed
and the phenomenon of black hole torques along the lines of \cite{1972ApJ...175..243P,LeTiec:2020bos,Goldberger:2020fot}.
Third, it would be interesting to compute the Love 
coefficients for charged spinning Kerr–Newman black holes~\cite{Newman:1965my}.
Fourth, one can carry out a systematic analysis of the frequency-dependent
Love numbers for spinning black holes.
Eventually, it will be important to understand 
if there is an extra symmetry of the Schwarzschild 
and Kerr spacetimes
which makes the conservative static black hole 
response vanish in four dimensions.
We leave these research directions 
for future work.

\textit{Note added.} While this paper was being prepared, 
Refs.~\cite{Chia:2020yla,Poisson:2020vap,Goldberger:2020fot} appeared. These papers have 
some overlap with our work in the interpretation
of dissipative response coefficients. In particular, 
we have independently obtained that the response coefficients
presented 
as ``Love numbers''
in Ref.~\cite{LeTiec:2020bos}
actually
correspond to purely non-conservative effects.

\paragraph{Acknowledgments}

This work is supported in part by the NSF award PHY-1915219 and by the BSF grant 2018068.
MI is partially supported by the Simons Foundation’s Origins of the Universe program.

\appendix

\section{Useful Mathematical Relations}
\label{app:math}

\subsection{Spherical Harmonics}

\paragraph{Scalar Spherical Harmonics.}

We use the following definition for the
(scalar) spherical harmonics
\be 
Y_{\ell m}(\theta,\phi)=\frac{(-1)^{\ell +\frac{|m|+m}{2}}}{2^\ell \ell!}
\left[\frac{2\ell+1}{4\pi}\frac{(\ell-|m|)!}{(\ell +|m|)!}\right]^{1/2}e^{im\phi}(\sin\theta)^{|m|}\left(\frac{d}{d\cos\theta}\right)^{\ell+|m|}(\sin\theta)^{2\ell}\,,
\ee
valid for $\ell\geq 0$, $-\ell<m<\ell$. These harmonics 
obey the following relations
\be
\Delta_{\mathbb{S}^2}Y_{\ell m}=-\ell(\ell+1)Y_{\ell m}\,,
\quad Y^*_{\ell m}(\textbf{n}) =(-1)^m Y_{\ell (-m)}(\textbf{n})\,, \quad \oint_{\mathbb{S}^2}d\Omega~ Y_{\ell m}Y^*_{\ell' m'} =\delta_{\ell \ell'}\delta_{mm'}\,,
\ee 
where $\Delta_{\mathbb{S}^2}$ is the two-sphere Laplacian.

\paragraph{Spin-Weighted Spherical Harmonics.}

One can introduce the following spin $s$-raising 
and spin $s$-lowering operators,
\be
\eth\equiv -\left(\d_\theta +\frac{i}{\sin\theta} \d_\phi
-s\frac{\cos\theta}{\sin\theta} \right) \,,\quad 
\bar{\eth}\equiv -\left(\d_\theta -\frac{i}{\sin\theta} \d_\phi
+s\frac{\cos\theta}{\sin\theta} \right) \,.
\ee
Applying these operators on the 
the
(scalar) spherical harmonics $Y_{\ell m}\equiv {}_{0}Y_{\ell m}$ one can define the spin-weighted 
spherical harmonics for $\ell \geq |s|$,
\be 
\begin{split}
&\eth({}_sY_{\ell m})=+\sqrt{(\ell-s)(\ell+s+1)}
{}_{s+1}Y_{\ell m}\,,\\
&\bar{\eth}({}_sY_{\ell m})=-\sqrt{(\ell+s)(\ell-s+1)}
{}_{s-1}Y_{\ell m}\,.
\end{split}
\ee
These harmonics obey the following relations
\be
% \Delta_{\mathbb{S}^2}{}_sY_{\ell m}=-\ell(\ell+1){}_sY_{\ell m}\,,
% \quad 
{}_sY^{*}_{\ell m}(\textbf{n}) =(-1)^{m+s}{}_{-s}Y_{\ell (-m)}(\textbf{n})\,, \quad \oint_{\mathbb{S}^2}d\Omega~ {}_sY_{\ell m}{}_{s}Y^*_{\ell' m'} =\delta_{\ell \ell'}\delta_{mm'}\,.
\ee 

\paragraph{Transverse Vector Spherical Harmonics.}

The transverse vector analog of scalar spherical harmonics 
% which transform as a vector under $SO(3)$, 
are defined 
as follows
\be
\label{eq:sphervec}
 \vec{Y}^T_{\ell m}\equiv -\frac{1}{\sqrt{\ell(\ell+1)}}\vec{r}\times \vec{\nabla} Y_{\ell m}\,,
 \quad 
 Y^T_{i\ell m}= -\frac{\sqrt{\text{det}g_3}\varepsilon_{ijk}}{\sqrt{\ell(\ell+1)}}x^j\nabla^k Y_{\ell m}\,,
\ee
where $g_3$ is the 3d metric, $\varepsilon_{ijk}$ 
is the three-dimensional Levi-Civita symbol ($\varepsilon_{123}=1$) and $\nabla^k$ is the corresponding covariant derivative.
Note that there is a sign difference 
between our definition and the one adopted in Ref.~\cite{Thorne:1980ru}.
We also stress that $\varepsilon_{ijk}$ denotes the fully-antisymmetric
\textit{symbol}, whereas $\epsilon_{ijk}$ stands for the anti-symmetric \textit{tensor}, 
\be 
\epsilon_{ijk}=\sqrt{g_3}\varepsilon_{ijk}\,,\quad 
\epsilon^{ijk}=\frac{\varepsilon^{ijk}}{\sqrt{g_3}}
\,.
\ee
The transverse spin-1 spherical harmonics are related to the spin-weighted spherical harmonics 
through
\be 
Y^T_{i\ell m}=\frac{i}{\sqrt{2}}({}_{-1}Y_{\ell m}m^i + {}_{+1}Y_{\ell m}{m^*}^i)\,.
\ee
Since $\vec{Y}^T_{\ell m}$ are orthogonal to the radial direction $n_i$, it is also convenient to use their projections onto $\mathbb{S}^2$, known as the Regge-Wheeler (RW) vector spherical harmonics~\cite{Regge:1957td,1978JMP....19.2441S}
\be 
\label{eq:vecspherdef}
{Y_{a}^{\rm RW}}_{\ell m}\equiv \frac{1}{\sqrt{\ell(\ell+1)}}
\sqrt{g_2}~\varepsilon_{ab}g_2^{bc}\nabla_{c}Y_{\ell m}\,,
\ee 
where $a=(\theta,\phi)$, $g^{ab}_2$ is the metric tensor on $\mathbb{S}^2$, $g_2\equiv \text{det}g_2$,
$\nabla_a$ is the covariant derivative
on $\mathbb{S}^2$, and we have introduced the flat-space 2-dimensional Levi-Civita symbol
\be
\varepsilon_{\theta \phi}=-\varepsilon_{\phi \theta}=1 \,,\quad
\varepsilon_{\theta\theta}=\varepsilon_{\phi\phi}=0\,.
\ee
In our conventions the RW and the vector harmonics
defined in Eq.~\eqref{eq:sphervec}
coincide in the orthonormal spherical coordinates basis.
The 2d transverse spherical harmonics satisfy~\cite{Hui:2020xxx,1978JMP....19.2441S}:
\be
\Delta_{\mathbb{S}^2}Y^{\rm RW}_{a \ell m}=-\left(\ell(\ell+1)-1\right)Y^{\rm RW}_{a\ell m} \,,\quad 
\oint_{\mathbb{S}^2}d\Omega~
g^{ab}_2Y^{\rm RW}_{a\ell m}Y^{\rm RW}{}^*_{b\ell' m'} =\delta_{\ell \ell'}\delta_{mm'}\,.
\ee

\paragraph{Symmetric Trace-Free Tensors.}
Finally, instead of the spherical harmonics it may be 
conveneient to use the basis of the symmetric trace-free tensors of rank $\ell$ (``STF-$\ell$ tensors'')~\cite{Thorne:1980ru}. These tensors generate an irreducible representation of $SO(3)$ and hence there exists a one-to-one
mapping between them and the spherical harmoncis. This mapping is realized via 
\be
\label{eq:STFtospher}
Y_{\ell m}={\mathscr{Y}^*}^L_{\ell m} n_{\langle L \rangle}\,,\quad \text{or}\quad  n^{\langle L \rangle}=\frac{4\pi \ell!}{(2\ell+1)!!}
\sum_{m=-\ell}^\ell Y_{\ell m}{\mathscr{Y}}^L_{\ell m}\,,
% \sum_{m=-\ell}^\ell 
\ee
where the constant STF tensors ${\mathscr{Y}}^L_{\ell m}$ satisfy
\be
{\mathscr{Y}}^L_{\ell m}= 
\frac{(2\ell+1)!!} {4\pi \ell!}\oint_{\mathbb{S}^2}d\Omega~
n_{\langle L \rangle}Y^*_{\ell m}\,,\quad 
{\mathscr{Y}}^L_{\ell (-m)}=(-1)^m{\mathscr{Y}^*}^L_{\ell m}\,.
\ee
Since ${\mathscr{Y}}^L_{\ell m}$ tensors form a basis 
for the ($2\ell+1$) dimensional vector space 
of the STF tensors
on $\mathbb{S}^2$, any STF tensor can be expanded over them as 
\be
\mathcal{F}^L=\sum_{m=-\ell}^\ell 
 {\mathscr{Y}^*}^L_{\ell m}\mathcal{F}_{\ell m}\,,\quad 
 \mathcal{F}_{\ell m}=
 \frac{4\pi \ell!}{(2\ell+1)!!}
{\mathscr{Y}}^L_{\ell m}
 \mathcal{F}_L\,.
\ee
All in all, any scalar function on $\mathbb{S}^2$ can be represented as 
\be
F(\theta,\phi)=\sum_{\ell =0}^\infty \sum_{m=-\ell}^\ell
\mathcal{F}_{\ell m} Y_{\ell m}=\sum_{\ell=0}^\infty  \mathcal{F}^L n_{\langle  L\rangle}\,.
\ee
Some other important identities are
\be 
\label{eq:epsY}
\begin{split}
&\varepsilon_{jpq}\mathscr{Y}^*{}^{\ell m}_{p (L-1)}\mathscr{Y}^{\ell m}_{q (L-1)}=-im
\frac{(2\ell+1)!!}{4\pi \ell!\ell} \hat{\xi}^0_j\,,\\
&\varepsilon_{jpq}\mathscr{Y}^*{}^{\ell m}_{p (L-1)}
\mathscr{Y}^{\ell (m+1)}_{q (L-1)}=-i
\frac{(2\ell+1)!!}{4\pi \ell!2\ell} 
\left[2(\ell-m)(\ell+m+1)\right]^{1/2}\hat{\xi}^{-1}_j\,,\\
&\varepsilon_{jpq}\mathscr{Y}^*{}^{\ell m}_{p (L-1)}
\mathscr{Y}^{\ell (m-1)}_{q (L-1)}=i
\frac{(2\ell+1)!!}{4\pi \ell!2\ell} 
\left[2(\ell+m)(\ell-m+1)\right]^{1/2}\hat{\xi}^{+1}_j\,,\\
&\varepsilon_{jpq}\mathscr{Y}^*{}^{\ell m}_{p (L-1)}\mathscr{Y}^{\ell (m+\mu)}_{q (L-1)}=0\quad \text{if}
\quad \mu\neq 0~\text{or}~\pm 1\,,
\end{split}
\ee
where $\hat \xi^0_j=\delta^3_j$, 
$\hat \xi^{\pm 1}_j=\mp (\delta^1_j\pm i \delta^2_j)/\sqrt{2}$.

\subsection{Gamma Function}

The Euler Gamma funciton is defined via 
\be
\Gamma(x+1)=\Gamma(x)x\,. 
\ee
We use several important relations in the main text
\be 
\label{eq:G2sinh}
|\Gamma(1+\ell+b i)|^2=\frac{\pi b}{\sinh(\pi b)}
 \prod_{n=1}^\ell(n^2+b^2) \quad 
 \text{for}\quad \ell\in \mathbb{N}\,,
\ee
as well as $\Gamma(z^*)=\Gamma^*(z)$, and Euler's reflection
formula,
\be
\Gamma(z)\Gamma(1-z)=\frac{\pi}{\sin(\pi z)}\,. 
\ee
We also need the Taylor expansions of the Gamma function
around its poles that correspond to natural values 
of the orbital number $\ell$. To obtain them, we shift the argument of the Gamma function as $\ell\to\ell +\varepsilon$,
$\varepsilon\ll 1$, which yields
\be
\frac{1}{\Gamma(2\ell+1)\Gamma(-2\ell) }
=-2\varepsilon\,,\quad \quad \quad  \frac{1}{\Gamma(-\ell)}=(-1)^{\ell+1}(\ell!)\varepsilon\,.
\ee 
These expressions lead to the following relations
\be
\begin{split}
&\frac{\Gamma(-2\ell-1)}{\Gamma(-\ell)}
=\frac{(-1)^{\ell+1}\ell!}{2(2\ell+1)!}\,,\quad 
\frac{\Gamma(-2\ell-1)}{\Gamma(-\ell-2)}
=\frac{(-1)^{\ell+1}(\ell+2)!}{2(2\ell+1)!}\,,\\
&\frac{\Gamma(-2\ell-1)}{\Gamma(-\ell-1)}
=\frac{(-1)^{\ell}(\ell+1)!}{2(2\ell+1)!}\,,\quad 
\frac{\Gamma(-2\ell)}{\Gamma(-\ell)}
=\frac{(-1)^{\ell}\ell!}{2(2\ell)!}\,.
\end{split}
\ee

\subsection{Gauss Hypergeometric Function}
The classic hypergeometric equation has the following form
\be 
\label{eq:hyper}
x(1-x)y''+(c -(1+a+b)x)y' - ab y = 0\,.
\ee
If $c\neq 0, -1,-2,...$, this equation has the 
following solution in terms of the Gauss hypergeometric function
\be
\label{eq:hyp2}
y = {}_{2}F_1(a,b,c,x)\equiv \sum_{n=0}^\infty \frac{\Gamma(n+a)\Gamma(n+b)}{\Gamma(a)\Gamma(b)}
\frac{\Gamma(c)}{\Gamma(n+c)}\frac{x^n}{n!}
\,. 
\ee
The other indepedent solution of Eq.~\eqref{eq:hyper}
is given by 
\be
y=x^{1-c}{}_{2}F_1(b-c+1,a-c+1,2-c,x)\,. 
\ee
This solution is singular at $z=0$.

If $c=-n$, where $n=0,1,2,...$, the (regular at $x=0$) solution to Eq.~\eqref{eq:hyper} takes the following form
\be
y = x^{1+n}{}_{2}F_1(a+n+1,b+n+1,n+2,x) \,.
\ee
If $a=-n$, $n=0,1,2,...$ and $c=-m$, $m=n,n+1,n+2,...$ the hypergeometric series truncates 
at order $m$. If $a+b-c<0$, the hypergeometric series converges at $|x|=1$. Otherwise it 
generically converges for $|x|<1$ (unless it is a polynomial). 

The hypergeometric function \eqref{eq:hyp2} can be analytically continued 
at $x=\infty$ via 
\be
\begin{split}
& {}_{2}F_1(a,b,c,x)=\frac{\Gamma(c)\Gamma(b-a)}{\Gamma(b)\Gamma(c-a)}(-x)^{-a} {}_{2}F_1(a,a+1-c,a+1-b,x^{-1})\\
& +\frac{\Gamma(c)\Gamma(a-b)}{\Gamma(a)\Gamma(c-b)}(-x)^{-b} {}_{2}F_1(b,b+1-c,b+1-a,x^{-1})\,,
\end{split}
\ee
and around $x=1$ via 
\be
\begin{split}
& {}_{2}F_1(a,b,c,x)=\frac{\Gamma(c)\Gamma(c-a-b)}{\Gamma(c-a)\Gamma(c-b)} {}_{2}F_1(a,b,a+b+1-c,1-x)\\
& +\frac{\Gamma(c)\Gamma(-c+a+b)}{\Gamma(a)\Gamma(b)}(1-x)^{c-a-b} {}_{2}F_1(c-a,c-b,c+1-a-b,1-x)\,.
\end{split}
\ee

\section{Calculation of Maxwell-Newman-Penrose Scalars}
\label{app:NPmax}

In this Appendix we compute 
the stationary 
electromagnetic field around the Kerr black hole
for a source located at spatial infinity.
This is the calculation relevant for the 
extraction of the response coefficients.
In the Newman-Penrose formalism, 
the electromagnetic tensor $F_{\mu \nu}$
is represented in terms of 3 complex scalar functions, 
\be
\Phi_0=F_{\mu \nu} l^\mu m^\nu\,,
\quad 
\Phi_1=\frac{1}{2}F_{\mu \nu} \left(l^\mu n^\nu +
{m^*}^\mu m^\nu \right)\,,\quad 
\Phi_2=F_{\mu \nu}
{m^*}^\mu n^\nu \,,\quad 
\ee
where $l^\mu,n^\mu,m^\mu$ are the NP null tetrades
and $m^*$ is the complex conjugate of $m^\mu$.
In what follows we will use the Boyer-Lindquist 
coordinates, in which the Kinnersley tetrades are given by Eq.~\eqref{eq:NPtetr}.
Instead of the usual scalars $\Phi_0,\Phi_1,\Phi_2$,
it is convenient to work in terms of the rescaled 
scalars,
\be
\tilde{\Phi}_0=\Phi_0\,,\quad  
\tilde{\Phi}_1=\frac{(r-ia\cos\theta)^2}{(r_+-r_-)^2}\Phi_1\,,\quad 
\tilde{\Phi}_2=\frac{(r-ia\cos\theta)^2}{(r_+-r_-)^2}\Phi_2\,.
\ee
The stationary (i.e. $\omega=0$) vacuum Maxwell 
equations take the following form in terms of the NP quantities~\cite{Teukolsky:1972my,Teukolsky:1973ha}:
\begin{subequations}
\begin{align}
\label{eq:NPM1}
&\sqrt{2}r_{sa}^2\left(\d_r+\frac{a}{\Delta}\d_\phi\right)\tilde{\Phi}_1
-(r-ia\cos\theta)\left(\d_\theta + \text{cot}\theta
-\frac{i}{\sin\theta}\d_\phi
\right)\tilde{\Phi}_0+ia \sin\theta \tilde{\Phi}_0=0\,,\\
\label{eq:NPM2}
&\sqrt{2}r_{sa}^2\left(\d_\theta+\frac{i}{\sin\theta}\d_\phi\right)\tilde{\Phi}_1
+(r-ia\cos\theta)\left(\d_r
-\frac{a}{\Delta}\d_\phi
\right)\Delta\tilde{\Phi}_0-\Delta \tilde{\Phi}_0=0\,,\\
\label{eq:NPM3}
&\frac{1}{\sqrt{2}}\left(\d_\theta-\frac{i}{\sin\theta}\d_\phi\right)\tilde{\Phi}_1
-(r-ia\cos\theta)\left(\d_r
+\frac{a}{\Delta}\d_\phi
\right)\tilde{\Phi}_2+\tilde{\Phi}_2=0\,,\\
\label{eq:NPM4}
&\frac{1}{\sqrt{2}}\left(\d_r-\frac{a}{\Delta}\d_\phi\right)\tilde{\Phi}_1
+(r-ia\cos\theta)\left(\d_\theta + \text{cot}\theta
+\frac{i}{\sin\theta}\d_\phi
\right)\frac{\tilde{\Phi}_2}{\Delta}-\frac{ia \sin\theta}{\Delta} \tilde{\Phi}_2=0\,.
\end{align}
\end{subequations}
Teukolsky has shown that $\tilde \Phi_0$ and $\tilde \Phi_2$
factorize in the Kerr background as
\be
 \tilde \Phi_2
 =\sum_{\ell m} a_{\ell m}R^{(2)}_{\ell m}(r)~{}_{-1}Y_{\ell m}(\theta,\phi)\,,\quad 
 \tilde \Phi_0
 =\sum_{\ell m} a_{\ell m}R^{(0)}_{\ell m}(r)~{}_{+1}
 Y_{\ell m}(\theta,\phi)\,,
\ee 
where the radial harmonic $R^{(2)}_{\ell m}$
satisfies Eq.~\eqref{eq:Teuksm1}. Applying the operator 
\[
\d_\theta-i\d_\phi/\sin\theta
\]
to Eq.~\eqref{eq:NPM1} and the operator 
\[
\d_r+a\d_\phi/\Delta
\]
to Eq.~\eqref{eq:NPM3}, we find Eq.~\eqref{eq:R0throughR2},
which means we have obtained both $\tilde{\Phi}_0$ and $\tilde{\Phi}_2$. 
It is important to express the radial function $R^{(0)}_{\ell m}$ as a second derivative over $R^{(2)}_{\ell m}$ 
because we will have to integrate over it 
to get $\tilde{\Phi}_1$.

The calculation of $\tilde{\Phi}_1$ is more intricate as it does not
factorize in $\theta$ and $x$.
The axial symmetry suggests 
the following ansatz
\be\tilde\Phi_1=\sum_{m=-\infty}^{\infty}
\left(\frac{x}{1+x}\right)^{-i\gamma m}e^{im\phi}
~
\tilde \Phi_{1m}(x,\theta)\,.
\ee
Plugging this
into Eq.~\eqref{eq:NPM1} and integrating over $x$ (which is related to the radial coordinate $r$) 
we obtain \eqref{eq:Phi1final} plus an integration
constant, which corresponds to black hole's charge. 
Since we consider the neutral black holes, we put this constant to zero. 

\section{Spin-1 Magnetic Love Numbers}
\label{app:mag}

Due to the presence of magnetic-electric duality
in four dimensions, we have anticipated that 
the electric and magnetic response coefficients would 
coincide in the Kerr background.
In this Appendix we explicitly show it.
To that end, we extract the magnetic response
from the Maxwell-Newman-Penrose scalar $\Phi_1$
and match them to the Wilson coefficients of the magnetic field worldline EFT.

Our first goal is to extract the Newtonian
response coefficients from the Kerr solution using Eq.~\eqref{eq:Lovemag}. We will match one particular component, 
$F_{\theta \phi}$. 
To proceed, we need to simplify the commutator $\nabla_{[a}Y^{\rm RW}_{b]~\ell m}$. 
A crucial observation is that\footnote{It can be shown 
that in general $2\nabla_{[a} \sqrt{g_2} \varepsilon_{b]c}\nabla^c =-\sqrt{g_2}\varepsilon_{ab}\nabla^2$.}
\be
\label{eq:laplac}
\begin{split}
\nabla_{[\theta} Y^{\rm RW}_{\phi]~\ell m}&=  \frac{1}{2\sqrt{\ell(\ell+1)}}
\left[
\nabla_\theta
\left(
 \sqrt{\text{det}g_2}\varepsilon_{\phi \theta}
 g^{\theta \theta}_2\nabla_\theta
 \right)
 -
 \nabla_\phi
\left(
 \sqrt{\text{det}g_2}\varepsilon_{\theta \phi}
 g^{\phi \phi}_2\nabla_\phi
\right)
\right]Y_{\ell m}\\
&=\frac{1}{2\sqrt{\ell(\ell+1)}}
\left[-\sqrt{\text{det}g_2}
\Delta_{\mathbb{S}^2}
\right]Y_{\ell m}=
\frac{\sin\theta}{2}
\sqrt{\ell(\ell+1)}Y_{\ell m}\,.
\end{split}
\ee
Thus, we have\be 
\label{eq:Ftphiexp}
\begin{split}
F_{\theta \phi}&
 =
\sum_{\ell=1}\sum_{m=-\ell}^\ell 
% (\ell(\ell+1))^{1/2}
\sin\theta
 Y_{\ell m}\beta_{\ell m} r^{1+\ell} \left[1-\frac{\ell+1}{\ell}
\tilde{k}^{(1)}_{\ell m}
% r_s^\ell
 \left(\frac{r}{r_s}\right)^{-2\ell-1}
% \left(1+\frac{r_s}{r}+...\right)
 \right]\,.
 \end{split}
\ee
To extract the magnetic Love numbers, we need to compare
this expression with
our formula for $F_{\theta \phi}$ that we have obtained 
by solving for $\Phi_1$. 
We have
\be
\label{eq:Fthetaphi}
\begin{split}
F_{\theta \phi}\Big|_{r\to \infty}
&=2\text{Im}\Phi_1~r^2\sin\theta \\&
=2r^2\sin\theta \frac{\sqrt{2}r_{sa}^2}{r^2}
\text{Im}\sum_{\ell=1}\sum_{m=-\ell}^\ell 
\frac{a_{\ell m}Y_{\ell m}}{(\ell(\ell+1))^{1/2}}\left[x\frac{d}{dx}(y_{\ell m})-y_{\ell m}\right]\Bigg|_{x=(r-r_+)/r_{sa}}\\
&=
% 2\sqrt{2}
% r_{sa}^2
\sin\theta 
% r_{sa}^2
\sum_{\ell =1}\sum_{m=-\ell}^\ell
\beta_{\ell m}
% \sqrt{\ell(\ell+1)}
Y_{\ell m}
% \frac{
% }{(\ell(\ell+1))^{1/2}}
F^{\theta \phi}_{\ell m}(r)\,,
\end{split}
\ee
where in the last line we have used Eq.~\eqref{eq:alphatoa} and we have also introduced the new function
\be
F^{\theta \phi}_{\ell m}=
r_{sa}^{\ell+1}x^{\ell+1}\cdot
\left[1
-x^{-2\ell-1}\frac{\ell+1}{\ell}\frac{ \Gamma (-2 \ell-1) \Gamma (\ell+2) \Gamma (\ell+2 i m\gamma +1)}{ \Gamma (1-\ell) \Gamma (2 \ell+1) \Gamma (2 i m\gamma -\ell)}
\right] \,.
\ee
Note that the sum $\sum_{\ell m}\beta_{\ell m}Y_{\ell m} (xy'_{\ell m}-y_{\ell m})$ is real. 

Matching this with Eq.~\eqref{eq:Ftphiexp} we obtain 
that the magnetic and electric response coefficients 
coincide 
in the Kerr background
\be
\tilde{k}^{(1)}_{\ell m}= k^{(1)}_{\ell m}\,.
\ee

\paragraph{Matching to the EFT} can be done using the
EFT electromagnetic action Eq.~\eqref{eq:eftEB}.
In analogy with the electric field 
we introduce an external background magnetic field source as 
\be 
\begin{split}
\bar A_j &=
% \sqrt{g_3}\epsilon_{jik}x^j \nabla^k=
% % \sqrt{g_3}\epsilon_{ji_1q}
% \bar{\beta}_{qi_2...i_{\ell}}
% % n^p 
% x^{i_1}...x^{i_{\ell}}
% =
\frac{r^{\ell}}{\sqrt{\ell(\ell+1)}}
\sum_{ m=-\ell}^\ell\bar{\beta}_{\ell m} Y^T_{j\ell m}
=-\frac{\sqrt{g_3}\varepsilon_{ji k}x^i \nabla^k}{\ell(\ell+1)}
\sum_{ m=-\ell}^\ell r^{\ell}\bar{\beta}_{\ell m} Y_{\ell m}\\
&=
-\frac{\sqrt{g_3}\varepsilon_{ji k}x^i \nabla^k}{\ell(\ell+1)}
\bar\beta_{i_1...i_\ell}x^{i_1...i_\ell}
\,,
\end{split}
\ee
where $\bar{\beta}_{i_1...i_{\ell}}$ is 
the STF tensor. By construction, our source $\bar A_j$
is manifestly harmonic $\nabla^2\bar A_j=0$
and transverse $\nabla^j \bar A_j=0$
and hence it satisfies the spatial part 
of the Maxwell equations $\nabla^\mu F_{\mu j}=0 $.
Expanding the Maxwell action 
to quadratic order in $A_j$
and solving perturbatively its equation of motion 
with the coupling to the source included, we obtain
\be 
\label{eq:blm}
\begin{split}
&A_j=\sum_{m=-\ell}^\ell \bar{\beta}_{\ell m}r^{\ell} Y^T_{j\ell m}
% \left[1
-\frac{\sqrt{g_3}\varepsilon_{ji'k'}x^{i'}\nabla^{k'}}{\ell(\ell+1)}{\tilde \lambda}^{(1)}{}^{i_1...i_{\ell}}_{i'_1...i'_{\ell}}
\bar{\beta}_{i_1...i_{\ell}} 
n^{\langle i'_1...i'_{\ell}\rangle} (-1)^\ell \frac{2^{\ell-2}}{\pi^{1/2}\Gamma(1/2-\ell)}r^{-\ell-1}
% \right]
\,.
\end{split}
\ee
Now we are in position to match the angular component
of the magnetic tensor $F_{\theta \phi}$.
As a first step we compute the response part 
of the Maxwell tensor $F^{\rm response}_{ab}$ 
from Eq.~\eqref{eq:blm} and use that $2\sqrt{g_3}\varepsilon_{[bjk}x^j\nabla^k\nabla_{a]}
=\sqrt{g_2}\varepsilon_{ab}r \nabla^2_{\mathbb{S}^2}$ in the spherical coordinate basis,
which yields
\be 
F^{\rm response}_{ab}=
-\frac{\sqrt{g_2}\varepsilon_{ab}r \nabla^2_{\mathbb{S}^2}}{\ell(\ell+1)}
{\tilde \lambda}^{(1)}{}^{i_1...i_{\ell}}_{i'_1...i'_{\ell}}
\bar{\beta}_{i_1...i_{\ell}} 
n^{\langle i'_1...i'_{\ell}\rangle} (-1)^\ell \frac{2^{\ell-2}}{\pi^{1/2}\Gamma(1/2-\ell)}r^{-\ell-1}\,.
\ee
This can be compared with Eq.~\eqref{eq:Fthetaphi}, which we rewrite for a single orbital harmonic $\ell$ as follows:
\be 
F_{\theta \phi}=
-\frac{\sqrt{g_2}\varepsilon_{\theta \phi}r\nabla^2_{\mathbb{S}^2}}{\ell(\ell+1)}\sum_{m=-\ell}^\ell\beta_{\ell m}
Y_{\ell m}F^{\theta \phi}_{\ell m}(r)/r\,.
\ee
Now we can rewrite the expression above 
using the Thorne STF tensors and arrive at the anticipated result
\be
\label{eq:chifinals1mag}
\tilde{\lambda}^{(1)}{}^{i'_1...i'_{\ell}}_{i_1...i_\ell} =
\frac{r_s^{2\ell +1}}{B_\ell }
\frac{4\pi \ell!}{(2\ell+1)!!}
\frac{-(\ell+1)}{\ell}
\sum_{m=-\ell}^\ell 
\tilde{k}^{(1)}_{\ell m}
{\mathscr{Y}^*}^{i_1...i_\ell}_{\ell m} 
% n^L
\mathscr{Y}^{i'_1...i'_{\ell}}_{\ell m}\,,
% \mathcal{E}_{L'}
% n_{i_1...i_\ell} \mathcal{E}_{i'_1...i'_{\ell}}
\ee
where $B_\ell$ a constant is given in Eq.~\eqref{eq:Al}.
This expression
coincides with the electric 
response coefficient tensor given in Eq.~\eqref{eq:chifinals1}
up to a factor $(\ell+1)/\ell$.

\section{Comment on the Near-Field Approximation}
\label{app:near}

In this appendix we discuss the validity
of the solution of the frequency-dependent 
radial Teukolsky equation~\eqref{eq:RTeuk}
in the near-field approximation.
We start with the Klein-Gordon 
equation in the Schwarzschild 
background.

\subsection{Scalar Field Example}

The differential equation defining the radial mode function of the scalar field in the Schwarzschild 
background takes
the following form~\cite{1973JETP...37...28S,1974JETP...38....1S}
\be
\label{eq:Rsch}
 x(1+x)R''+ (2x+1)R' +\left[(r_s \omega)^2\frac{(1+x)^4}{x(1+x)} - \ell(\ell+1) \right]R=0\,.
\ee
The near field approximation amounts to replacing~\cite{1973JETP...37...28S,1974JETP...38....1S,Page:1976df}
\be 
\label{eq:NZlead}
\frac{(r_s \omega)^2(1+x)^4}{x(1+x)} \to \frac{(r_s \omega)^2}{x(1+x)}\,.
\ee
The corrections to the r.h.s. term are small as long as 
\be
(r_s \omega) x\ll (\ell+1) \,.
\ee
The near-zone approximation can be systematically formulated by an introduction 
of a formal expansion parameter $\alpha$ such that
\be \label{eq:x4}
(1+x)^4\to 
(1+\alpha x)^4 =  1 + \mathcal{O}(\alpha x)\,.
\ee
The final expressions have to be evaluated at $\alpha=1$.
Then the corrections beyond Eq.~\eqref{eq:NZlead}
can be systematically computed order-by-order in $\alpha$.
However, we note that this does not correspond to a 
low-frequency expansion with the small parameter $r_s\omega \ll 1$.
Indeed, sufficiently far from the horizon, 
i.e. for $x= \mathcal{O}(1)$, we have 
\be
 (r_s \omega)^2\sim (r_s \omega)^2 x^2 \,,
\ee
and hence keeping the term $(r_s \omega)^2$ while 
neglecting the term $(r_s \omega)^2 x^2$ in Eq.~\eqref{eq:NZlead}
is not justified by the smallness of  $r_s\omega$.

Now let us write Eq.~\eqref{eq:Rsch}
in the zeroth order near zone approximation ($\alpha x \ll 1$),
\be 
 x(1+x)R''+ (2x+1)R' +\left[\frac{(r_s \omega)^2}{x(1+x)} - \ell(\ell+1) \right]R=0\,.
\ee
The solution consistent with the purely 
incoming boundary condition 
at the horizon is given by 
\be
R=\text{const}\cdot \left(\frac{x}{1+x}\right)^{i\omega r_s} {}_2F_1(\ell+1,-\ell,1+2i r_s\omega,-x) \,.
\ee
If we now analytically continue 
this solution for $x\gg 1$, we will obtain,
\be 
\label{RexpanSch}
\begin{split}
&R=
  \text{const}\cdot \left(\frac{x}{1+x}\right)^{i\omega r_s} \Bigg(\frac{\Gamma(1+2ir_s\omega)\Gamma(2\ell+1)}{\Gamma(\ell+1)\Gamma(1+\ell+2ir_s\omega)}
x^{\ell}\cdot
{}_2F_1\left(-\ell,-\ell-2ir_s\omega,-2\ell,-x^{-1}\right) \\
& +
\frac{\Gamma(1+2ir_s\omega)\Gamma(-2\ell-1)}{\Gamma(-\ell)\Gamma(-\ell+2ir_s\omega)}
x^{-\ell-1}\cdot
{}_2F_1\left(\ell+1,\ell+1-2ir_s\omega,2\ell+2,-x^{-1}\right)\Bigg)\\
&
\xrightarrow[x\to \infty]{}
% \overrightarrow{x\to \infty }
% \quad \frac{\Gamma(1+2im\g)\Gamma(2\ell+1)}{\Gamma(\ell+1)\Gamma(1+\ell+2im\g)}
r_{s}^\ell
x^{\ell}\left(
1 + 
k_{\ell }^{(0)\text{NF}}
% \frac{\Gamma(1+2im\g)\Gamma(-2\ell-1)}{\Gamma(-\ell)\Gamma(-\ell+2im\g)}
x^{-2\ell-1}\right) 
% \quad \text{at} \quad x\to \infty 
\,,
\end{split}
\ee
where the coefficient $k_{\ell }^{(0)\text{NF}}$
might be interpreted as a frequency-dependent 
response coefficient,
\be
\label{eq:komega}
k_{\ell }^{(0)\text{NF}}\equiv \frac{\Gamma(-2\ell-1)\Gamma(\ell+1)\Gamma(1+\ell+2ir_s\omega)}{\Gamma(2\ell+1)\Gamma(-\ell)\Gamma(-\ell+2ir_s\omega)} \,.
\ee
Indeed, for $\omega = 0$ this expression reproduces
the scalar Love number for the
Schwarzschild black hole.
However, strictly speaking, we cannot use Eq.~\eqref{eq:komega}
for the Love number matching 
because there may be other frequency-dependent contributions
that have been omitted in the near zone approximation.
This will be shown shortly when 
we compare Eq.~\eqref{eq:komega}
with the accurate solution
to the Teukolsky equation.

\subsection{Teukolsky Equation in the Near-Field
Approximation}

Now we compute
the solution of the frequency-dependent 
radial Teukolsky equation~\eqref{eq:RTeuk}
for the mode function $R$ in the 
potential region (near-field zone), characterized by 
\be 
x(r_+-r_-)\ll (\ell+1)/\omega\,.
\ee
for a rotating black hole and perturbation
of a generic spin $s$.
We additionally expand over
the small parameter
$\omega M\sim \omega r_s\ll 1$.
Using that $\omega a\leq \omega M \ll 1$, we can approximate $A=(\ell-s)(\ell+s+1)$ and
hence
the radial differential equation can be written in the following simple form~\cite{1974JETP...38....1S,Page:1976df},
\be 
\label{eq:teuknf}
\begin{split}
&\Bigg[s^2+s-\ell^2-\ell+ 
\frac{Q^2 +isQ (2 x+1)}{x (x+1)}
+(s+1) (2 x+1) \frac{d}{dx}+x (x+1) \frac{d^2}{dx^2}\Bigg]R(x)=0\,.
\end{split}
\ee 
This is the same equation as \eqref{eq:teuks1}, but 
with $\gamma m$ replaced by $Q$.
Hence, the required solution is given by Eq.~\eqref{eq:teuk2}
with $\gamma m \to Q$. If we now formally
analytically continue this solution for $x>1$, 
we can obtain the following expression
for response coefficients
\be 
\label{eq:Lgent}
\begin{split}
&k_{\ell m}^{(s)\text{NF}} =
\frac{\Gamma (-2 \ell-1) \Gamma (\ell-s+1) \Gamma (\ell +2 i Q+1)}{\Gamma (-\ell-s) \Gamma (2 i Q-\ell)\Gamma (2 \ell +1)}
% r_{sa}^{2\ell+1}
\left(\frac{r_{+}-r_{-}}{r_s}\right)^{2\ell+1}\\
% \frac{\Gamma (-2 \ell-1) \Gamma (L-s+1) \Gamma (L+2 i a m+1)}{\Gamma (2 L+1) \Gamma (-s-L) \Gamma (-L+2 i a m)}
&=(-1)^{s+1}\frac{i}{2}\left(m\chi-2 r_+ \omega \right)
\frac{(\ell+s)!(\ell-s)!}{(2\ell)!(2\ell+1)!}
 \prod_{n=1}^\ell\left[n^2(1-\chi^2)+\left(m\chi-2r_+ \omega \right)^2\right]\\
 &=(-1)^{s+1}ir_+ \left(m\Omega- \omega \right)
\frac{(\ell+s)!(\ell-s)!}{(2\ell)!(2\ell+1)!}
 \prod_{n=1}^\ell\left[n^2(1-\chi^2)+
 4r_+^2\left(m\Omega- \omega \right)^2\right]\,.
\end{split}
\ee 
where we used black hole's angular velocity 
$\Omega\equiv a/(r_+^2+a^2)=a/(2Mr_+)$.
We can see that this expression is not invariant under time reversal transformations $\omega \to -\omega$,
$m\to -m$, which implies that the near-field 
response is purely 
dissipative. However,
this result is uncertain up to other frequency-dependent corrections.
To estimate these corrections, let us 
use the perturbed angular eigenvalues \eqref{eq:Aeig}
instead of the usual ones. 
We have: 
\be
\tilde{\nu}=\ell+\Delta \ell=\ell+a\omega\frac{2ms^2}{\ell(\ell+1)(2\ell+1)}+... 
\ee
Now we can easily find a solution to Eq.~\eqref{eq:teuknf} with $\ell$ replaced by $\nu$. 
It is given by 
\be
R\Big|_{x\to\infty}=\text{const}\times 
x^{\tilde{\nu}}\left(1+
x^{-2{\tilde{\nu}}-1}\frac{\Gamma (-2 {\tilde{\nu}}-1) \Gamma ({\tilde{\nu}}-s+1) \Gamma (\nu +2 i Q+1)}{\Gamma (-{\tilde{\nu}}-s) \Gamma (2 i Q-{\tilde{\nu}})\Gamma (2 {\tilde{\nu}} +1)}
\right)\,.
\ee
Therefore, the relevant response coefficient 
reads
\be 
\begin{split}
k_{\ell m}^{(s)\text{NF}} =
\frac{\Gamma (-2 {\tilde{\nu}}-1) \Gamma ({\tilde{\nu}}-s+1) \Gamma ({\tilde{\nu}} +2 i Q+1)}{\Gamma (-{\tilde{\nu}}-s) \Gamma (2 i Q-{\tilde{\nu}})\Gamma (2 {\tilde{\nu}} +1)}\left(\frac{r_{sa}}{r_s}\right)^{2\ell+1}
(1-2\Delta \ell \ln x)\,,
\end{split}
\ee
where the logarithm comes from the Taylor expansion
of $x^{-2{\tilde{\nu}}-1}$,
\[
x^{-2{\tilde{\nu}}-1}=x^{-2\ell-1}(1-2\Delta \ell \ln x)\,.
\]
After some simplifications we obtain
\be
\begin{split}
k_{\ell m}^{(s)\text{NF}}=&
(-1)^{s+1} \sin(2i Q \pi -\Delta \ell \pi)
 \frac{2i Q}{\sin(2i Q \pi)}
\frac{(\ell-s)!(\ell+s)!}{2(2\ell)!(2\ell+1)!}
 \prod_{n=1}^\ell (n^2+4 Q^2) \\
 &\times (r_{sa}/r_s)^{2\ell+1}(1-2\Delta \ell \ln x) +\mathcal{O}(\Delta \ell^2)\,.
 \end{split}
\ee
For $\Delta \ell = 0$ the response coefficients reduce to Eq.~\eqref{eq:Lgent}.
However, we can see that the near field approximation
misses
$\mathcal{O}(\omega a)$
and $\mathcal{O}(\omega a \ln x)$ corrections.
This can be confirmed by an 
explicit comparison with a solution obtained 
in a small-frequency expansion
of the Teukolsky equation~\cite{Mano:1996vt,Mano:1996mf,Mano:1996gn,Sasaki:2003xr}.

\subsection{Comparison with the Low-Frequency Solution}

The systematic treatment of the 
Teukolsky equation in the
low-frequency limit~\cite{Mano:1996vt,Mano:1996mf,Mano:1996gn,Sasaki:2003xr}
gives a solution which is somewhat different 
from the near-field expression, c.f. Eq~\eqref{eq:Lgent}
and Eq.~\eqref{eq:ksf}. Importantly, the relevant 
response coefficients are not purely imaginary in this case.
To see this, let us expand Eq.~\eqref{eq:ex}
to linear order in $\epsilon$, while keeping
all powers of
$\tilde Q$. 
Because of the presence of the simple pole at $\epsilon=0$, is important that we also expand 
the renormalized angular momentum $\nu = \ell+\Delta \ell$,
where $\ell$ is an integer number satisfying $\ell\geq |s|$,
and $\Delta \ell =\mathcal{O}(\epsilon^2)$.
We have 
\be
\begin{split}
&\frac{\Gamma (-2 \nu-1)}{\Gamma (2 \nu+1)} =
\frac{1}{2\Delta \ell (2\ell+1)!(2\ell)!}+\mathcal{O}(\epsilon^0)\,,\\
&\frac{\Gamma(\nu+1-i\epsilon-s)}{\Gamma(-\nu-s-i\epsilon)}=(-1)^{\ell+s+1}
 (i\epsilon+\Delta \ell)(\ell-s)!(\ell+s)!+\mathcal{O}(\epsilon^3)\,,\\
 &\frac{\Gamma(\nu+2i \tilde Q+1)}{\Gamma(-\nu+2i\tilde Q)}
 =(-1)^\ell \sin(2i\tilde Q \pi -\Delta \ell \pi)
 \frac{2i\tilde Q}{\sin(2i\tilde Q \pi)}\prod_{n=1}^\ell (n^2+4\tilde Q^2)
 +\mathcal{O}(\epsilon^4)\,,
\end{split}
\ee
This gives 
\be 
\label{eq:vkall}
\begin{split}
&\varkappa^{(s)}_{\nu m}=\Bigg[
\frac{-2\tilde Q \epsilon}{\Delta \ell}
+\frac{2 i\tilde Q}{\sinh(2 \tilde Q \pi)}
\left(
\sinh(2\pi \tilde Q)-\epsilon \pi \cosh(2\pi \tilde Q)
\right)
-\Delta \ell 
\frac{(2\pi \tilde Q)\cosh(2\pi \tilde Q)}{\sinh(2\pi \tilde Q)}
\Bigg]\\
&\times (-1)^{s+1}
\frac{(\ell-s)!(\ell+s)!}{2(2\ell+1)!(2\ell)!}
\left(\prod_{n=1}^\ell (n^2+4\tilde Q^2) \right)\,.
\end{split}
\ee
We see that our response coefficient has a pole at $\Delta \ell = 0$.
When we match the EFT result to the GR calculation, we use
only finite parts, and hence this singular contribution
can be ignored.\footnote{It is also worth stressing that 
the full GR solution is regular,
the pole in Eq.~\eqref{eq:vkall} in fact gets canceled
by a similar singularity that is contained in the source series.
To see this, we have to get back to
the original solution \eqref{eq:ex} and  
regularize the hypergeometric function
that is attached to the source solution $\propto x^\nu$ as follows
\be 
\begin{split}
&{}_2F_1(-\nu+2i\tilde{Q},-\nu+s+i\epsilon,-2\nu,-x^{-1})
\\
&=(-x)^{-2\ell-1}
\frac{\Gamma(\ell+1+2i\tilde{Q})\Gamma(\ell+s+1+i\epsilon)
}{\Gamma(2\ell+2)\Gamma(-\ell+2i\tilde{Q})\Gamma(-\ell+s+i\epsilon)}
\frac{-1}{(2\ell)!2\Delta \ell}\\
&\times
{}_2F_1(\ell+1+2i\tilde{Q},\ell+s+1+i\epsilon,2\ell+2,-x^{-1})
+\mathcal{O}(\epsilon^0)\,,
\end{split}
\ee
which exactly cancels the divergence that we have encountered 
in the term $x^{-2\nu-1}\varkappa^{(s)}_{\nu m}$, 
see Eq.~\eqref{eq:vkall}.} 
However, it is important to note 
that this term also generates a finite logarithmic contribution,
\be 
\begin{split}
 \varkappa^{(s)}_{\nu m}x^{-2\nu-1}&=
\varkappa^{(s)}_{\nu m}
x^{-2\ell-1}(1-2\Delta \ell \ln x+\mathcal{O}(\epsilon^2))\\
&=x^{-2\ell-1}\Bigg[
4 \tilde Q \epsilon \ln x
+\frac{2 i\tilde Q}{\sinh(2Q\pi)}
\left(
\sinh(2\pi \tilde Q)-\epsilon \pi \cosh(2\pi \tilde Q)
\right)
% -\delta \ell 
% \frac{(2\pi \tilde Q)\cosh(2\pi \tilde Q)}{\sinh(2\pi \tilde Q)}
\Bigg]\\
&~~~~\times (-1)^{s+1}
\frac{(\ell-s)!(\ell+s)!}{2(2\ell+1)!(2\ell)!}
\left(\prod_{n=1}^\ell (n^2+4\tilde Q^2) \right) 
+\mathcal{O}(\epsilon^2)
\,.
\end{split}
\ee
Using that 
\be
 \sinh(2\pi \tilde Q)-\epsilon \pi \cosh(2\pi \tilde Q)=
 \sinh(2\pi \tilde Q -\epsilon \pi )+\mathcal{O}(\epsilon^2)\,,
\ee
we obtain the first correction in Eq.~\eqref{eq:ksf}. We see that this expression
coincidently matches the near-zone result Eq.~\eqref{eq:Lgent} 
at linear order 
in $\omega$ and zeroth order in $\omega \Omega$. 
However, the near-field approximation does not correctly 
capture $\mathcal{O}(\epsilon^2,\epsilon \Omega)$ corrections
and their logarithmic running.

Extracting the other contributions 
from Eq.~\eqref{eq:kappas} is straightforward. 
Collecting everything together we arrive at Eq.~\eqref{eq:ksf}.

\bibliographystyle{JHEP}
\bibliography{short}
\end{document}